\documentclass[12pt]{iopart}

\usepackage{graphicx}

\newcommand{\vc}[1]{\textbf{\em #1}}
\newcommand{\pder}[2]{\frac{\partial #1}{\partial #2}}

\begin{document}
\title[Turbulence in collisionless plasmas]{Turbulence in collisionless plasmas: statistical analysis from numerical simulations with pressure anisotropy}

\author{Grzegorz Kowal$^{1,2,3}$, D A Falceta-Gon\c calves$^{3,4}$ and A Lazarian$^3$}
\address{$^1$Instituto de Astronomia, Geof\'\i sica e Ci\^encias Atmosf\'ericas, Universidade de S\~ao Paulo, Rua do Mat\~ao 1226, 05508-900, S\~ao Paulo, Brazil}
\address{$^2$ Obserwatorium Astronomiczne, Uniwersytet Jagiello\'nski, Orla 171, 30-244 Krak\'ow, Poland}
\address{$^3$ Department of Astronomy, University of Wisconsin, 475 North Charter Street, Madison, WI 53706, USA}
\address{$^4$N\' ucleo de Astrof\' isica Te\' orica, Universidade Cruzeiro do Sul, Rua Galv\~ ao Bueno 868, 01506-000, S\~ao Paulo, Brazil}
\ead{kowal@astro.iag.usp.br}

\begin{abstract}
In the past years we have experienced an increasing interest in understanding of
the physical properties of collisionless plasmas, mostly because of the large
number of astrophysical environments, e.g. the intracluster medium (ICM),
containing magnetic fields which are strong enough to be coupled with the
ionized gas and characterized by densities sufficiently low to prevent the
pressure isotropization with respect to the magnetic line direction.  Under
these conditions a new class of kinetic instabilities arises, such as firehose
and mirror ones, which were extensively studied in the literature.  Their role
in the turbulence evolution and cascade process in the presence of pressure
anisotropy, however, is still unclear.  In this work we present the first
statistical analysis of turbulence in collisionless plasmas using three
dimensional double isothermal magnetohydrodynamical with the Chew-Goldberger-Low
closure (CGL-MHD) numerical simulations.  We study models with different initial
conditions to account for the firehose and mirror instabilities and to obtain
different turbulent regimes.  We found that the CGL-MHD subsonic and supersonic
turbulence show small differences comparing to the MHD models in most of the
cases.  However, in the regimes of strong kinetic instabilities the statistics,
i.e., the probability distribution functions (PDF) of density and velocity are
very different.  In subsonic models the instabilities cause an increase in the
dispersion of density, while the dispersion of velocity is increased by a large
factor in some cases.  Moreover, the spectra of density and velocity show
increased power at small scales explained by the high growth rate of the
instabilities.  Finally, we calculated the structure functions of velocity and
density fluctuations in the local reference frame defined by the direction of
magnetic lines.  The results indicate that in some cases the instabilities
significantly increase the anisotropy of fluctuations.  These results, even
though preliminary and restricted to very specific conditions, show that the
physical properties of turbulence in collisionless plasmas, as those found in
the ICM, may be very different from what has been largely believed.
Implications can range from interchange of energies to cosmic rays acceleration.
\end{abstract}

\section{Introduction}

Magnetized and low density (weakly collisional) plasmas are known to present
anisotropic pressures with respect to the magnetic field orientation, which can
survive considerably long times compared to the dynamical timescales of certain
systems.  In astrophysical environments such pressure anisotropy may be
generated by several different processes, such as kinetic pressure of cosmic
rays, supernovae explosions, stellar winds, or anisotropic turbulent motions
(see Quest and Shapiro 1996).

Under certain conditions gyrotropic plasmas give rise to new wave modes and
instabilities, which cannot be studied by standard isotropic magnetohydrodynamic
(MHD) model (Hasegawa 1969, Wang and Hau 2003, Passot and Sulem 2006).  For
instance, Hau and Wang (2007) showed that gyrotropic magnetohydrodynamic
equations closed by the Chew-Goldberger-Low laws (CGL-MHD) lead to a positive
density $n$ {\it versus} magnetic field strength $B$ correlations for the slow
magnetosonic mode under certain conditions in contradiction to the standard MHD
model.  It was commonly believed that the detected absence of slow modes in
several astrophysical sites was related to the strong damping of these waves.
However, in the collisionless plasmas it could be explained by wrong
identification of the positive $n-B$ correlations for fast magnetosonic modes.

The pressure anisotropies give rise to plasma instabilities depending on the
anisotropy ratio, e.g., $p_\parallel > p_\perp$ and $p_\perp > p_\parallel$ for
the firehose and mirror instabilities, respectively.  They are responsible for
the growth of the magnetic energy and acceleration of particles.  The
predictions of the CGL-MHD model, including new plasma instabilities, are also
important in weakly magnetized environments, since even a weak magnetic field is
enough to change the motion of the charged particles and therefore increase the
pressure anisotropy.

As an example of the possible applications Sharma {\em et al.} (2003) and Sharma
{\em et al.} (2006) showed the importance of the collisionless plasma approach
in protostellar disks.  They studied the role of the kinetic instabilities in
the magneto-rotational instability (MRI) showing that the transport of angular
momentum in the disk may be efficiently increased under certain circumstances.

The intracluster medium (ICM) is possibly the most suitable environment for
studies of gyrotropic plasma effects (Schekochihin {\em et al.} 2005).
Considering typical parameters of $n \sim 10^{-3}$ cm$^{-3}$, $T\sim10^7$ K and
$B\sim1\ \mu$G (Ensslin and Vogt 2006), it is possible to show that the
cyclotron frequency ($\Omega$) is much larger than the collision frequency
($\nu_{ii}$).  Under such conditions, the plasma fluctuations with wave numbers
$k\geq$1 kpc will be subject to different processes related to the pressure
anisotropy.  The turbulent cascade, for example, may be modified by the new wave
modes and instabilities resulting in a different picture of the energy budget in
these environments.  It may be particularly important in understanding of the
cooling flow and the cosmic rays acceleration processes.

Schekochihin {\em et al.} (2008) showed that in the ICM the anisotropy-induced
firehose and mirror instabilities may grow non-linearly up to the saturation at
$\delta B/B \sim 1$.  This growth is very fast.  They estimated the increase of
power at small scales.  As a consequence, the excess of small scale fluctuations
of $B$ and the energy transport at these environments may drastically change the
picture.  In certain cases the small scales may also be considered as
collisionless, what may be important, for instance, in the development of
turbulent cascade.  For wavelengths smaller than the Larmor radius the kinetic
treatment of plasma is necessary, as shown by Howes {\em et al.} (2006).
However, in this work we focus on the large scale so the CGL-MHD approximation
may be used instead.

The aim of this work is to provide an extensive statistical analysis of the MHD
turbulence in collisionless plasmas and to study the role of the different
instabilities in the evolution of the system.  To accomplish that we performed
the first 3-dimensional simulations focusing on the evolution of turbulence in
the presence of pressure anisotropy.  The description of the model as well as
the presentation of the governing equations is given in
Section~\ref{sec:cgl-mhd} with additional discussion of the instabilities and
the double-isothermal approximation.  In Section~\ref{sec:setup} we describe the
numerical simulations.  The results and the extensive statistical analysis
including the derivation of probability distribution functions, spectra and
structure functions of the fluctuations of density and velocity are presented in
Section~\ref{sec:results}.  We discuss the most important results in
Section~\ref{sec:discuss} followed by Section~\ref{sec:conclusions}, where we
draw main conclusions.

\section{The Double-Isothermal CGL-MHD Approximation}
\label{sec:cgl-mhd}

\subsection{Governing Equations}

In the fluid approximation, a gyrotropic plasma can be described by the Chew
{\em et al.} (1956) magnetohydrodynamic (CGL-MHD henceforth) equations expressed
in the conservative form as follows:
\numparts
\begin{eqnarray}
 \pder{\rho}{t} + \nabla \cdot \left( \rho \vc{v} \right) & = & 0, \label{eq:mass} \\
 \pder{\rho \vc{v}}{t} + \nabla \cdot \left[ \rho \vc{v} \vc{v} + \left( \mathsf{P} + \frac{B^2}{8 \pi} \right) I - \frac{1}{4 \pi} \vc{B} \vc{B} \right] & = & \vc{f}, \label{eq:momentum} \\
 \pder{\vc{B}}{t} - \nabla \times \left( \vc{v} \times \vc{B} \right) & = & 0, \label{eq:magnetic_flux}
\end{eqnarray}
\endnumparts
where $\rho$ and $\vc{v}$ are the plasma density and velocity, respectively,
$\vc{B}$ is the magnetic field, $\mathsf{P} = p_\perp \hat{I} + (p_\parallel -
p_\perp) \hat{b} \hat{b}$ is the pressure tensor, $\hat{b}=\bf{B}/|\bf{B}|$ is
the unit vector along the magnetic field, $p_\parallel$ and $p_\perp$ are the
pressure components parallel and perpendicular to $\hat{b}$, respectively, and
$\vc{f}$ represents the forcing term.

The above set of equations is completed by the description of the parallel and
perpendicular pressure components.  To avoid the complexities related to
explicitly calculated processes that may be important for the description of the
energies, e.g., the anisotropic heat conduction and the emission and absorption
of the radiation, we can make use of the double-polytropic equations instead, as
suggested by Chew {\em et al.} (1956).  Tests of this approximation using the
solar magnetosheath data confirmed its validity under the form (see Hau and
Sonnerup 1993):
\numparts
\begin{eqnarray}
 \frac{d}{dt} \left( \frac{p_\perp}{\rho B^{\gamma_\perp - 1}} \right) & = & 0, \label{eq:pr_prim} \\
 \frac{d}{dt} \left( \frac{p_\parallel B^{\gamma_\parallel - 1}}{\rho^{\gamma_\parallel}} \right) & = & 0, \label{eq:pp_prim}
\end{eqnarray}
\endnumparts
where $\gamma_\perp$ and $\gamma_\parallel$ are the polytropic exponents for the
perpendicular and parallel pressures, respectively.  These equations, according
to Hau (2002), can be expressed in the conservative form, as follows:
\numparts
\begin{eqnarray}
 \pder{S_\perp}{t} + \nabla \cdot \left( S_\perp \vc{v} \right) & = & 0, \\
 \pder{S_\parallel}{t} + \nabla \cdot \left( S_\parallel \vc{v} \right) & = & 0,
\end{eqnarray}
\endnumparts
where $S_\perp = p_\perp B^{1 - \gamma_\perp}$, $S_\parallel = p_\parallel
\left( B / \rho \right)^{\gamma_\parallel-1}$, and $B = | \vc{B} |$ is the
strength of magnetic field.  In this form, the double-polytropic equations can
be solved numerically similarly as the continuity equation.

\subsection{Double-Isothermal Closure}

Under the double-isothermal closure we have $\gamma_\parallel = \gamma_\perp =
1$ and the equation of state is described by two relations, $p_\perp = a_\perp^2
\rho$ and $p_\parallel = a_\parallel^2 \rho$, where $a_\perp$ and $a_\parallel$
are constants and represent speeds of sound along the perpendicular and parallel
directions to the magnetic field, respectively.  In such situation the
conservative form of the momentum equation for the double-isothermal CGL-MHD
model can be rewritten as follows,
\begin{eqnarray}
\frac{\partial \rho \bf{v}}{\partial t} + \nabla \cdot \left[ \left( a_\perp^2 \rho +
\frac{B^2}{8 \pi} \right) \hat{I} - \left( 1 - \alpha \right) \bf{B} \bf{B} \right] = \vc{f},
\end{eqnarray}
where $\alpha \equiv \frac{1}{2} \left( \beta_\parallel - \beta_\perp \right)
\equiv \frac{1}{2} \beta_\perp \left( \xi - 1 \right)$, $\xi \equiv p_\parallel
/ p_\perp$ is the pressure anisotropy ratio, and $\beta_\parallel$ and
$\beta_\perp$ are the plasma betas in the parallel and perpendicular directions
to the local field, respectively.

The main consequence of the double-isothermal closure is that the pressure
anisotropy is kept constant, i.e. $\xi = \mathrm{const}$, independently of the
evolution of the kinetic instabilities that may arise.  Therefore, the stability
condition is fulfilled by the local decrease of density and the increase of
magnetic pressure and not due to the local changes of the pressure tensor.

Furthermore, this work is focused on studying the differences of the turbulent
regimes in the collisionless plasmas and the standard MHD turbulence using as
the reference the studies done by Kowal {\em et al.} (2007), Kowal and
Lazarian (2010), where the authors presented an extensive statistical analysis
of density, velocity and magnetic field distributions in the isothermal MHD
simulations.  For this reason we chose the double-isothermal CGL-MHD model as
the natural extension of the isothermal MHD in order to understand the
importance of pressure anisotropy and its consequences in the turbulent plasmas.

\subsection{Firehose and Mirror Instabilities}

While in the MHD model the dispersion relations do not exhibit any
instabilities, in the CGL-MHD equations the term of the pressure anisotropy
introduces significant changes to the dynamical system.  The detailed
linearization and wave analysis of the double-isothermal CGL-MHD equations is
provided in the Appendix.  Here we describe the final dispersion formulas
resulting from the presence of pressure anisotropy.  For example, in the case of
the incompressible Alfv\'en mode, the dispersion relation can be written as
\begin{equation}
\left( \omega^2 / k^2 \right)_A = c_A^2 \left( 1 - \alpha \right) \cos^2 \theta, \label{eq:alfven_dispersion}
\end{equation}
where $\alpha \equiv \frac{1}{2} \left( \beta_\parallel - \beta_\perp \right)$
and $\theta$ is the angle between the mean field and the wave vector of the
perturbation.  Equation~(\ref{eq:alfven_dispersion}) becomes negative when
$\beta_\parallel - \beta_\perp > 2$, what results in the occurrence of the
firehose instability and the growth of the magnetic field fluctuations.  As a
consequence, the field lines bend and the magnetic pressure increases reducing
the parameter $\alpha$ to its saturation value $\alpha \approx 2$.  This is
known as the kinetic Alfv\'en firehose instability.  Two other instabilities
related to the slow mode are called the compressible firehose and mirror
instabilities and occur when $p_\parallel > p_\perp$ or $p_\perp > p_\parallel$,
respectively, and the dispersion relation for magnetosonic waves becomes
negative (see Appendix).

\begin{figure}
 \centering
 \includegraphics[width=0.48\columnwidth]{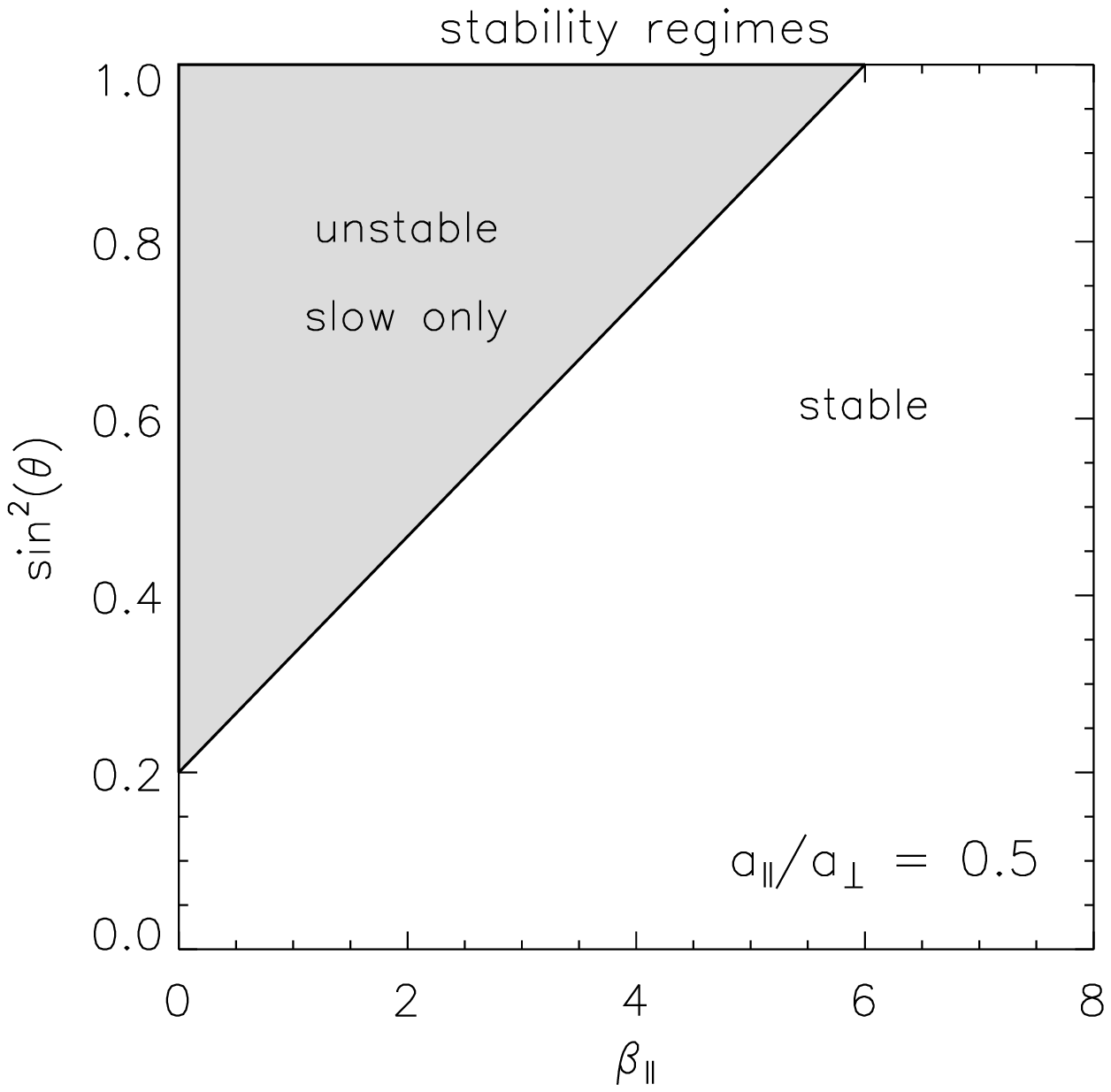}
 \includegraphics[width=0.48\columnwidth]{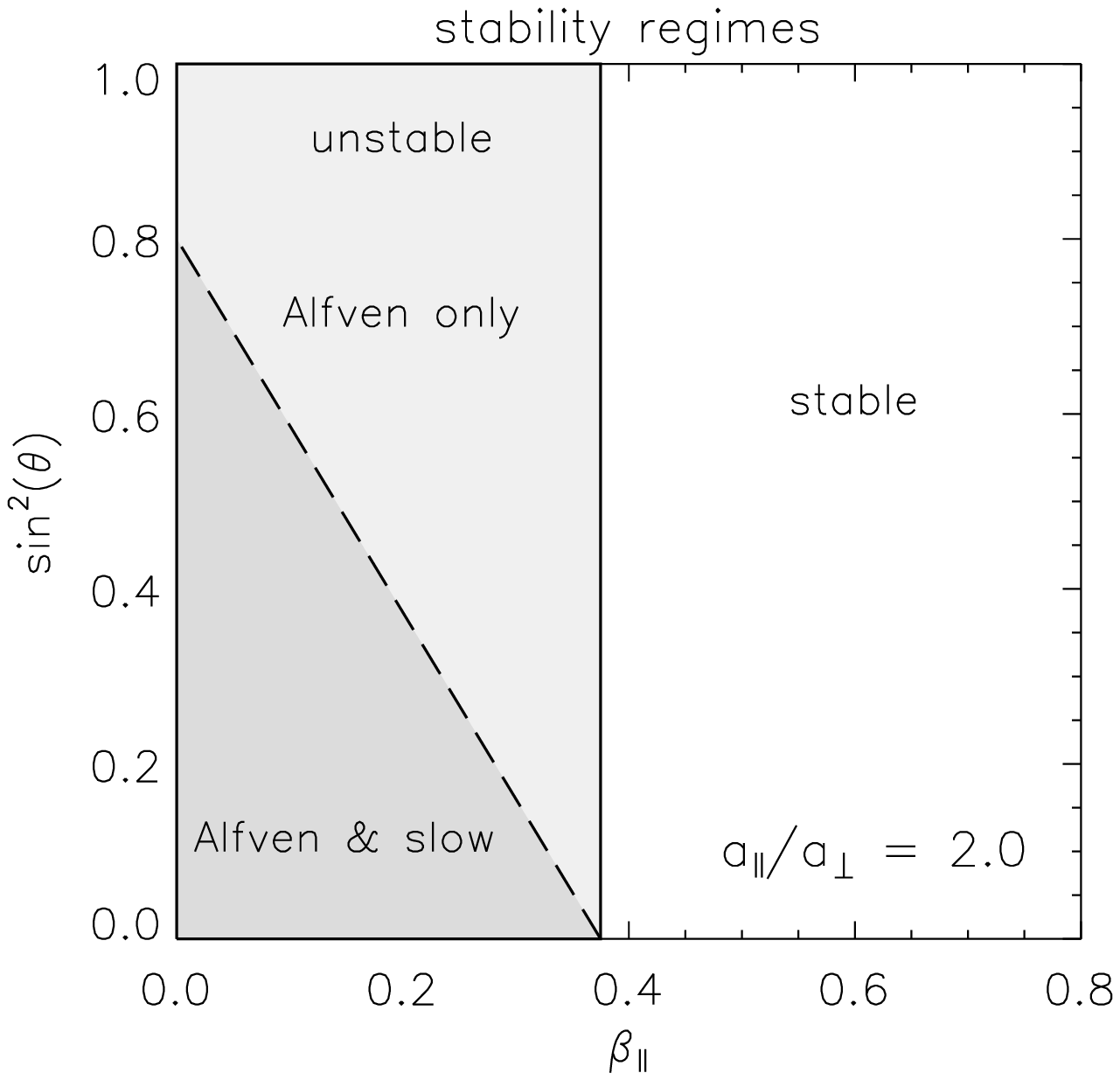}
 \caption{Stability conditions for models with the sound speed ratio
$a_\parallel / a_\perp = 0.5$ (left panel) and $a_\parallel / a_\perp = 2.0$
(right panel).  On the horizontal axis we show the plasma beta parameter related
to the parallel pressure and defined as $\beta_\parallel \equiv p_{mag} /
p_\parallel = \frac{1}{2} c_A^2 / a_\parallel^2$, where $p_{mag} \equiv
\frac{1}{2} |B|^2$ is the magnetic pressure.  On the vertical axis we show the
squared sinus of the angle between the directions of the perturbation and mean
field. \label{fig:stability}}
\end{figure}
In Figure~\ref{fig:stability} we show the stability regimes for two cases of the
pressure anisotropy studied in this paper, $a_\parallel / a_\perp = 0.5$ in the
left panel and $a_\parallel / a_\perp = 2.0$ in the right one (corresponding to
$p_\parallel / p_\perp = 0.25$ and $p_\parallel / p_\perp = 4.0$, respectively),
plotted as functions of the plasma beta $\beta_\parallel$ related to the
parallel pressure and the squared sinus of the angle between the directions of
the perturbation propagation and magnetic field.  When the perpendicular
pressure dominates (left panel) only the mirror instability corresponding to the
slow mode can occur.  When perturbations propagate in directions almost
perpendicular to the magnetic field the range of unstable $\beta_\parallel$
grows.  On the contrary, when the direction of propagation is close to the
direction of magnetic field the instability does not occur.  In the case of
dominating parallel pressure (right panel) the plasma becomes unstable due to
two firehose instabilities related to the incompressible Alfv\'en and
compressible slow modes, but the Alfv\'en mode firehose instability extends over
larger region of the parameter space up to $\beta_\parallel = 0.375$
independently of the angle between the perturbation propagation and magnetic
field directions.

Both instabilities have the growth rate $\gamma \equiv \mathsf{Im}(\omega)$
larger at smaller scales, i.e., increasing with the wave number $k$.  This
dependence introduces a stiffness in the numerical integration since the micro
instabilities, which may be due to the numerical imprecision, tend to grow very
fast and destroy the configuration of the studied problem.  As a consequence,
the pressure anisotropy tends to disappear quickly and the problem of interest
cannot be studied.  Sharma {\em et al.} (2006) studying the magneto-rotational
instability (MRI) in protostellar disks pointed out this important problem.
They bypassed it by implementing a quasi-stability condition for the
computational cells where the pressure anisotropy was larger than the threshold
for instability.  Physically, it can be understood as an almost
``instantaneous'' evolution of the system to the quasi-stable condition.  Under
such condition all anisotropic effects are kept at the large scales and the
problem of interest could be analyzed.  The downside of this method is that it
artificially removes the free energy of pressure anisotropy without, as a
counterpart, increasing the kinetic and/or magnetic energies.  That is because,
in the physical sense, the instability increases the magnetic energy and/or
accelerate the gas.

In the case of turbulence this kind of artificial removal of the energy at small
scales may not be the best approach since the small scale structures in
turbulence are generated by the cascade process operating in the turbulent
models and therefore by artificially influencing the dissipation of energy we
can distort its physical meaning and obtain incorrect conclusions. Nevertheless,
under the double-isothermal approximation the pressure anisotropy is kept even
after the growth of the small scale perturbations saturates, so the stability
condition at large scales may lay in the unstable region even after the
saturation at small scales.  Therefore, under certain conditions, the
double-isothermal approximation is valid and might be an interesting area of
studies of the evolution of turbulence in a collisionless plasma.

\section{Numerical Simulations}
\label{sec:setup}

The simulations of turbulence in collisionless plasma were performed solving the
set of double-isothermal CGL-MHD equations in a conservative form given by
equations~(\ref{eq:mass})-(\ref{eq:magnetic_flux}) with an addition term
$\vc{f}$ in the motion equation representing the turbulence driving.  The
numerical integration of the system evolution governed by the CGL-MHD equations
were performed using the second order shock-capturing Godunov-scheme code (Kowal
{\em et al.} 2007, 2009, Kowal and Lazarian 2010).  We incorporated the field
interpolated constrained transport (CT) scheme (see, e.g., T\'oth, 2000) into
the integration of the induction equation to maintain the $\nabla \cdot \bi{B} =
0$ constraint numerically, the general Harten-Lax-van Leer Riemann solver
(Einfeldt {\em et al.} 1991) to obtain the numerical fluxes.  The time
integration was done with the second order Runge-Kutta method (see e.g. Press
{\em et al.} 1992).  On the right-hand side, the source term $\bi{f}$ represents
a random solenoidal large-scale driving force.  The rms velocity $\delta v$ is
maintained to be approximately unity, so that $\bi{v}$ can be viewed as the
velocity measured in units of the rms velocity of the system and $\bi{B}/\sqrt{4
\pi \rho}$ as the Alfv\'{e}n velocity in the same units.  The time $t$ is in
units of the large eddy turnover time ($\sim L/\delta v$) and the length in
units of $L$, the scale of the energy injection.  The magnetic field consists of
the uniform background field and a fluctuating field: $\bi{B}= \bi{B}_0 +
\bi{b}$.  Initially, $\bi{b}=0.0$.  We use units in which the Alfv\'{e}n speed
$c_A = B_0 / \sqrt{4 \pi \rho}=1.0$ and $\rho=1.0$ initially.

For our calculations, similar to our earlier studies (Kowal {\em et al.} 2007,
Kowal and Lazarian 2010), the sound speeds and the strength of the external
field $B_0$ are the controlling parameters defining the sonic Mach number ${\cal
M}_s = \langle \delta v / a \rangle$ and the Alfv\'enic Mach number ${\cal M}_A
= \langle \delta v / c_A \rangle$, respectively.  The angle brackets $\langle
\rangle$ signify the averaging over the volume.  ${\cal M}_s < 1$ and ${\cal
M}_s > 1$ define subsonic and supersonic regimes, respectively, and ${\cal M}_A
< 1$ and ${\cal M}_A > 1$ define another two regimes, sub-Alfv\'enic and
super-Alfv\'enic, respectively.  Since these two parameters are independent we
can analyze, e.g., supersonic sub-Alfv\'enic turbulence, which signifies that
${\cal M}_s > 1$ and ${\cal M}_A < 1$.  In the case of the CGL-MHD simulations
the regimes are defined by two sonic Mach numbers corresponding to the parallel
and perpendicular sounds speeds.

\begin{table}
\caption{\label{tb:models} Description of the performed simulations of the double-isothermal CGL-MHD turbulence with the resolution 512$^3$. Initial density for all models was set to 1.0.}
\begin{indented}
\item[]\begin{tabular}{@{}ccccccc}
\br
Model & $B_{0}$ & $a_\parallel$ & $a_\perp$ & $\xi\equiv a_\parallel^2/a_\perp^2$ & $\mathcal{M}_s \equiv \langle |v| / a_\perp \rangle$ & $\mathcal{M}_A \equiv \langle |v| / c_A \rangle$ \\
\mr
1 & 1.0 & 1.0 & 2.0  & 0.25 & 0.7 & 0.7 \\
2 & 1.0 & 1.0 & 0.5  & 4.00 & 0.7 & 0.7 \\
3 & 0.1 & 0.1 & 0.2  & 0.25 & 7.0 & 2.0 \\
4 & 0.1 & 0.1 & 0.05 & 4.00 & 7.0 & 2.0 \\
5 & 0.1 & 1.0 & 0.5  & 4.00 & 0.7 & 2.0 \\
6 & 1.0 & 0.1 & 0.2  & 0.25 & 7.0 & 0.7 \\
\br
\end{tabular}
\end{indented}
\end{table}

We drove turbulence solenoidally at wave scale $k$ equal to about 2.5 (2.5 times
smaller than the size of the box).  This scale defines the injection scale in
our models.  We did not set the viscosity and diffusion explicitly in our
models.  The scale at which the dissipation starts to act is defined by the
numerical diffusivity of the scheme.

We performed six three-dimensional CGL-MHD simulations using the resolution
$512^3$ for different initial conditions, as shown in Table~\ref{tb:models}. We
simulated the clouds up to $t_{\rm max} \sim 6$, i.e.\ 6 times longer than the
dynamical timescale, to ensure a full development of the turbulent cascade. The
computational time required for each CGL-MHD simulation with intermediate
resolution was equivalent to a MHD $512^3$, as performed by Kowal {\em et al.}
(2007) and Kowal and Lazarian (2010).

Models 1 and 2 are examples of weak turbulence and belong to subsonic and
subAlfv\'enic regimes.  The difference between these models is the pressure
anisotropy accounting for the mirror and firehose instabilities. the comparison
of both models gives us an insight into the different evolution of turbulence in
these two cases.  Similarly we calculated Models 3 and 4 for strong turbulence
resulting in the supersonic and superAlfv\'enic regimes.  Again, we tested these
regimes for different instabilities.  Model 5 belongs to the subsonic and
superAlfv\'enic regime representing the physical conditions of the ICM, and can
be used as reference for further studies on this subject.  Finally, Model 6
belongs to the supersonic and subAlfv\'enic turbulent regime and is the only
simulation initiated with magnetic pressure larger than the thermal one.  In
this model all cells are initially stable but as the turbulence develops
unstable conditions can occur.  In the following sections we present the results
obtained for each model, as well as a direct comparison with the standard MHD
turbulent simulations with similar initial conditions.

\section{Results}
\label{sec:results}

\subsection{Distribution of Density and Velocity}

\begin{figure*}[tbh]
 \center
 \includegraphics[width=0.32\textwidth]{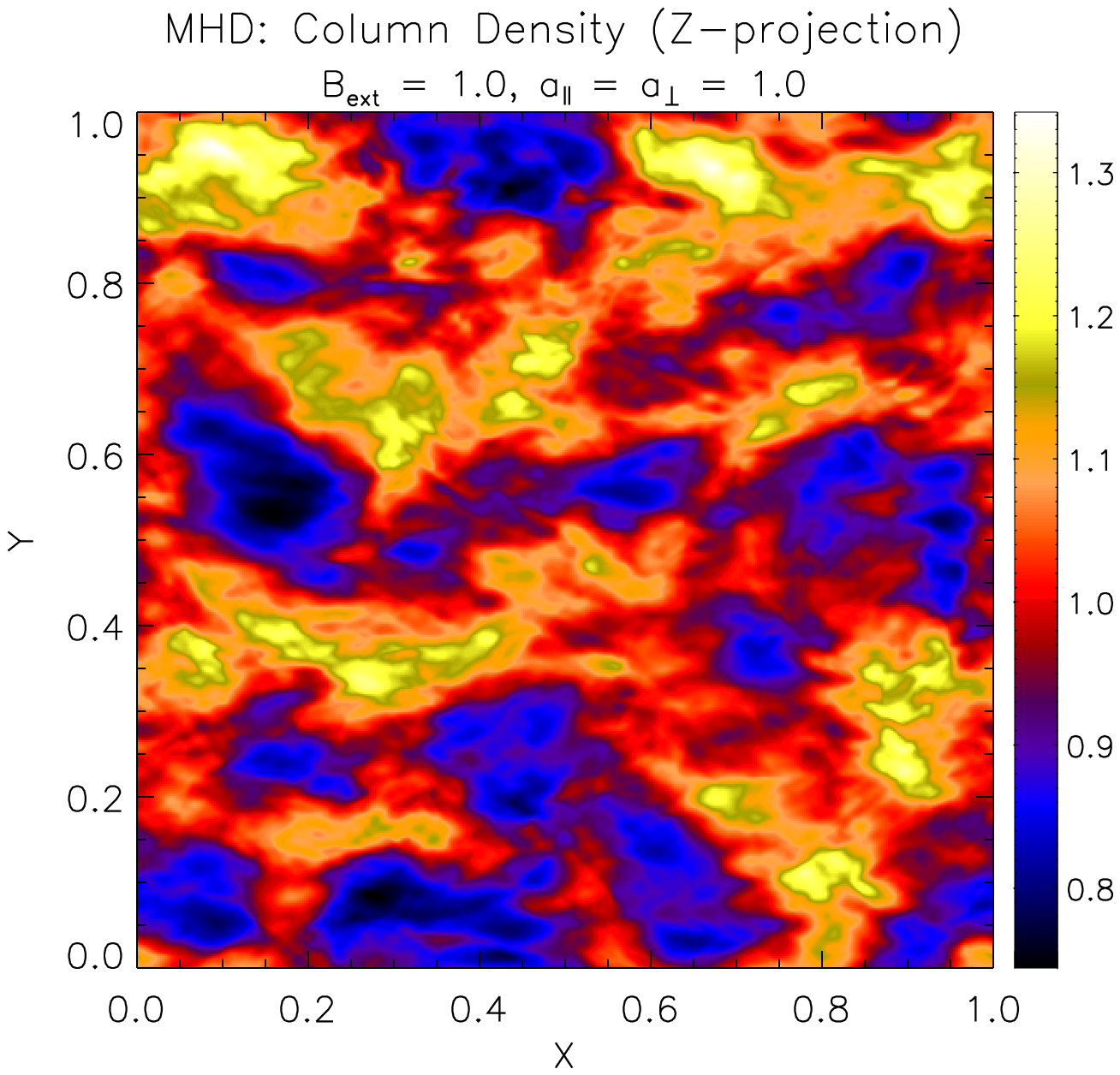}
 \includegraphics[width=0.32\textwidth]{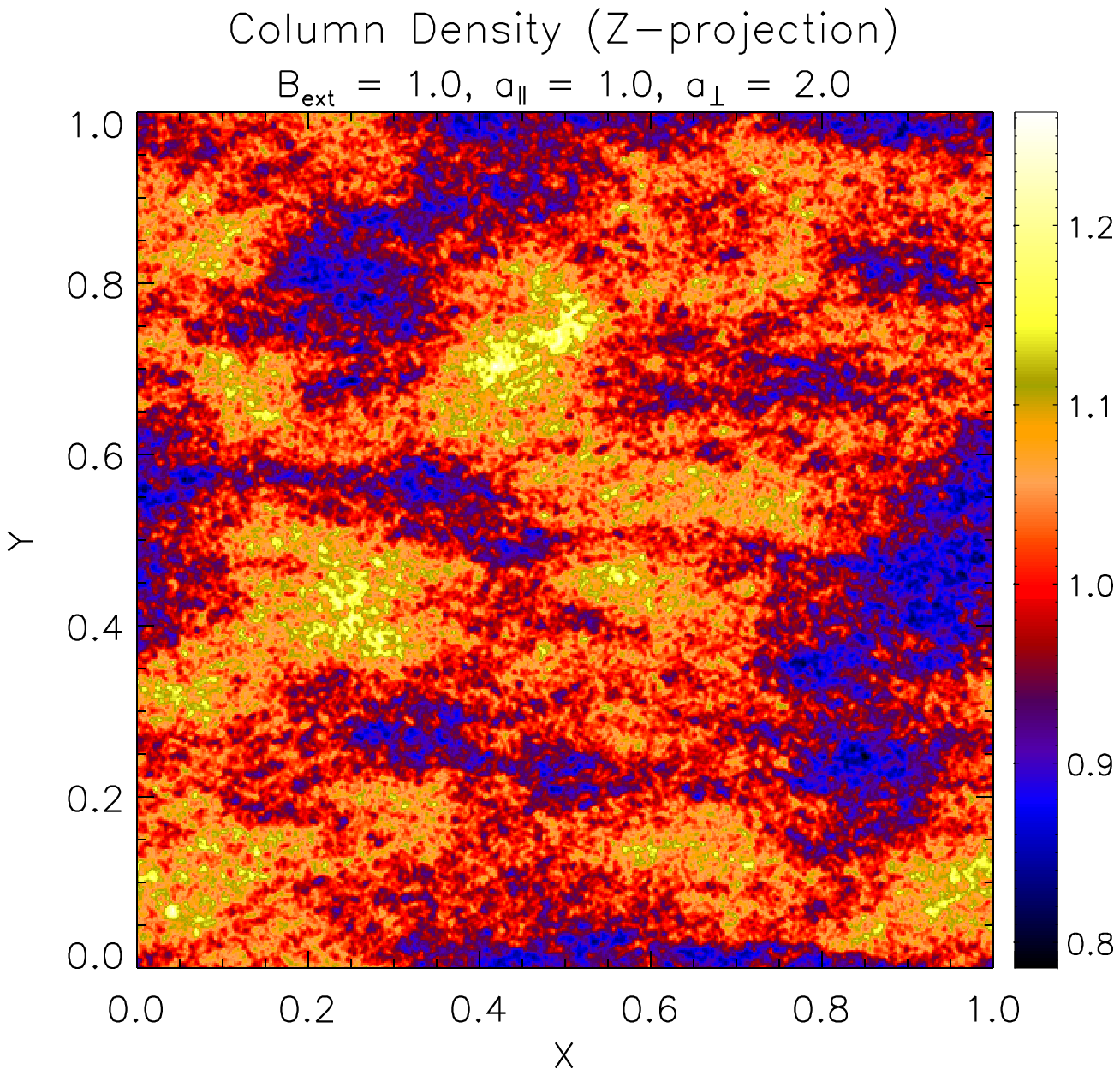}
 \includegraphics[width=0.32\textwidth]{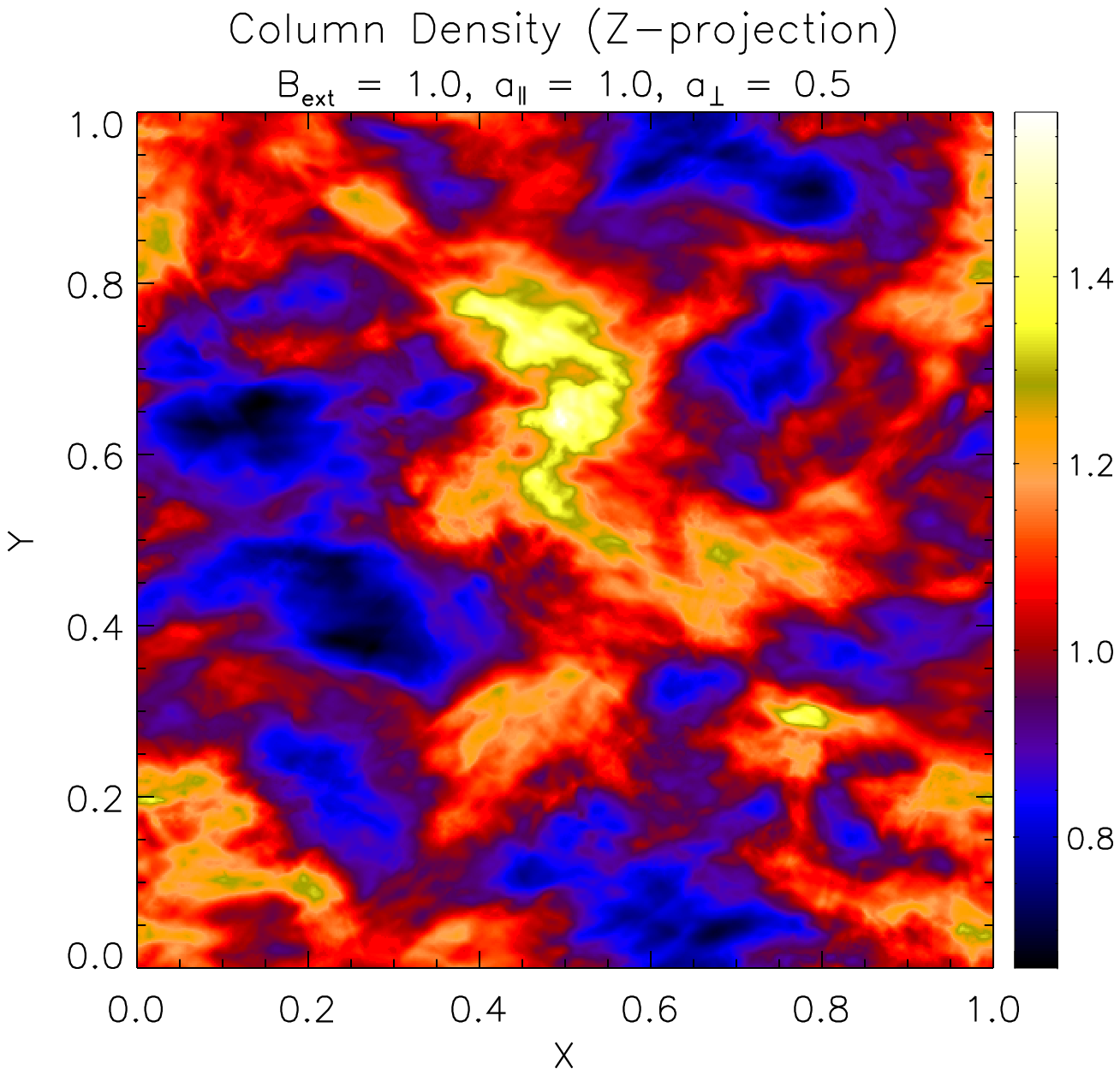}
 \includegraphics[width=0.32\textwidth]{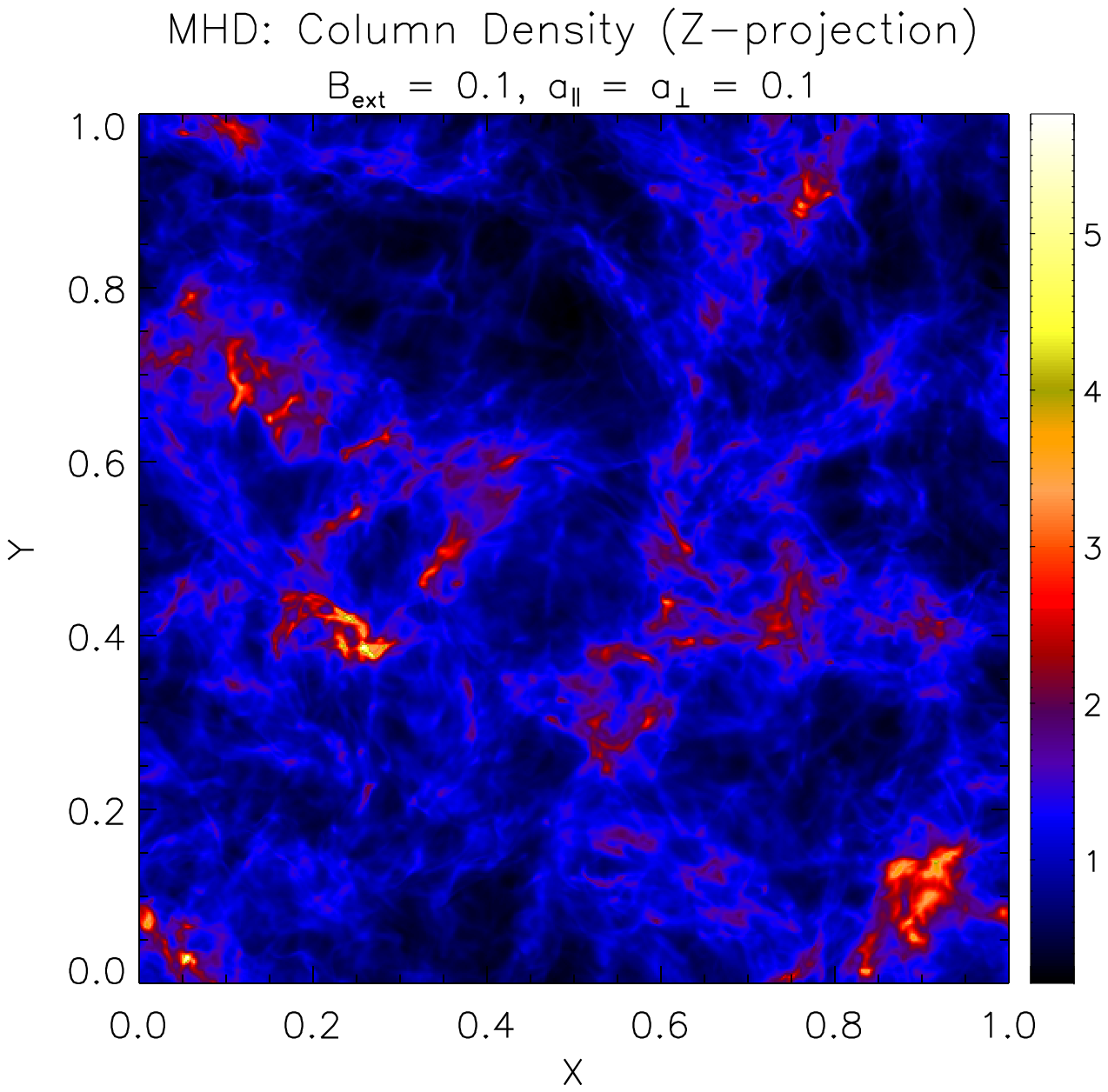}
 \includegraphics[width=0.32\textwidth]{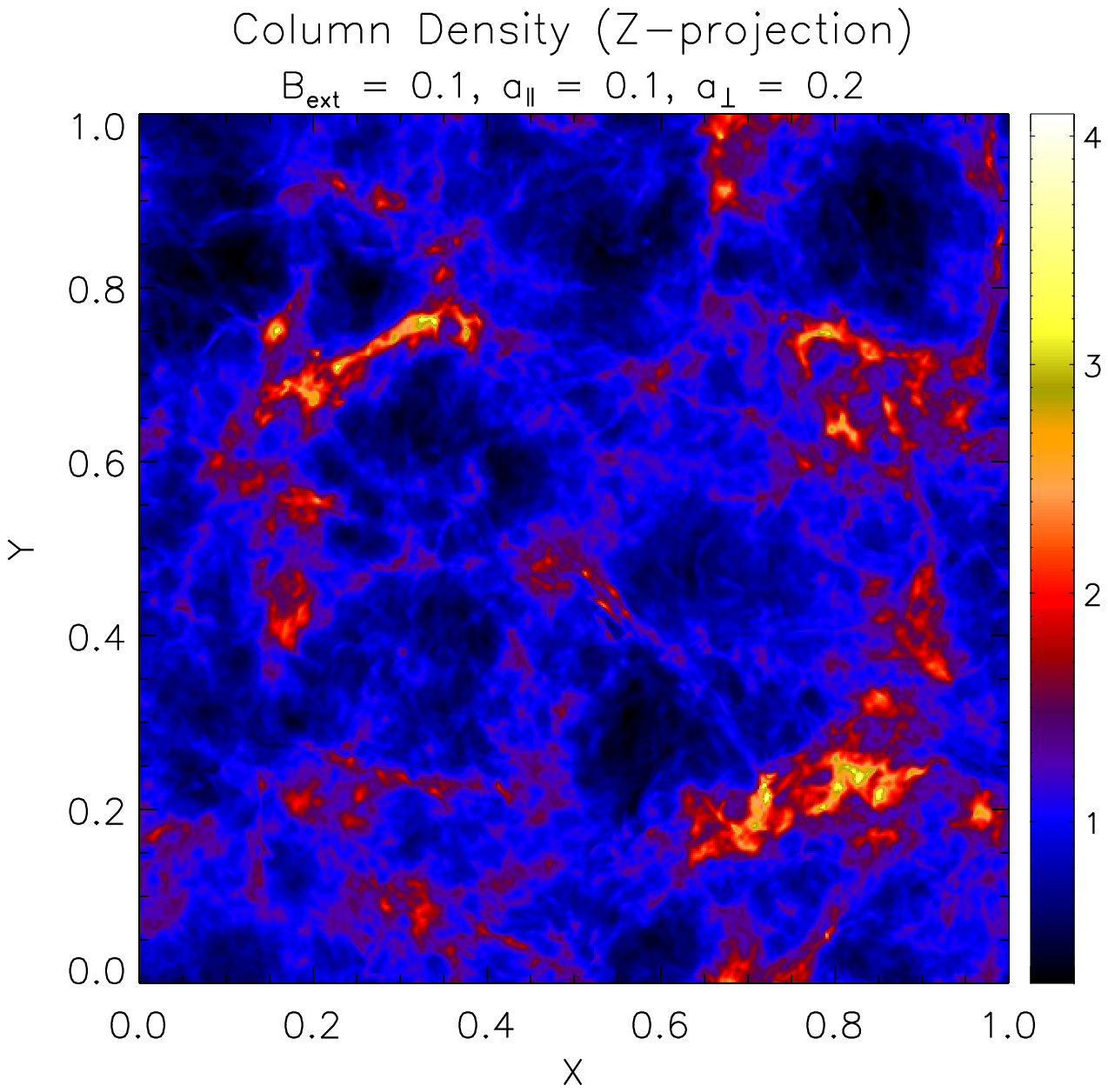}
 \includegraphics[width=0.32\textwidth]{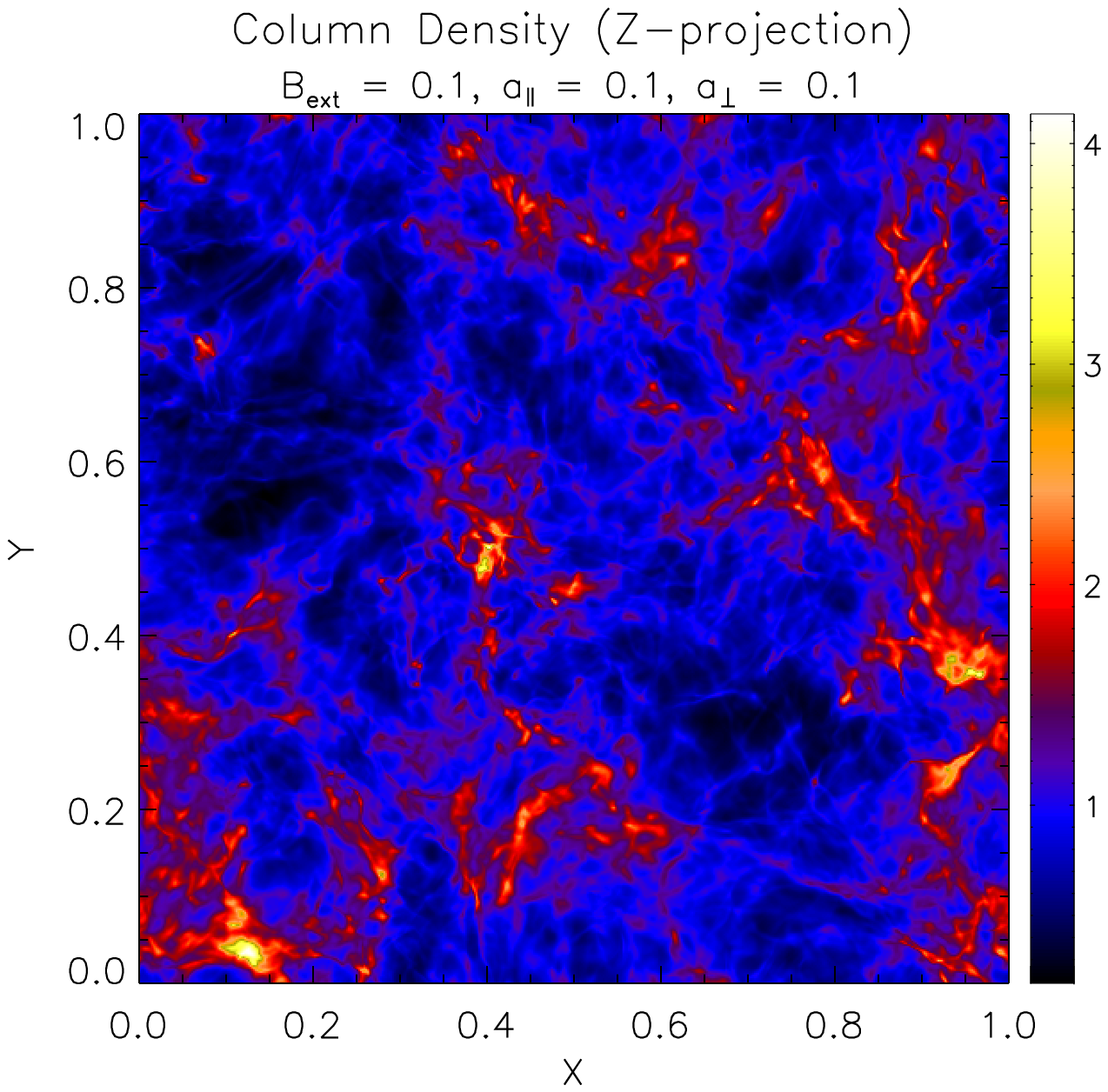}
 \includegraphics[width=0.32\textwidth]{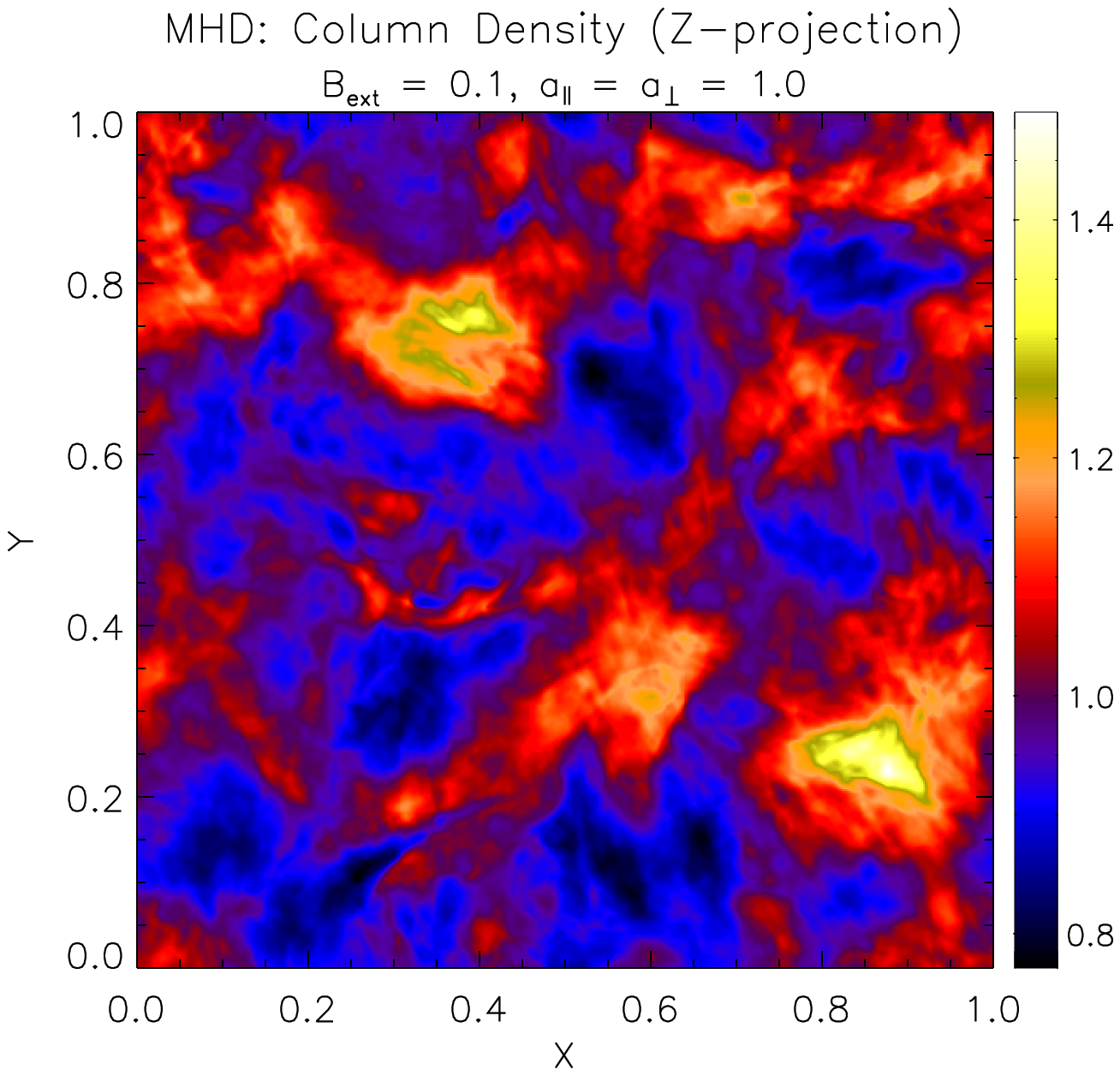}
 \includegraphics[width=0.32\textwidth]{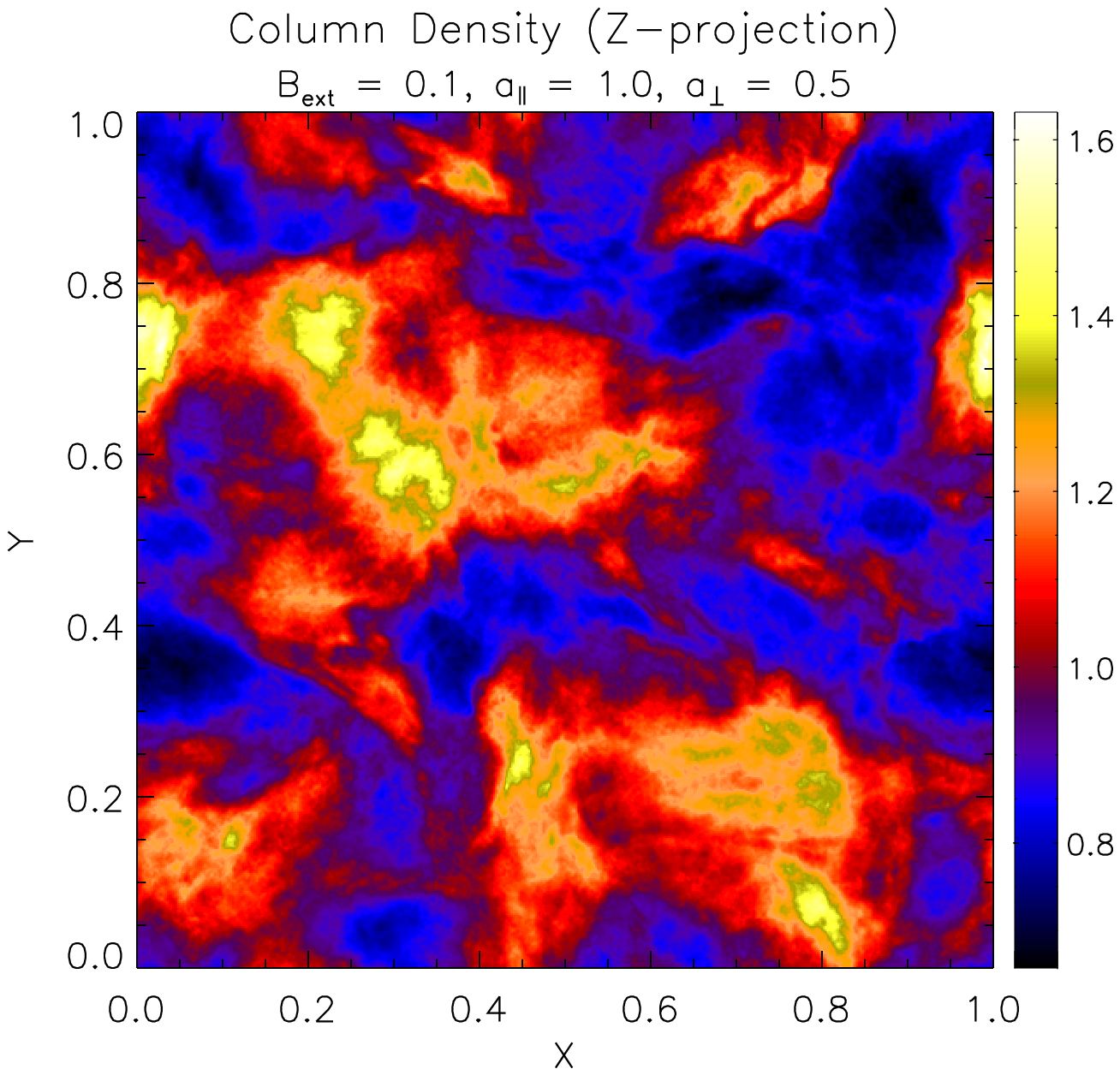} \\
 \includegraphics[width=0.32\textwidth]{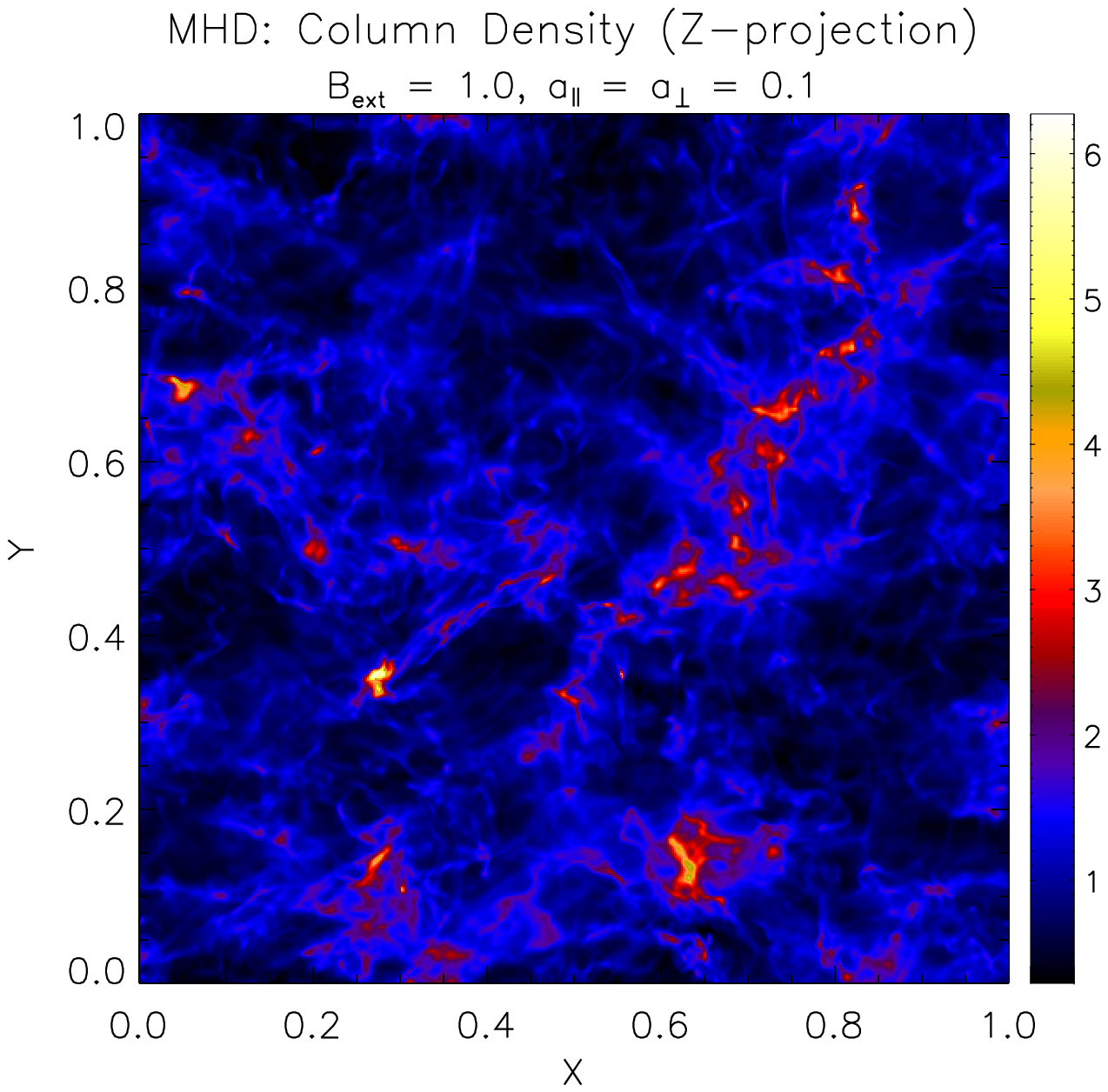}
 \includegraphics[width=0.32\textwidth]{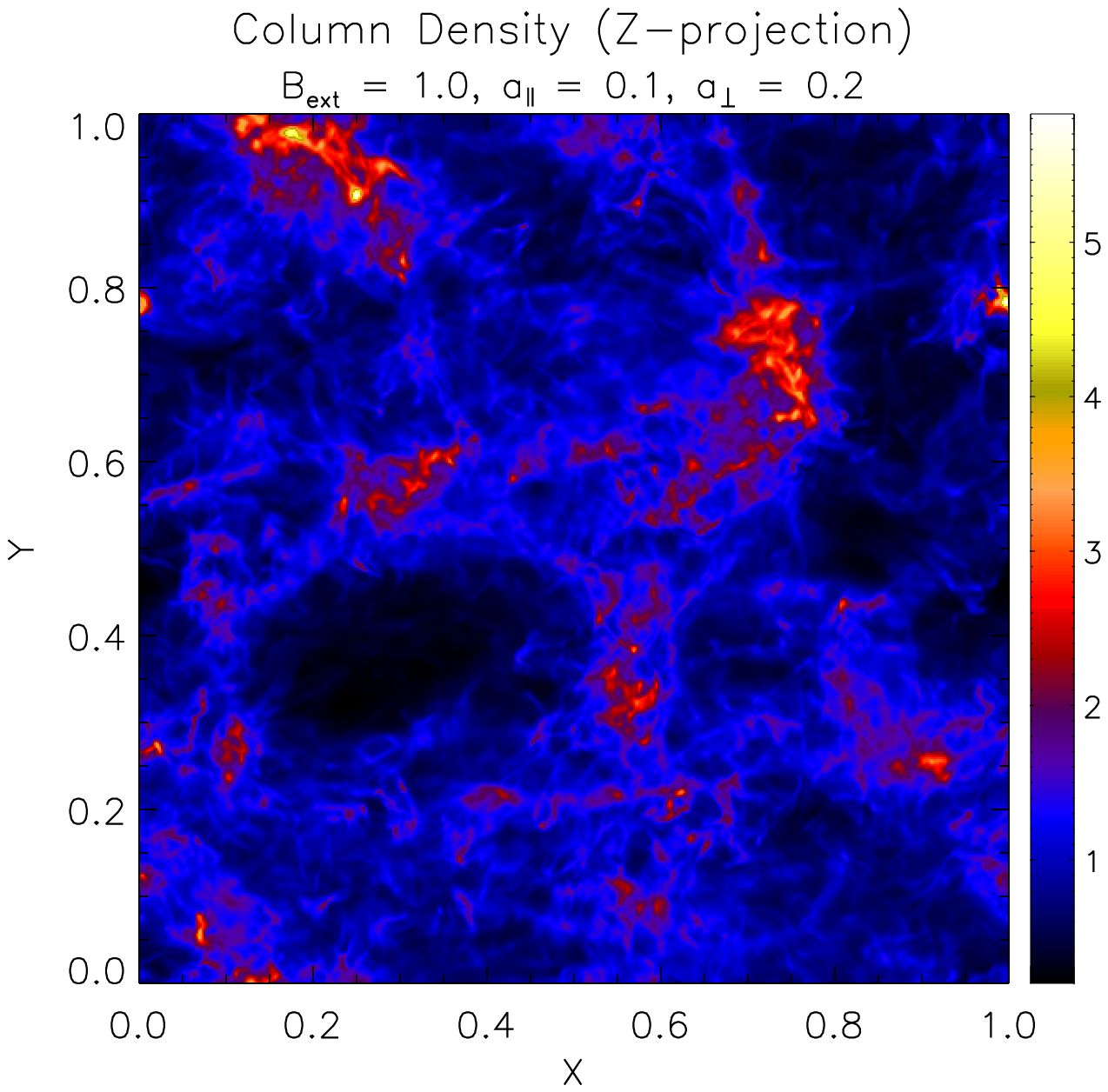}
 \caption{Column density for the different MHD and CGL-MHD models corresponding to the models presented in Table~\ref{tb:models}. \label{fig:column_density}}
\end{figure*}

The results obtained for the distribution of density and velocity show strong
differences between the CGL-MHD and standard MHD models.  However, the kinetic
instabilities play a role in the evolution of turbulence under specific
conditions.  In Figure~\ref{fig:column_density} we present the column density
obtained for each CGL-MHD model, as as well as the MHD models for comparison.
Each case is presented as labeled in Table 1, with the MHD case shown in the
left, and the CGL-MHD in the right.

From the column density maps it is possible to distinguish models 2 and 5.  Both
models are subsonic, where initially we set $p_\parallel/p_\perp = 2$, which
resulted in strong firehose instabilities.  Here, the firehose instability is
responsible for a deformation of the magnetic field lines.  The curved magnetic
lines tend to slow down and trap the flowing gas.  Since the growth rate is
larger at small scales we expect this effect to create more granulated maps.
This is not seen in model 4 because $\rho \delta v^2 > p_\parallel ~ \delta B^2$
and, therefore, the turbulence is able to destroy the configuration of the
growing instability.  The same occurs for model 3.  Even though it is not
visible in the column density maps, the kinetic instabilities are responsible
for changes in the statistics of the turbulence, as addressed below. For models
1 and 6, in which $p_\parallel/p_\perp = 0.5$, the column density maps show mild
differences.  In these cases we have a mirror instability operating, which is
responsible for changes in the velocity distribution.  For model 1 with a weak
turbulence, the mirror instability is responsible for the acceleration of the
gas resulting in the increase of the {\it effective} sonic Mach number.  In
model 6, where initially all cells were stable, the evolution of turbulence
causes the instability in most of the computational domain.  The instability is
responsible here for slowing the gas and reducing the {\it effective} sonic Mach
number.

\begin{figure*}[tbh]
 \center
 \includegraphics[width=0.32\textwidth]{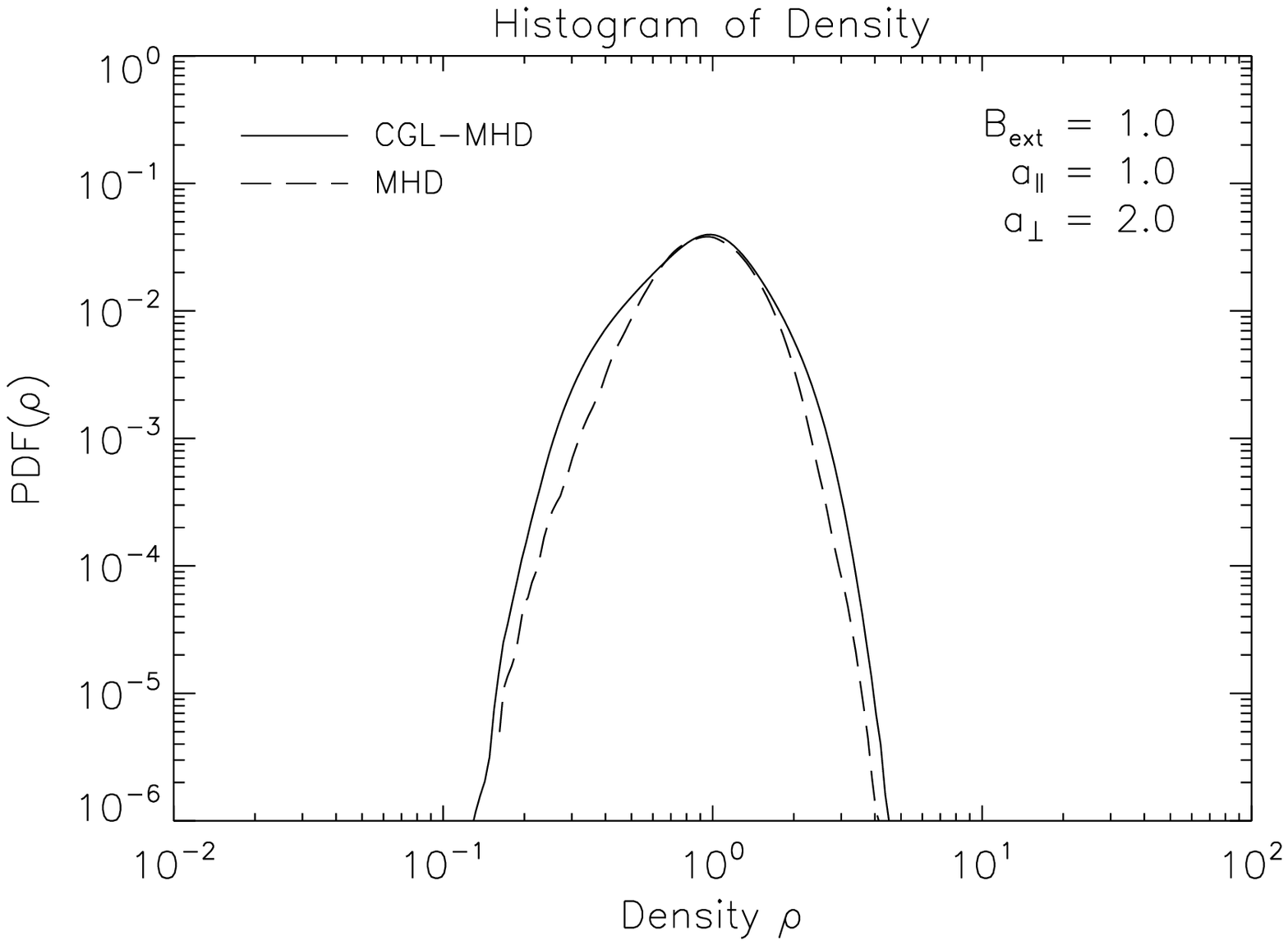}
 \includegraphics[width=0.32\textwidth]{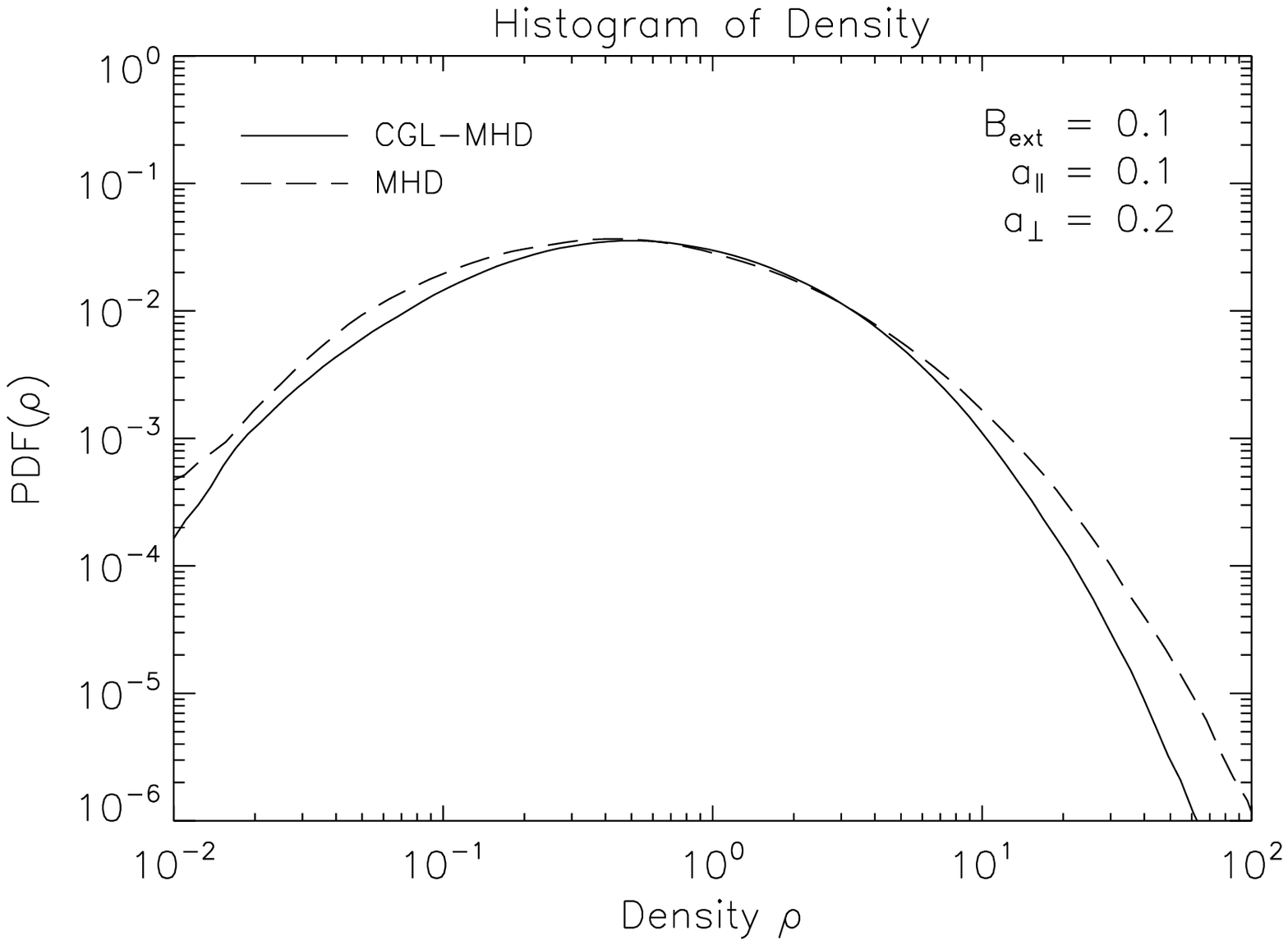}
 \includegraphics[width=0.32\textwidth]{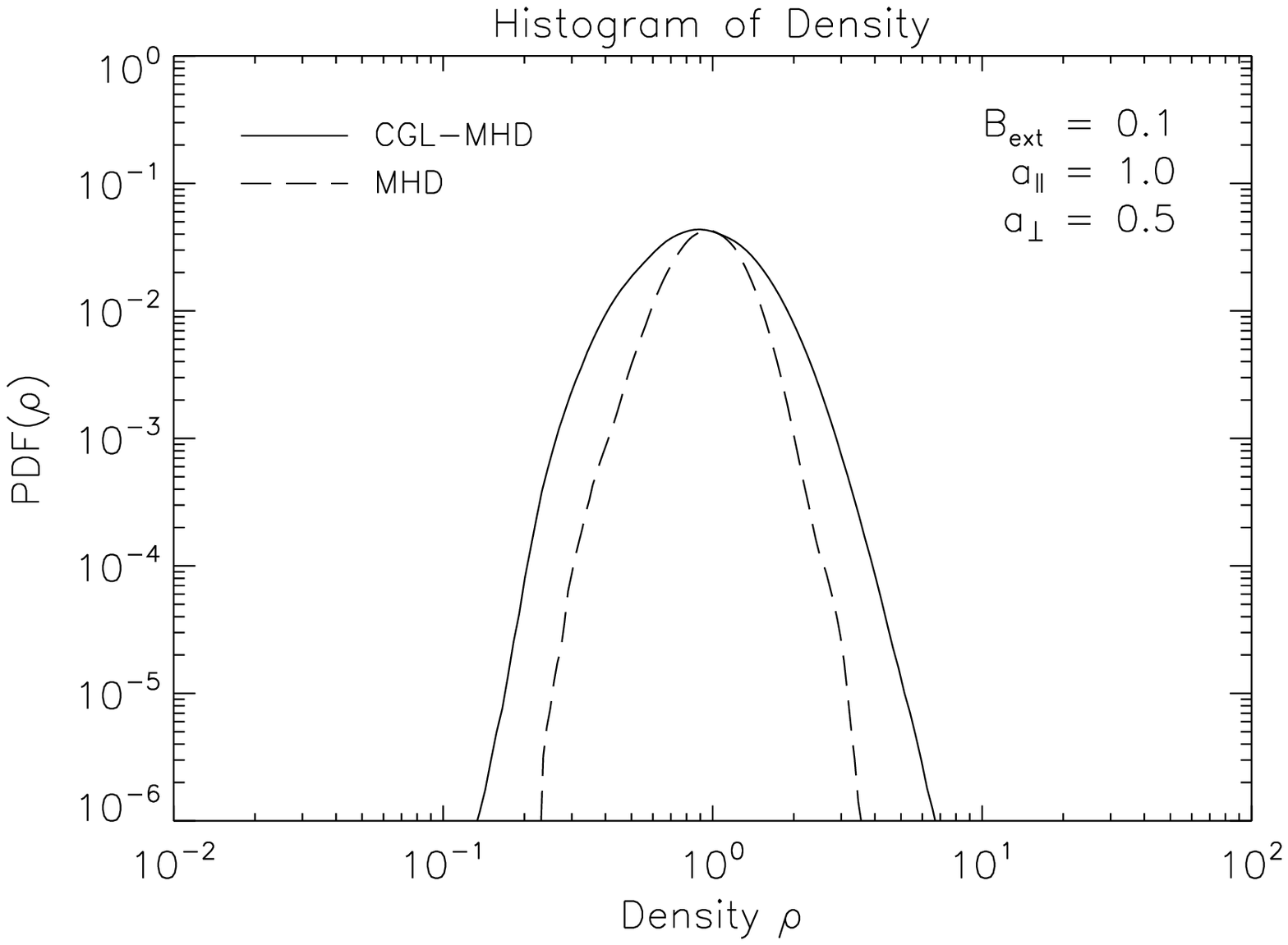}
 \includegraphics[width=0.32\textwidth]{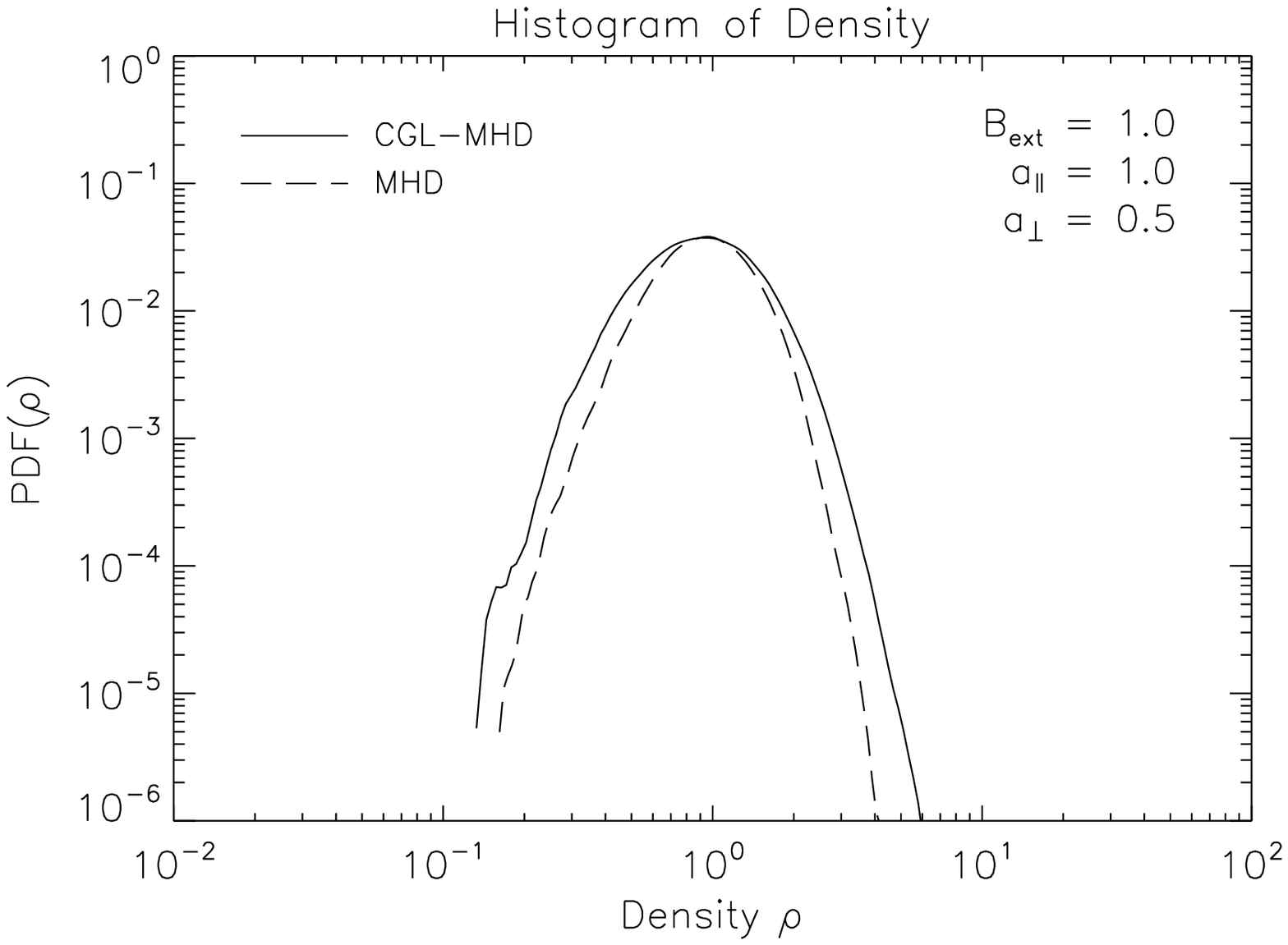}
 \includegraphics[width=0.32\textwidth]{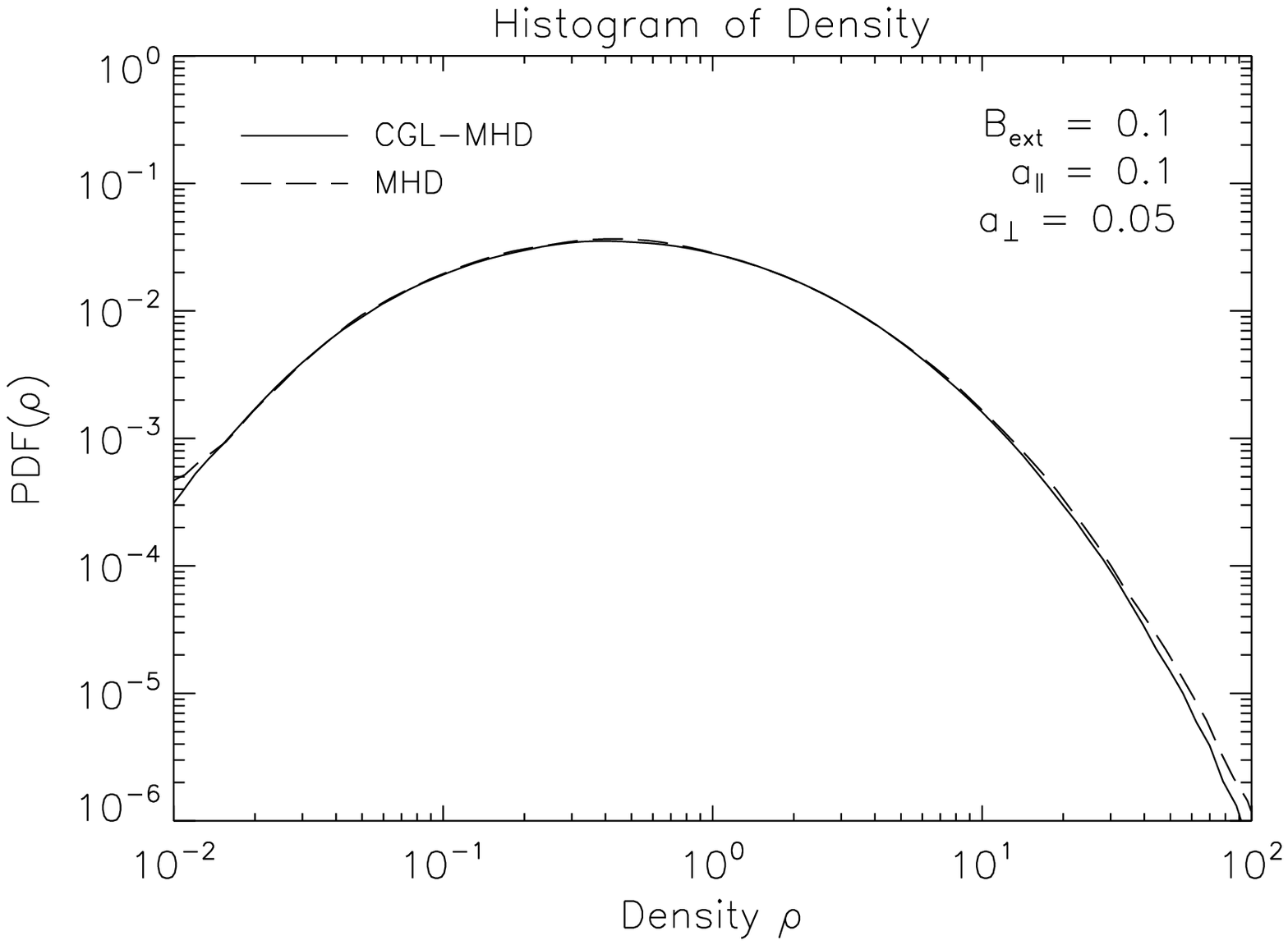}
 \includegraphics[width=0.32\textwidth]{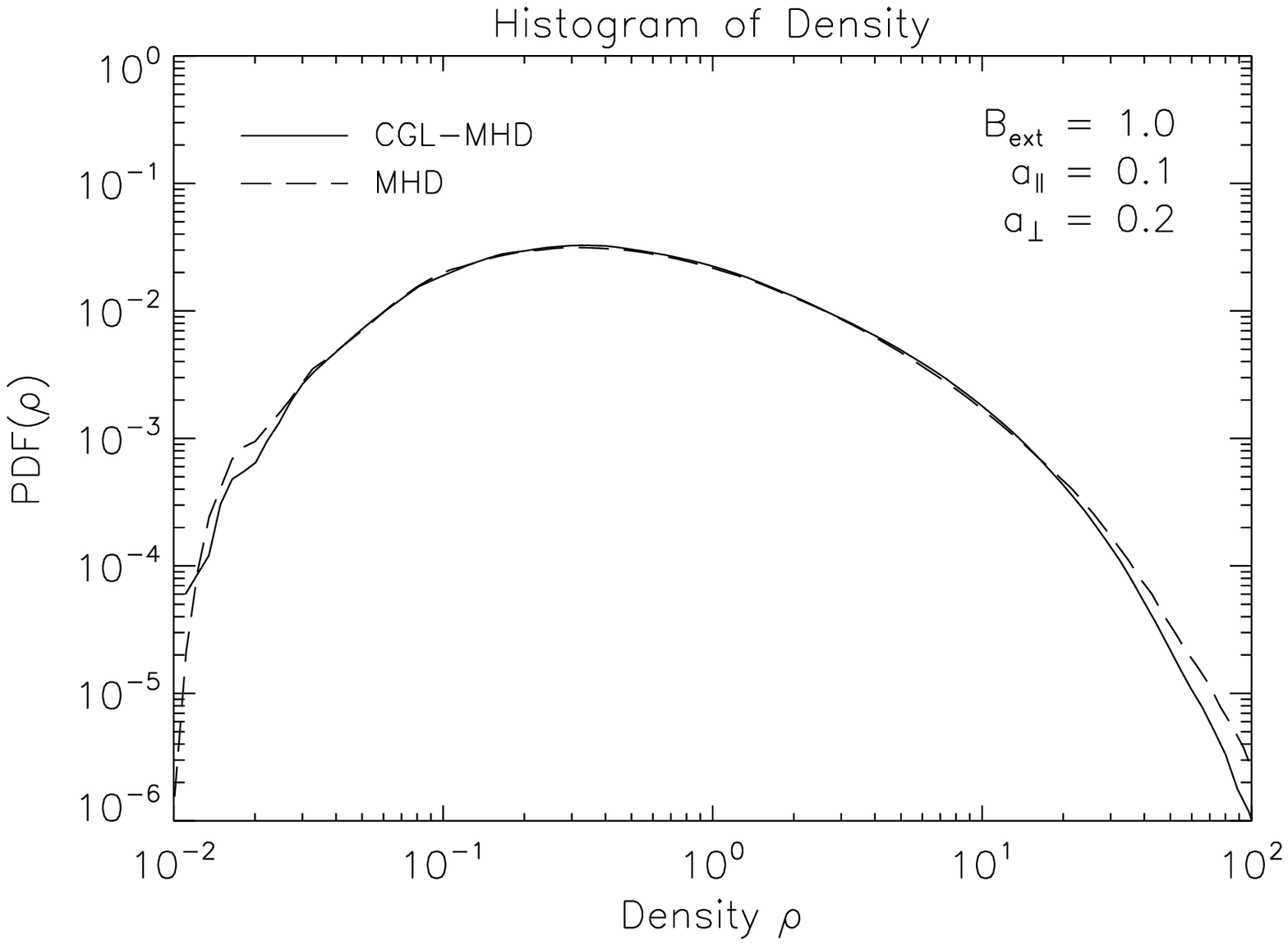}
 \caption{PDFs of density for the studied models. The left column shows models 1 and 2, the middle one shows models 3 and 4, and the right column shows models 4 and 5, according to Table~\ref{tb:models}.  For each case both, the CGL-MHD (solid lines) and MHD (dashed lines) models are shown for comparison. \label{fig:dens_pdfs}}
\end{figure*}

\begin{figure*}[tbh]
 \center
 \includegraphics[width=0.32\textwidth]{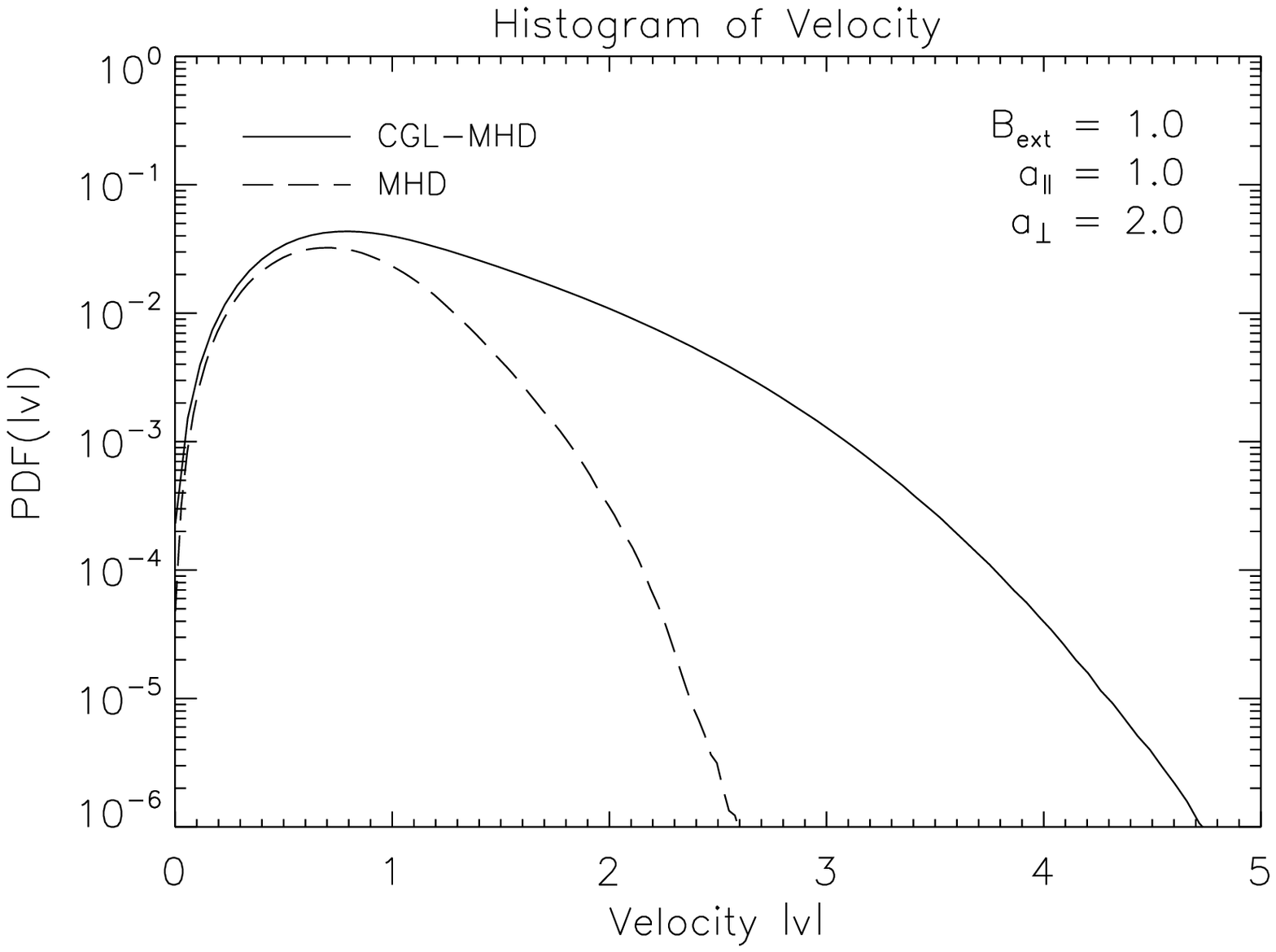}
 \includegraphics[width=0.32\textwidth]{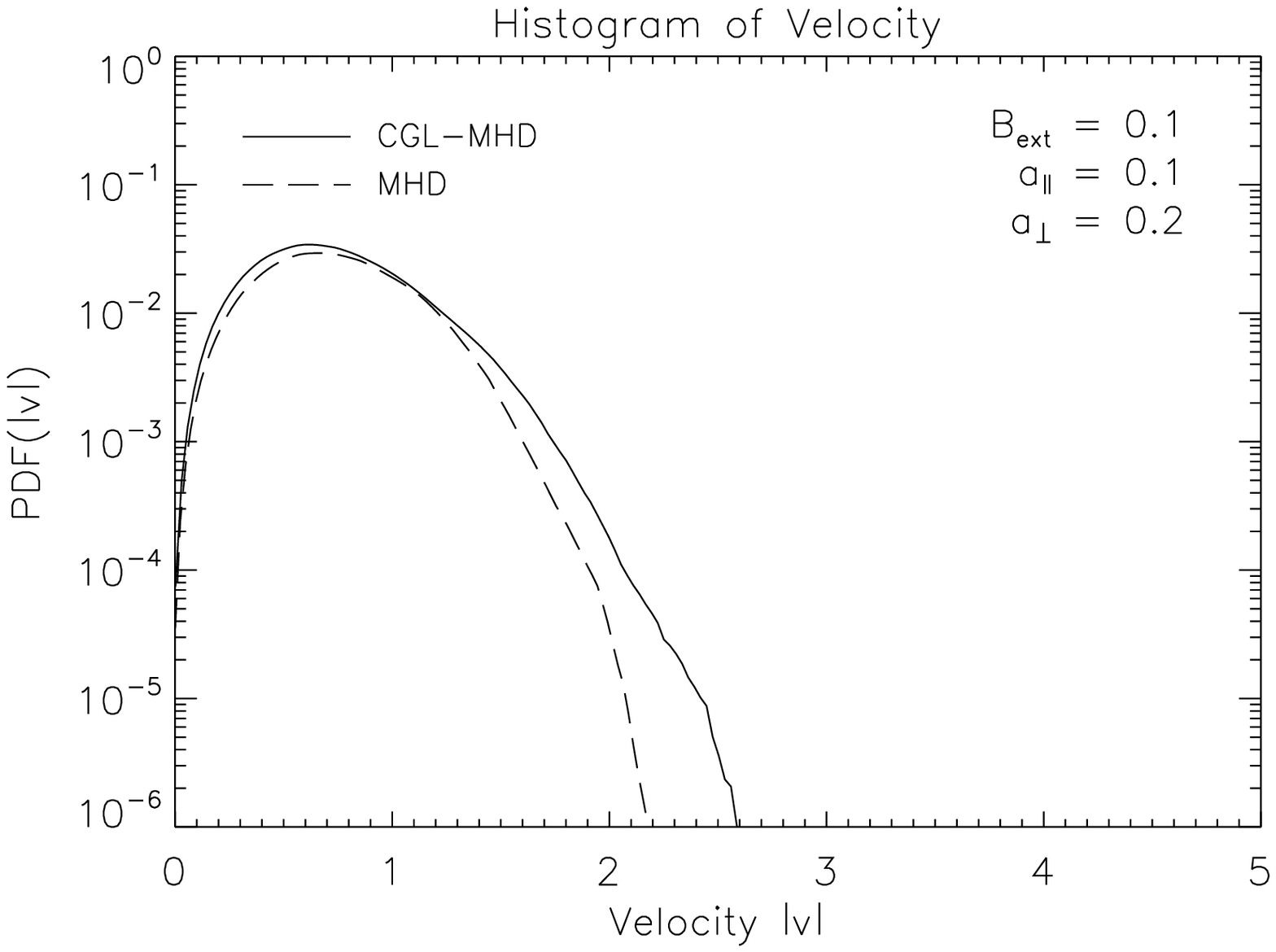}
 \includegraphics[width=0.32\textwidth]{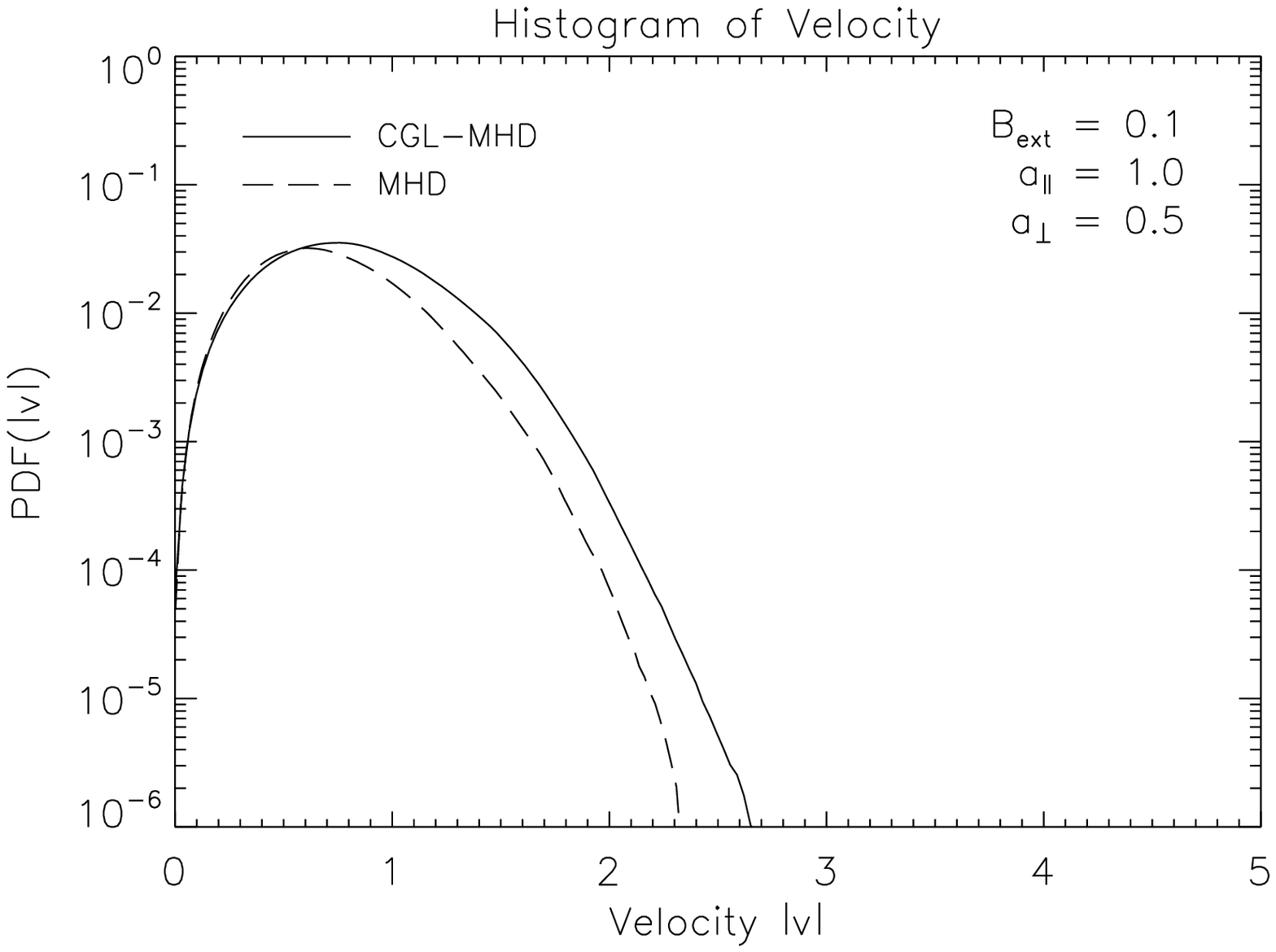}
 \includegraphics[width=0.32\textwidth]{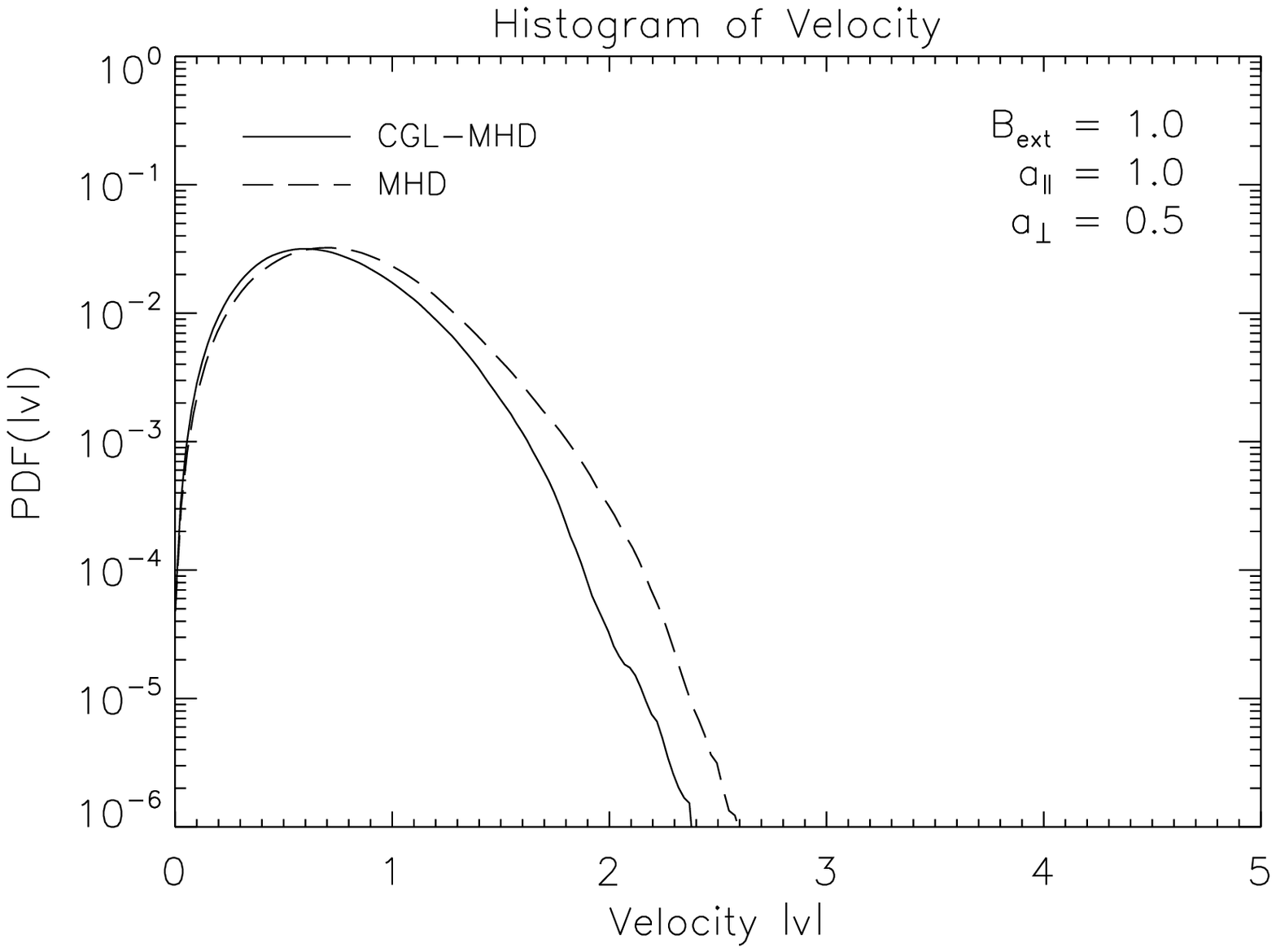}
 \includegraphics[width=0.32\textwidth]{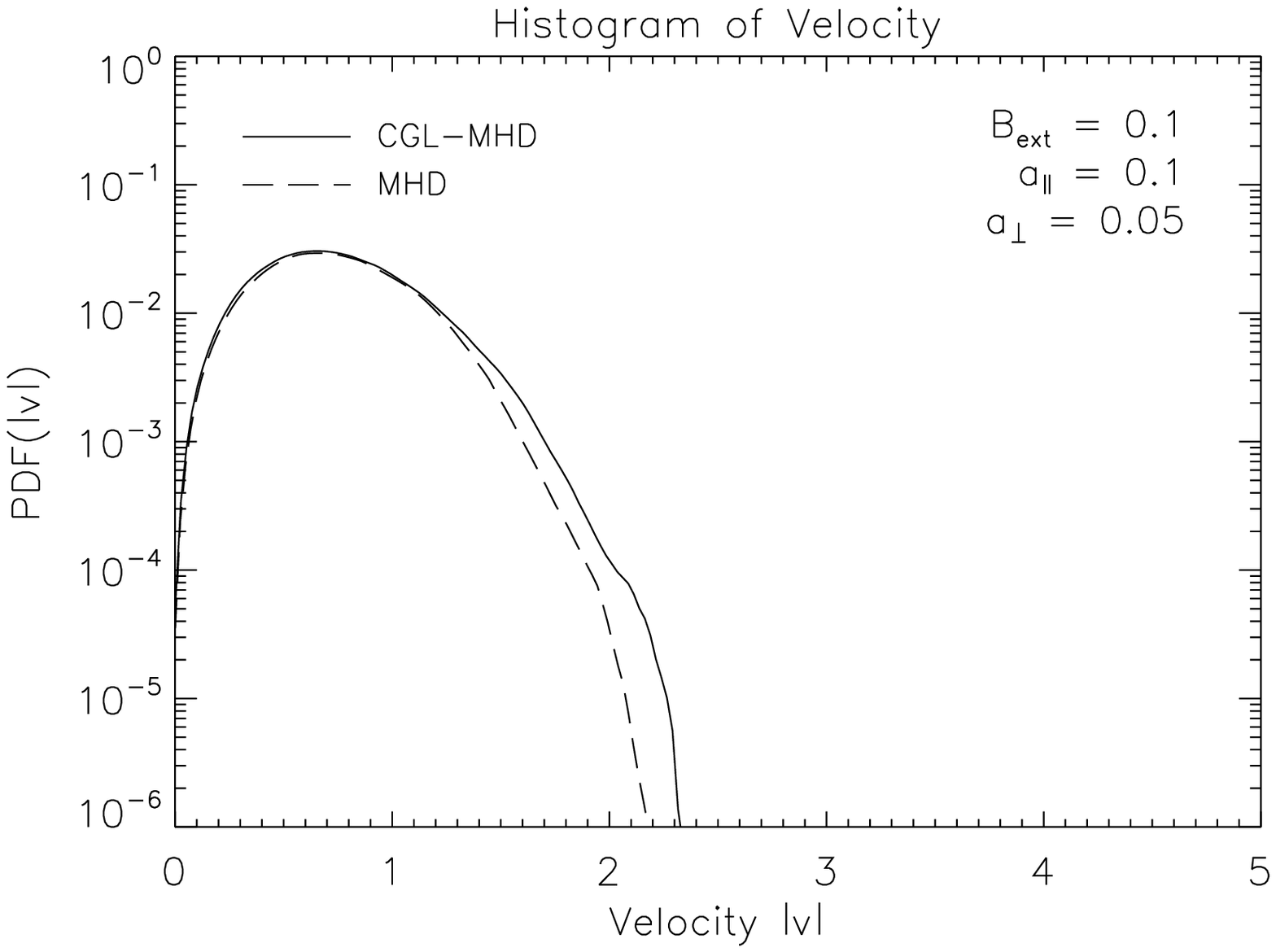}
 \includegraphics[width=0.32\textwidth]{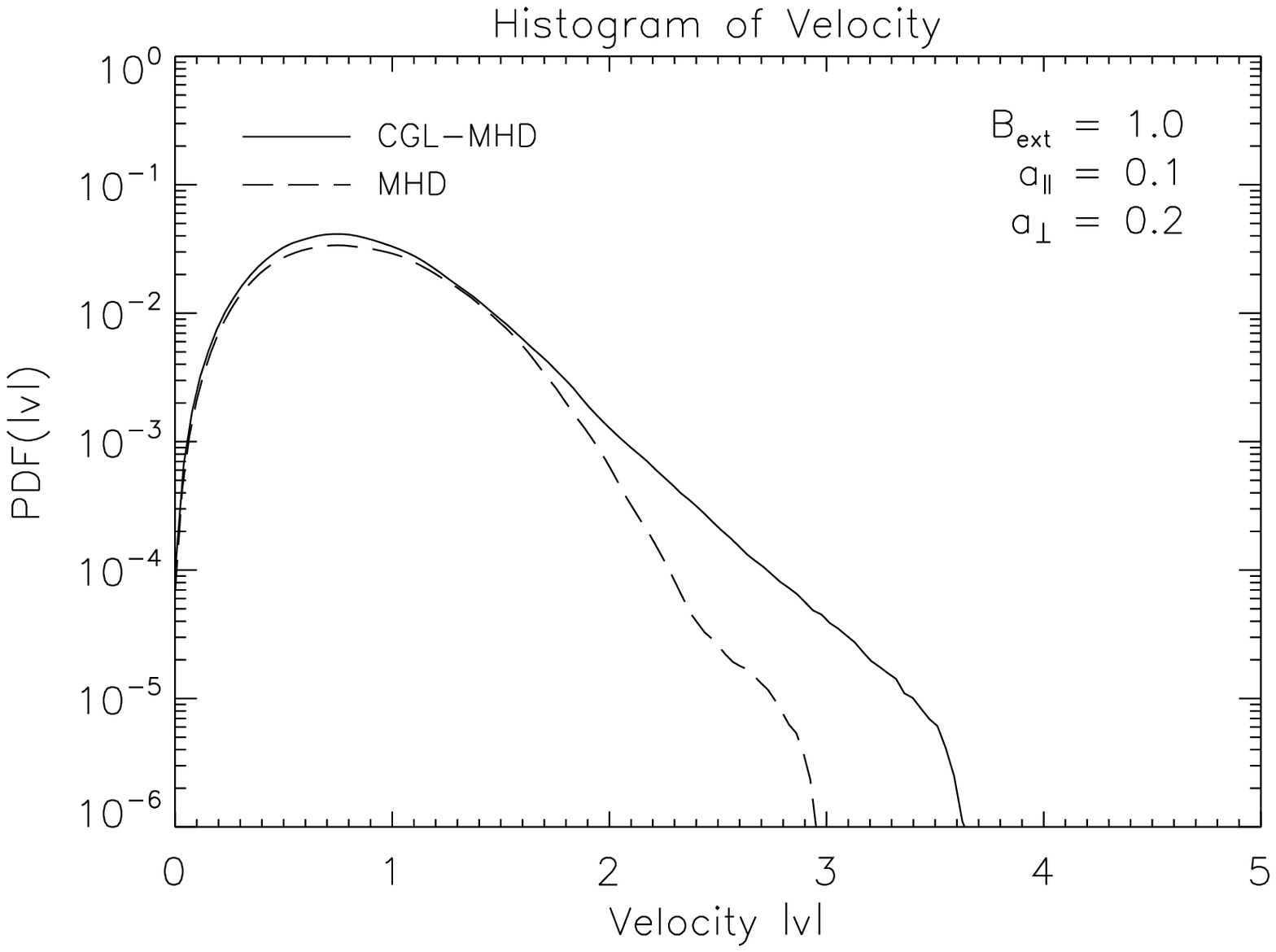}
 \caption{PDFs of velocity for the studied models. The left column shows models 1 and 2, the middle one shows models 3 and 4, and the right column shows models 4 and 5, according to Table~\ref{tb:models}.  For each case both, the CGL-MHD (solid lines) and MHD (dashed lines) models are shown for comparison. \label{fig:velo_pdfs}}
\end{figure*}

In Figure~\ref{fig:dens_pdfs} we present the probability distribution functions
(PDFs) of density obtained from each model.  In general, the distribution of gas
exhibits an increasing contrast with increasing sonic Mach number.  This result
is in agreement with the MHD models.  However, the kinetic instabilities cause
even larger density contrast in the models with weak turbulence, compared to the
MHD models.  As stated above, in models 1 and 2 the instabilities could freely
grow without being suppressed by the turbulent motions of the gas.  The PDFs of
density for the remaining models show no difference when compared to the MHD
models.  Surprisingly, in model 5 the firehose instability is responsible for
the granulated map of column density, but the PDF of density remains very
similar to the MHD case.

The PDFs of velocities are shown in Figure~\ref{fig:velo_pdfs}.  Again, the
supersonic and super-Alfvenic models (3 and 4) show small differences compared
to the standard MHD case.  On the other hand, the subsonic and sub-Alfvenic
models (1 and 2) present distorted velocity PDFs, with an increase in the high
velocity tail.  Both instabilities are responsible for more effective
acceleration of the plasma, broadening the distribution of velocities.  In the
case of the firehose instability, the weakly magnetized model presents more
evident changes, as noted in model 5.  Here, since the magnetic field is weak
the perpendicular flows are able to destroy the magnetic field configuration.
Therefore, the magnetic breaking is not important, what results in flows with
higher velocities.  The same process is responsible for the changes in the PDF
of model 6.  Interestingly, for model 6 we obtained narrower PDF with lower
velocities when compared to the MHD case.  As the turbulence develops and
unstable cells arise the local magnetic field grows.  The strong magnetic
breaking takes place resulting in a low velocity distribution.

\subsection{Power Spectra of Density and Velocity}

\begin{figure*}[tbh]
 \center
 \includegraphics[width=0.49\textwidth]{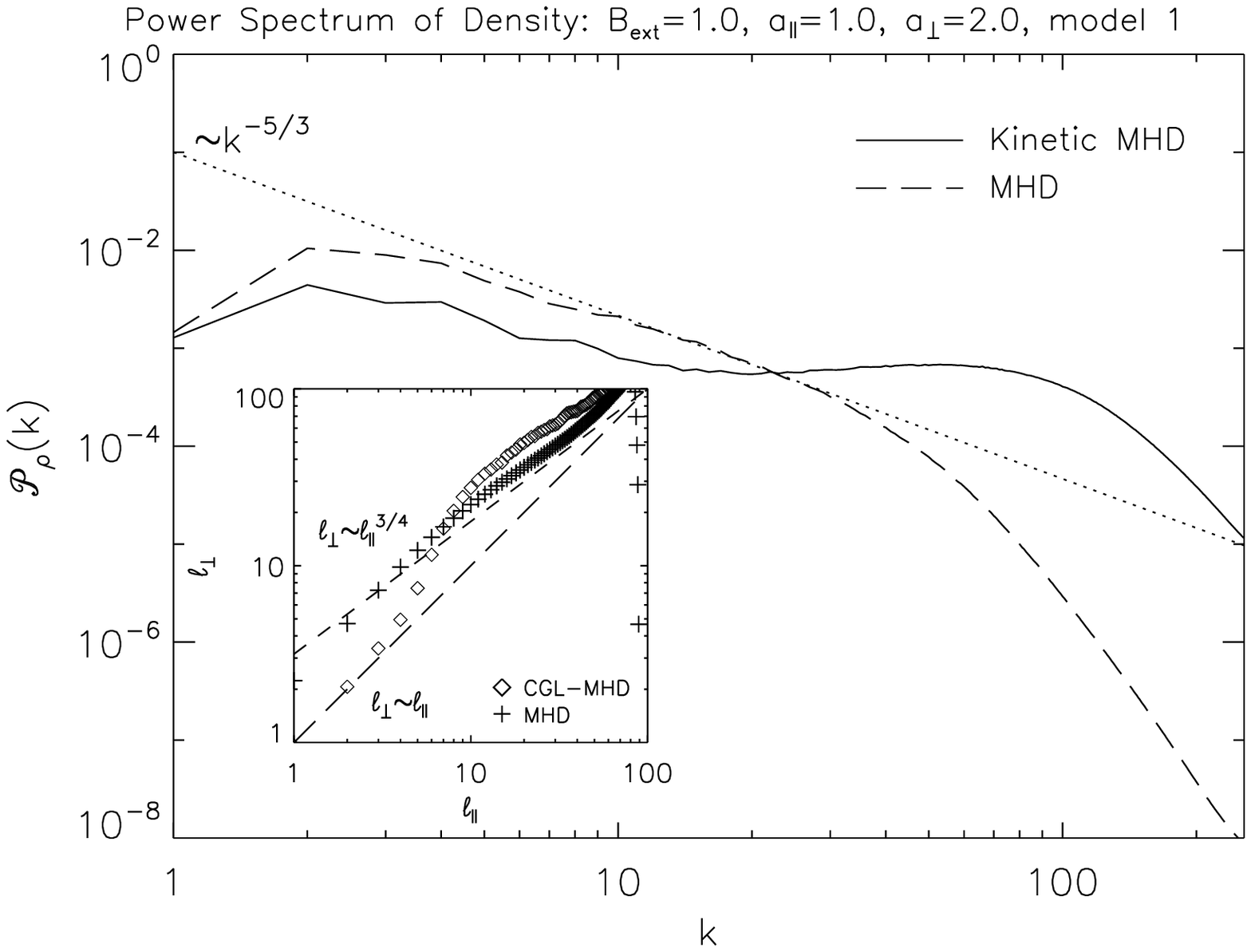}
 \includegraphics[width=0.49\textwidth]{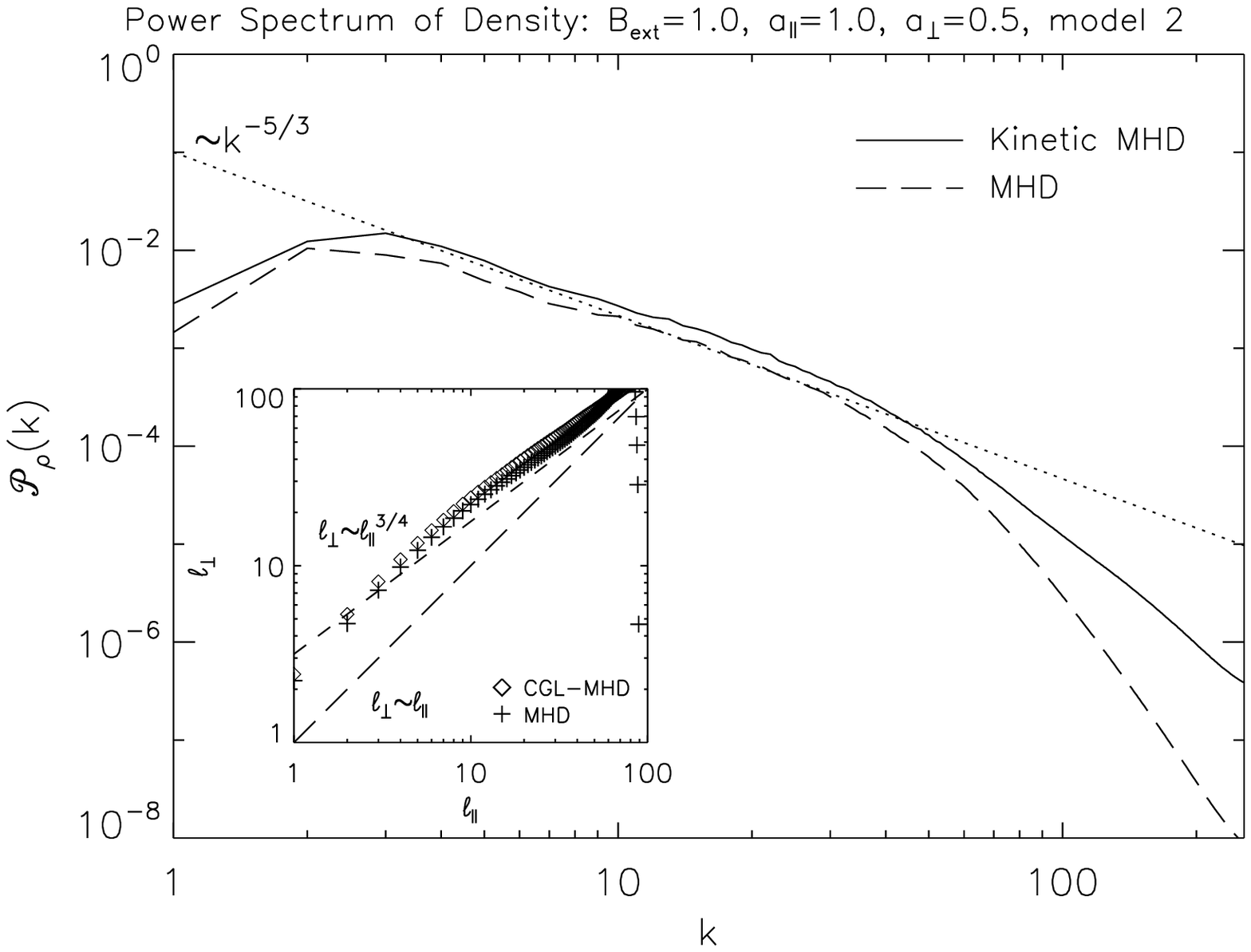}
 \includegraphics[width=0.49\textwidth]{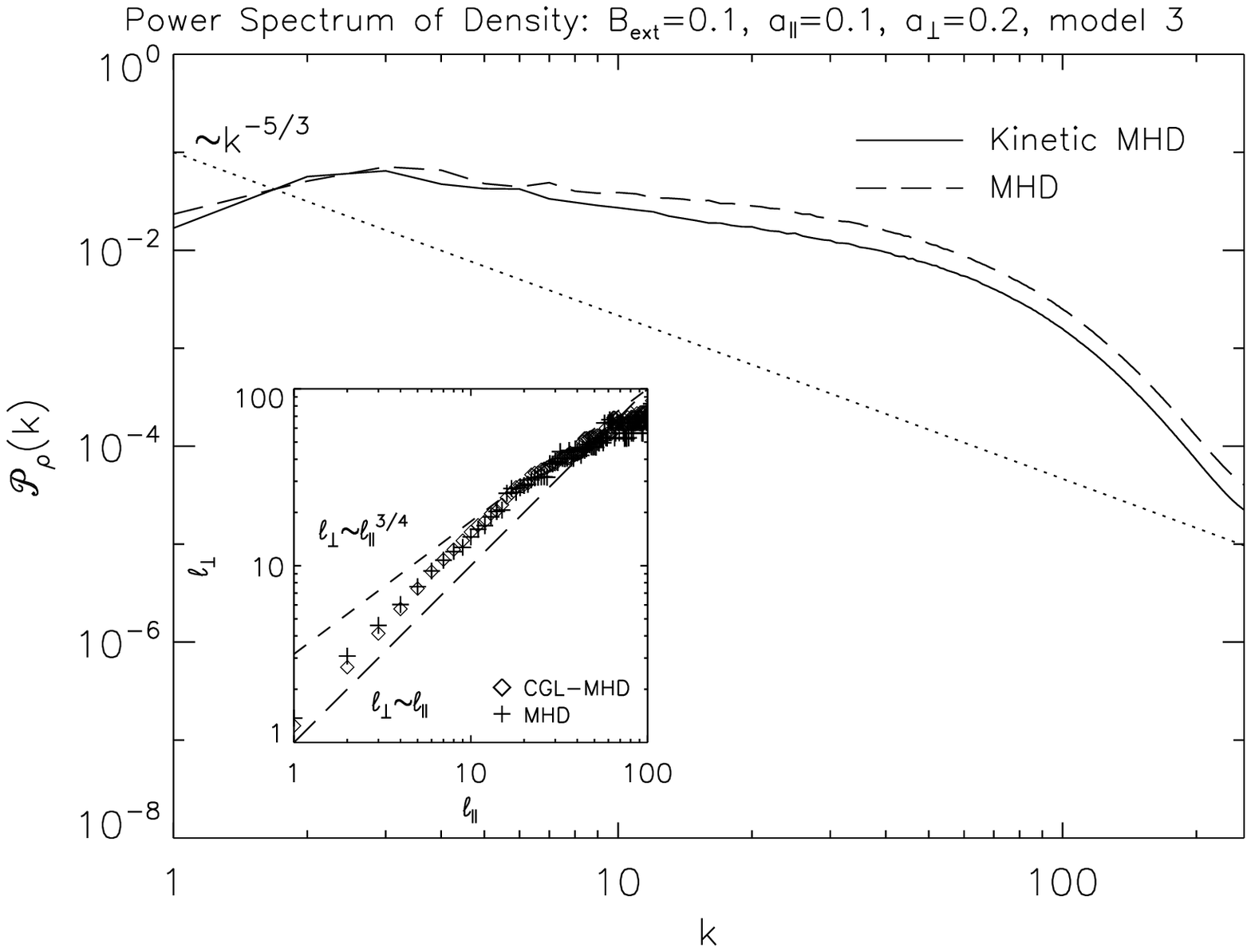}
 \includegraphics[width=0.49\textwidth]{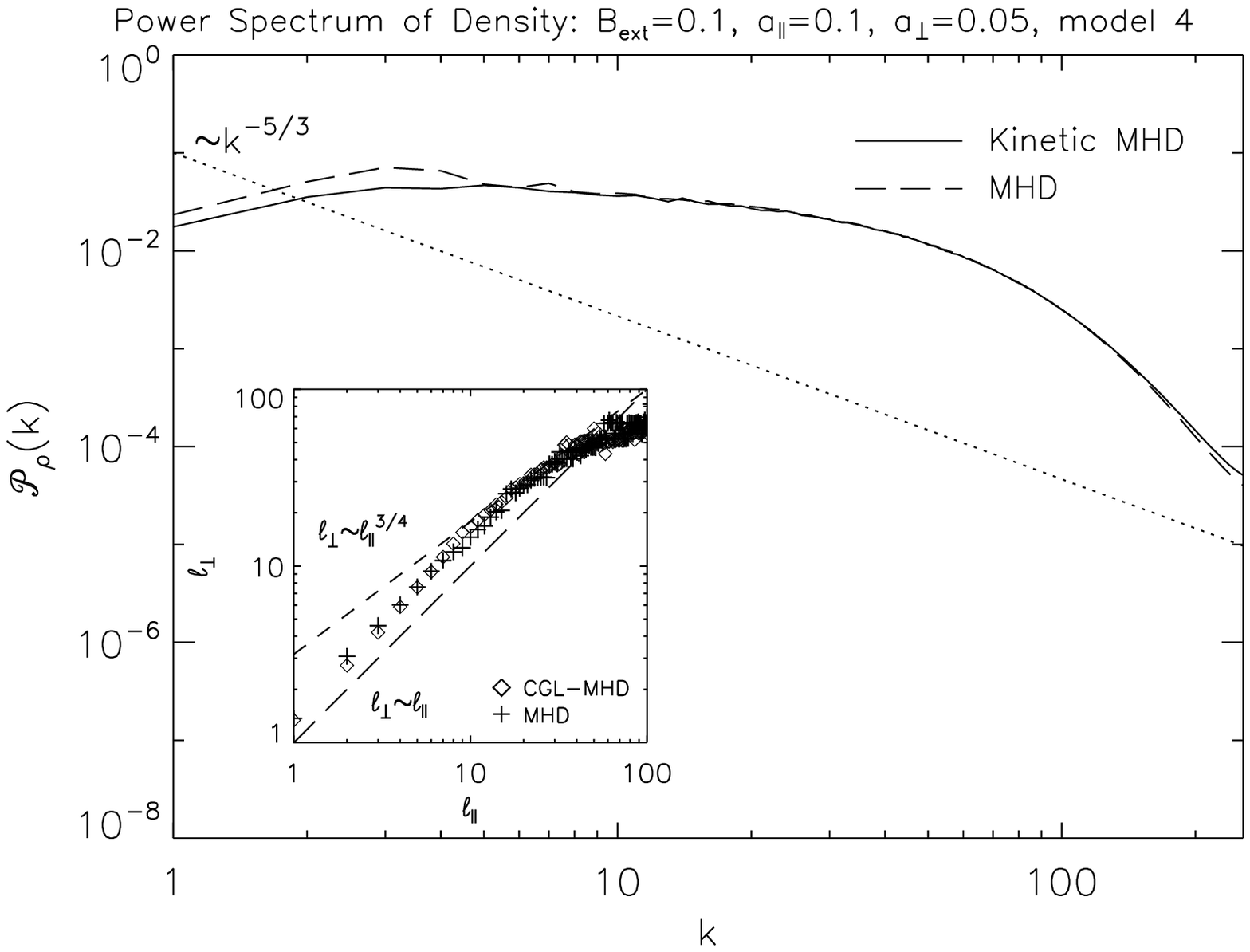}
 \includegraphics[width=0.49\textwidth]{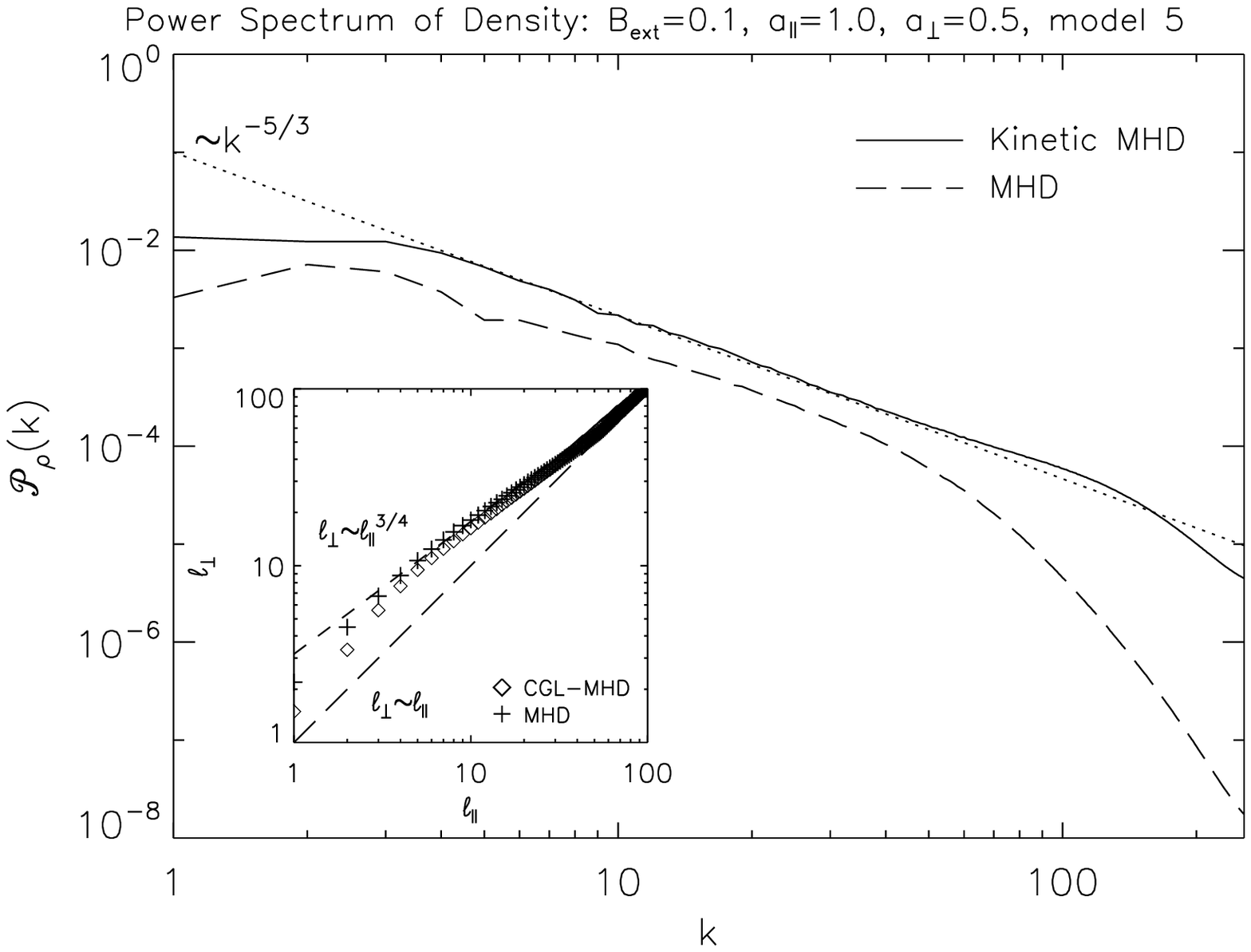}
 \includegraphics[width=0.49\textwidth]{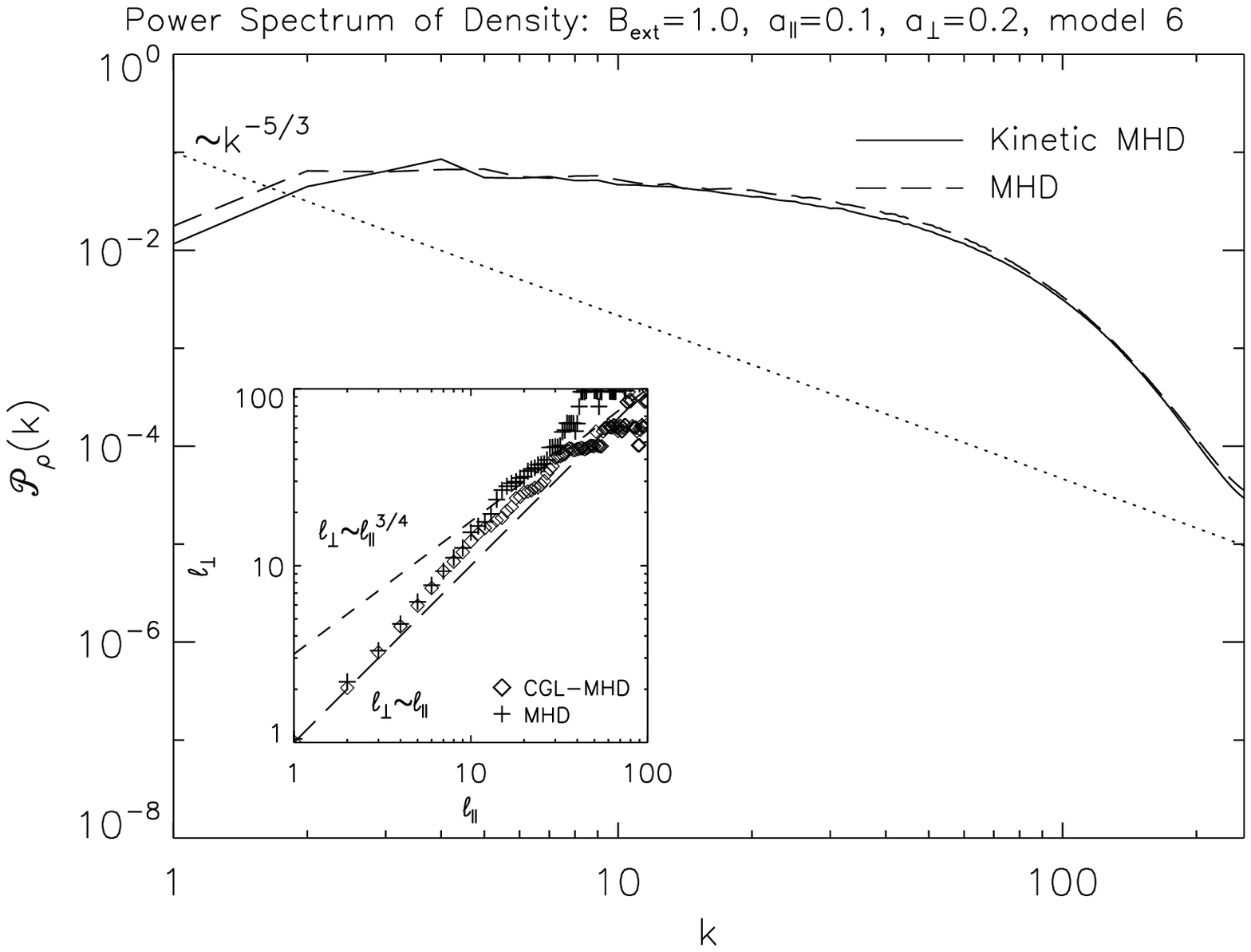}
 \caption{Spectra of density for the studied models, models 1 to 3 in the upper
row and models 4 to 6 in the lower row, according to Table~\ref{tb:models}.  For
each case both, the MHD (solid lines) and CGL-MHD (dashed lines) models are
shown for comparison. \label{fig:dens_spec}}
\end{figure*}

\begin{figure*}[tbh]
 \center
 \includegraphics[width=0.49\textwidth]{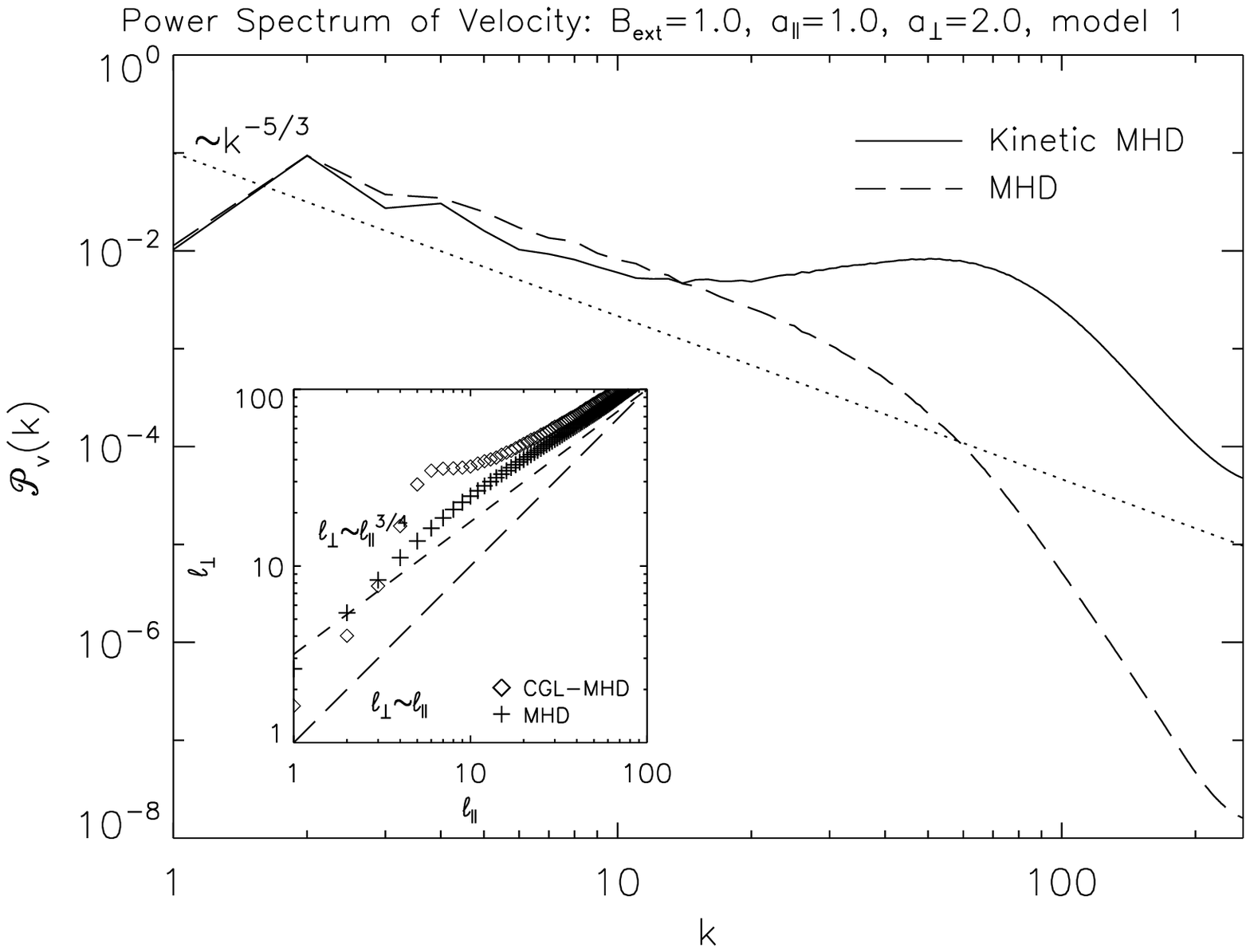}
 \includegraphics[width=0.49\textwidth]{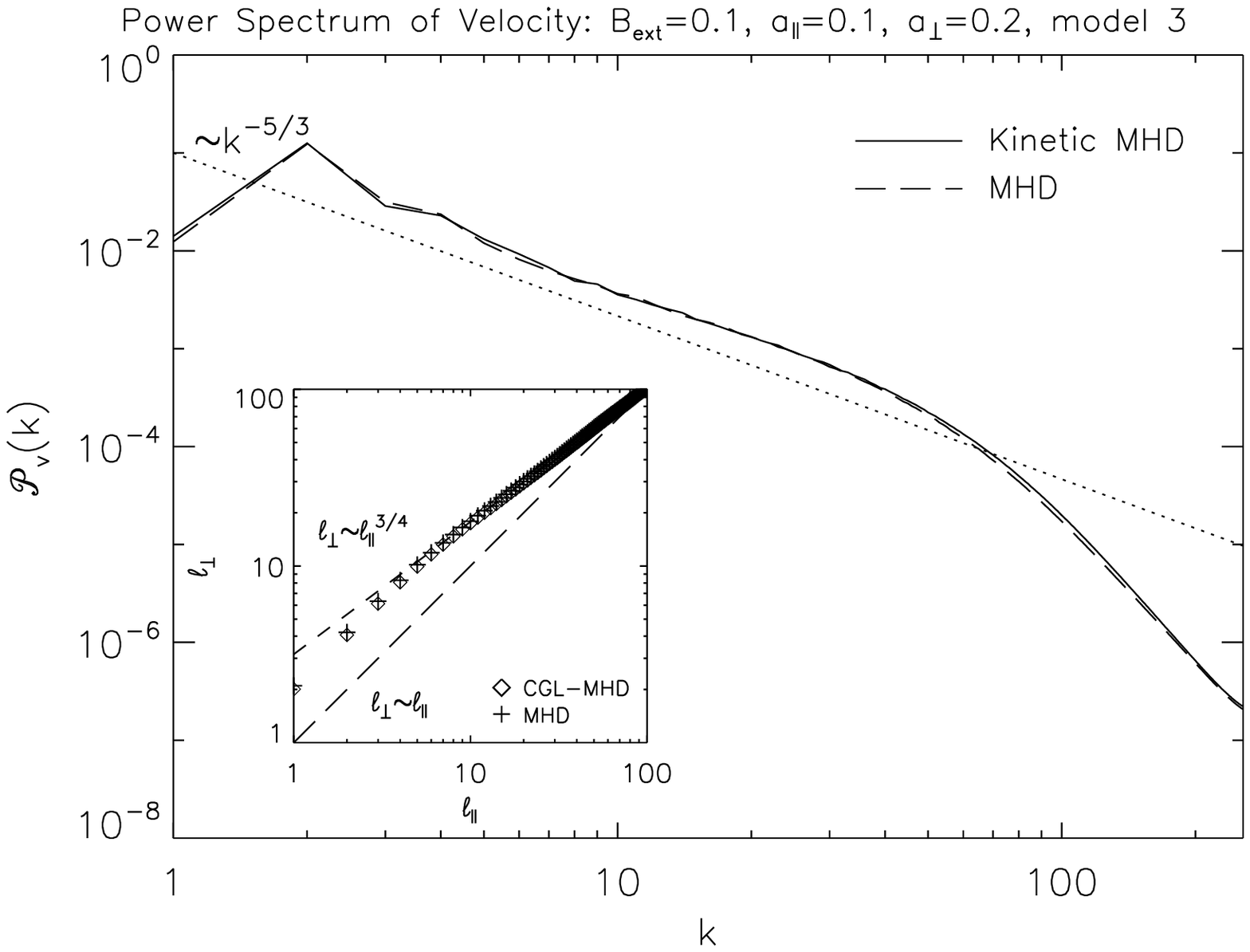}
 \includegraphics[width=0.49\textwidth]{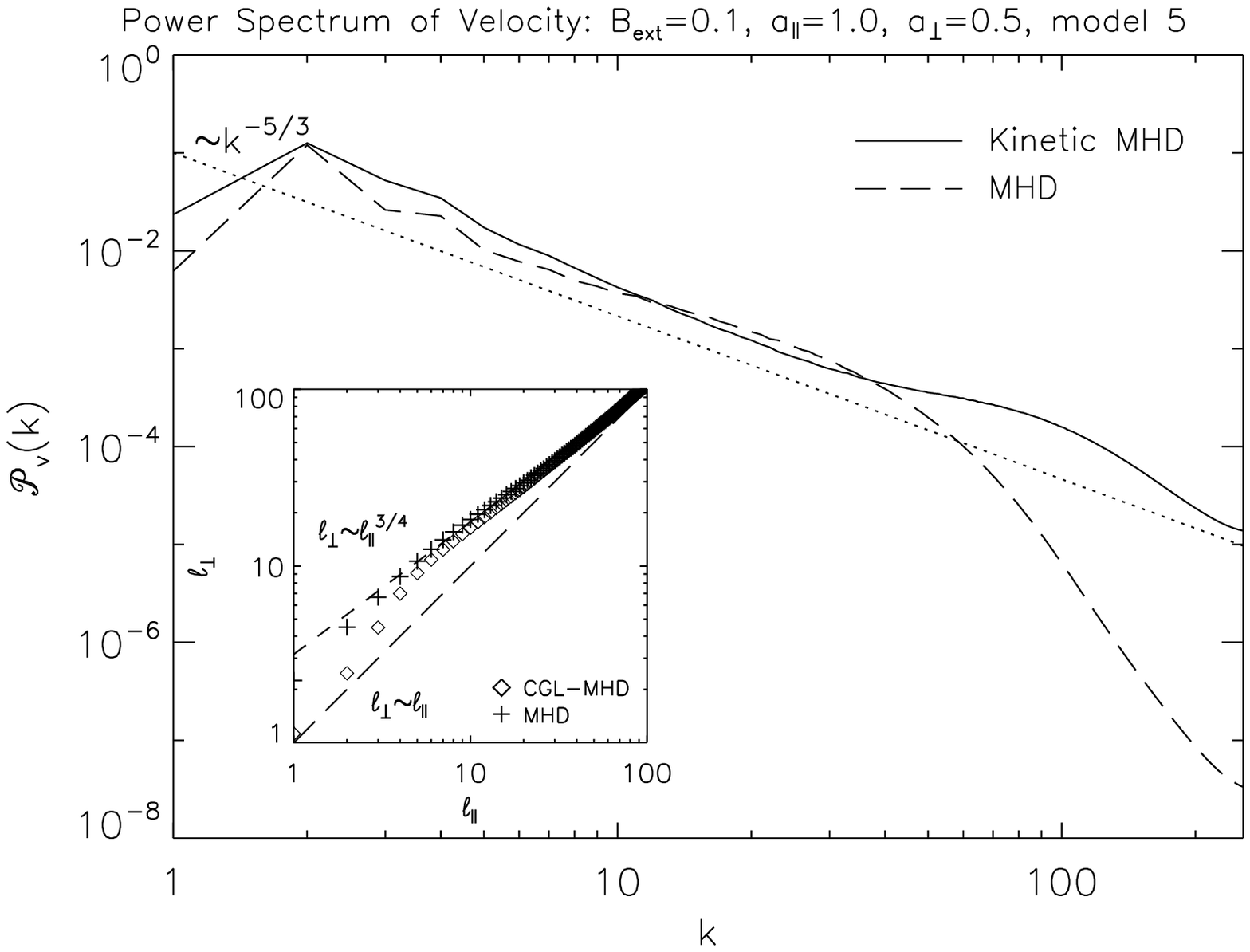}
 \includegraphics[width=0.49\textwidth]{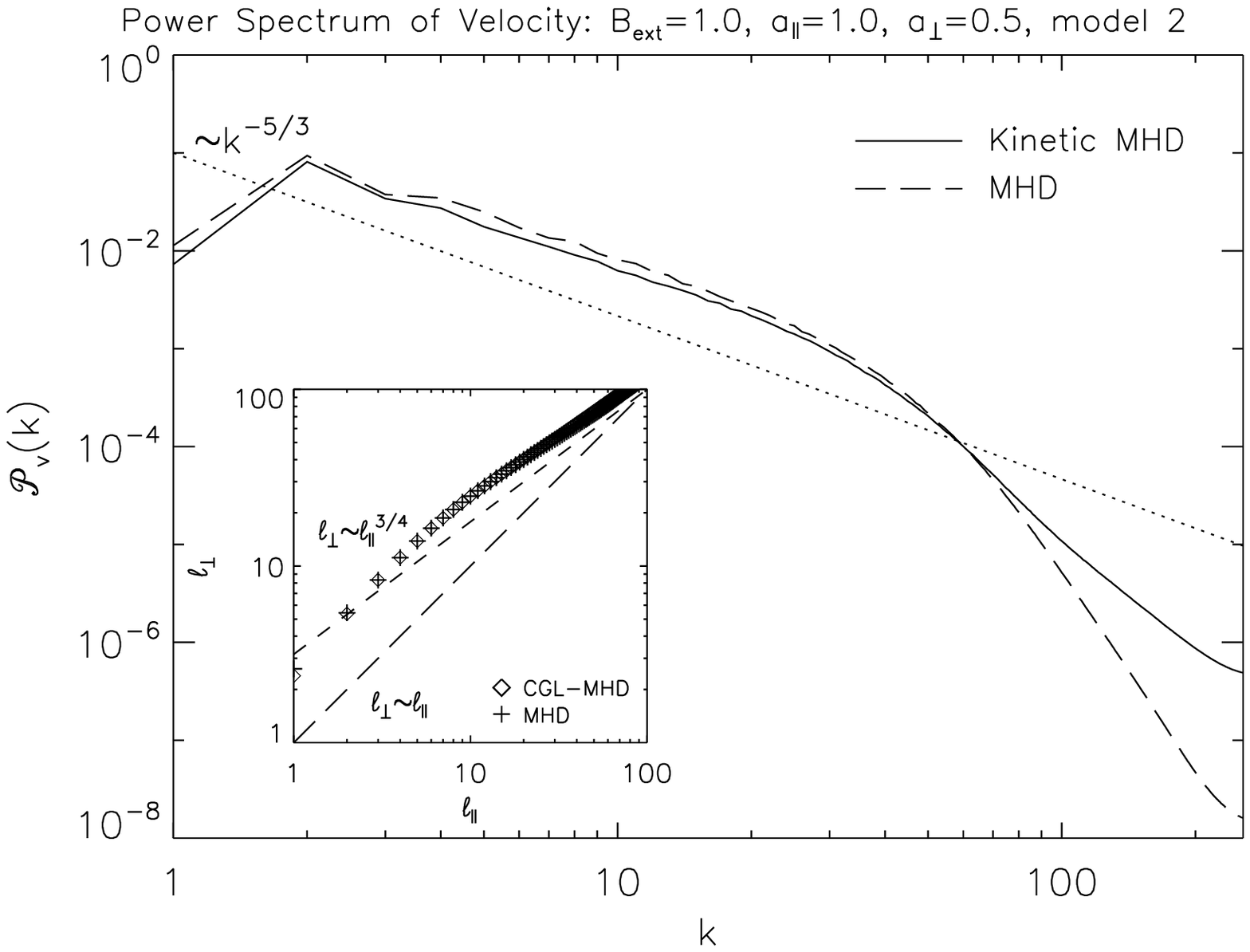}
 \includegraphics[width=0.49\textwidth]{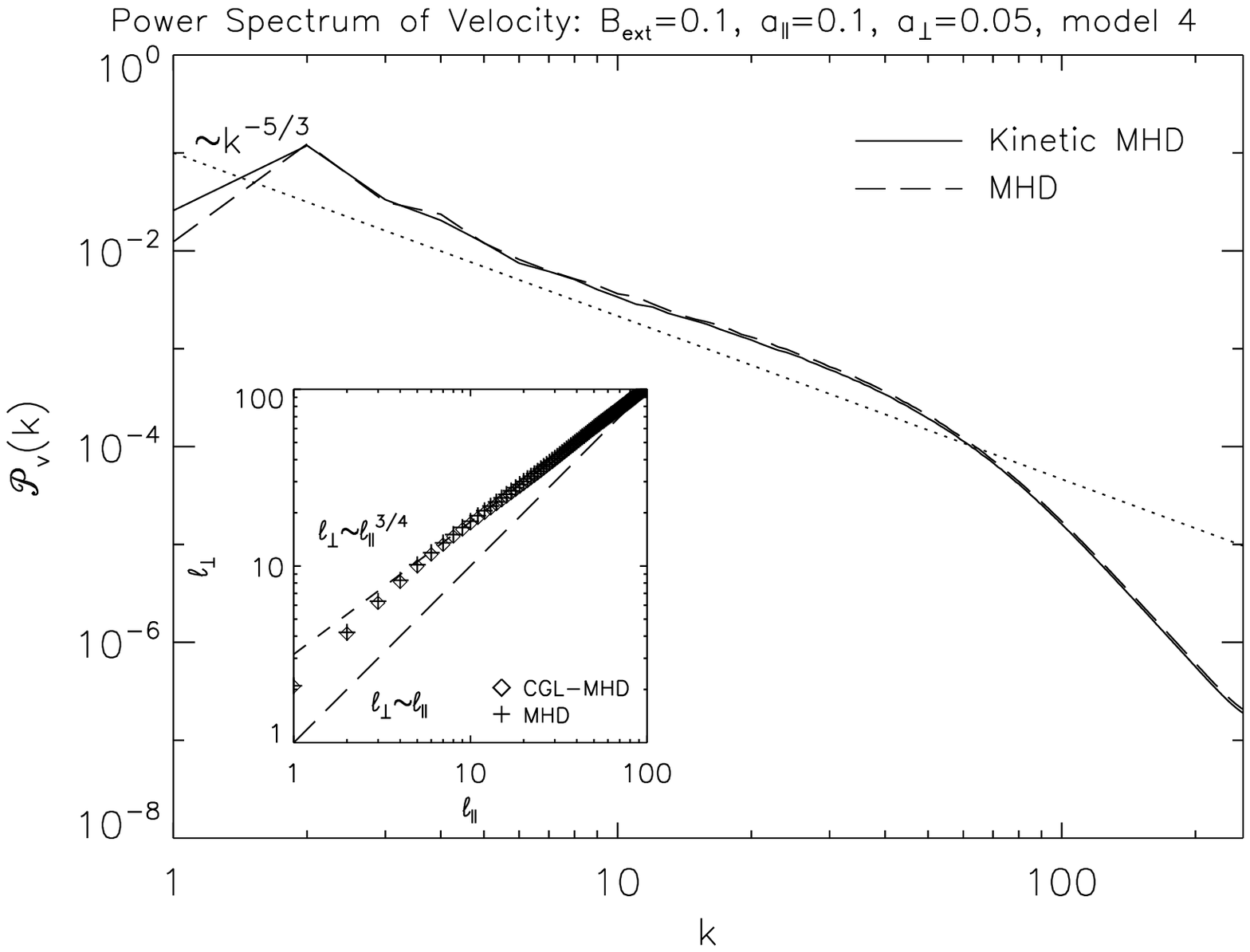}
 \includegraphics[width=0.49\textwidth]{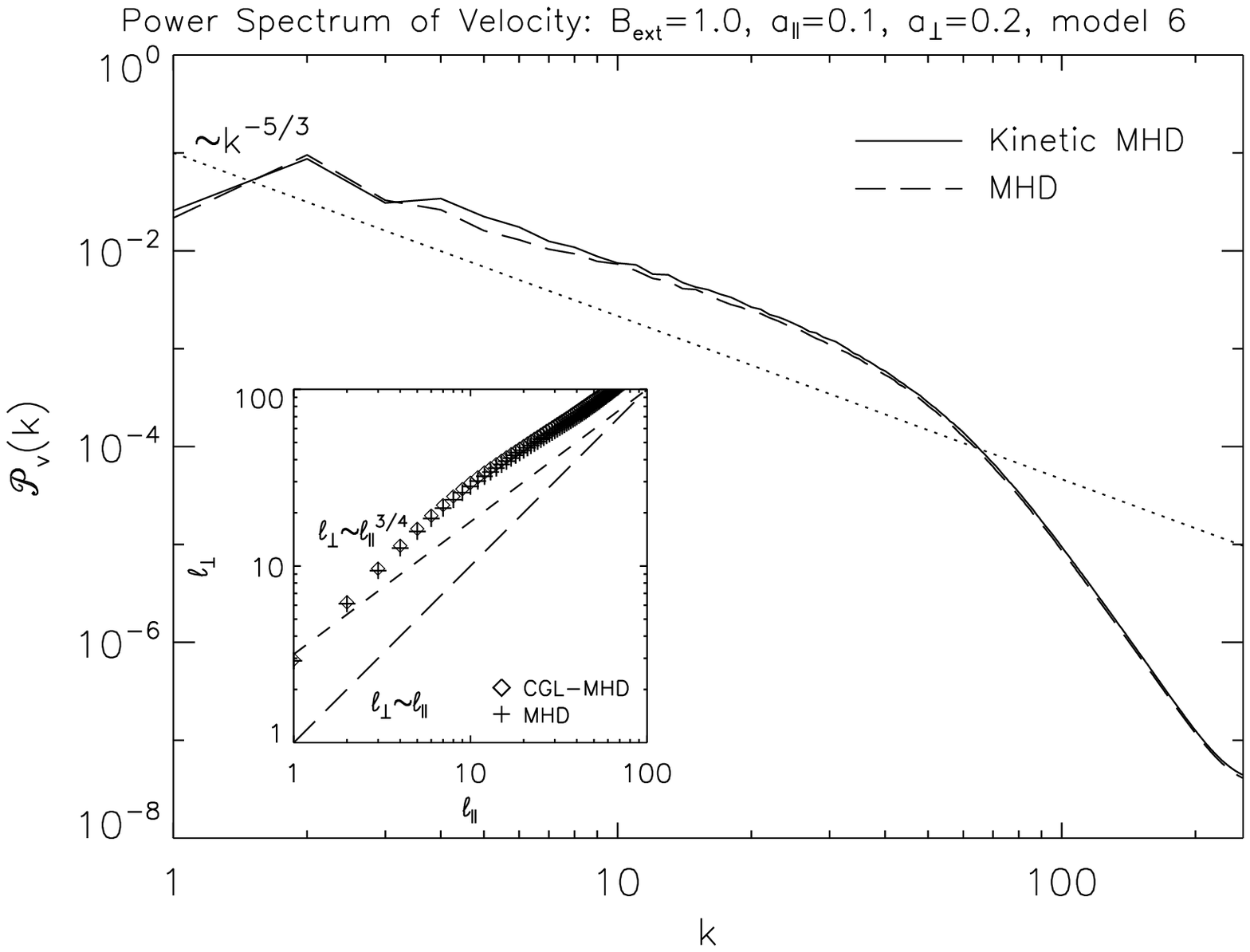}
 \caption{Spectra of velocity for the studied models, models 1 to 3 in the upper
row and models 4 to 6 in the lower row, according to Table~\ref{tb:models}.  For
each case both, the MHD (solid lines) and CGL-MHD (dashed lines) models are
shown for comparison. \label{fig:velo_spec}}
\end{figure*}

In order to characterize and study the changes in the energy cascade, as well as
the correlations between different scales in CGL-MHD models, we analyze the
power spectra.  In Figures~\ref{fig:dens_spec} and \ref{fig:velo_spec} we
present the spectra of fluctuations of density and velocity for all models.
Within each plot, we also present the anisotropy of the spectra for parallel and
perpendicular directions with respect to the global magnetic field.

At the center, we show the plots for Models 3 and 4.  As also shown previously,
there is no substantial difference between the CGL-MHD and the MHD simulations.
The density spectra present similar slope ($\alpha \sim -0.5$) at the inertial
range, while a slope of $\sim -2$ is obtained for the velocity spectra.
Regarding the anisotropy ($l_\parallel \propto l_\perp ^\zeta$), we found the
typical relations $\zeta \sim 1$ at small scales, but the Goldreich and Sidhar
(1995) slope $\zeta \sim 2/3$ at larger scales, for both MHD and CGL-MHD
simulations.  A similar result in spectra and anisotropy of density was obtained
for Model 6, with exception of a slight increase in the spectrum at small
scales.

Clearly, differences appear between the spectra of the MHD and CGL-MHD
simulations in Models 1, 2 and 5.  In these cases, it is noticeable the power
excess at large values of $k$ due to the fast growth of the instabilities in
these scales.  The slopes of velocity and density spectra for the MHD models
range between -1.7 to -2.0 at the inertial range, while CGL-MHD simulations show
positive slopes ($\sim +1$ for Models 1 and 5) in velocity spectra, and flat
density spectra ($\alpha \sim 0$).  Regarding the spectral anisotropy, as an
interesting result, we see that the firehose instability reduces the
anisotropies regarding the global magnetic field lines.  A more detailed study
of the anisotropy of density and velocity perturbations regarding the local
magnetic field is given below.

\subsection{Structure functions}

In the previous section we showed, from power spectra of density and velocity
fluctuation, that the instabilities in CGL-MHD models efficiently transport
power from large to small scales, where their concentration is increased.
Interesting modifications from MHD turbulence was also shown from the spectral
anisotropy regarding the global magnetic field. However, since the instabilities
are dominant at small scales, mapping the structure functions regarding the
local magnetic field seems to be a better approach if we want to determine the
role of the instabilities on the isotropization of fluctuations.

The second order structure function of a given parameter $f$ is defined as:
\begin{equation}
{\rm SF}(l) = \left< \left| f \left({\bf r}+{\bf l }\right) - f \left( {\bf r} \right) \right|^2 \right>,
\end{equation}
where $\bf{r}$ is the referenced position and $\bf{l}$ is the distance
calculated along the magnetic field line. The SF is calculated by randomly
choosing a large number of referenced positions for each studied correlation
length $l$. Here, to account for the importance of magnetic field fluctuation at
small scales, we calculate the correlation length $l$ along the field lines,
i.e. in the magnetic field reference frame.

\begin{figure*}[tbh]
 \center
 \includegraphics[width=0.32\textwidth]{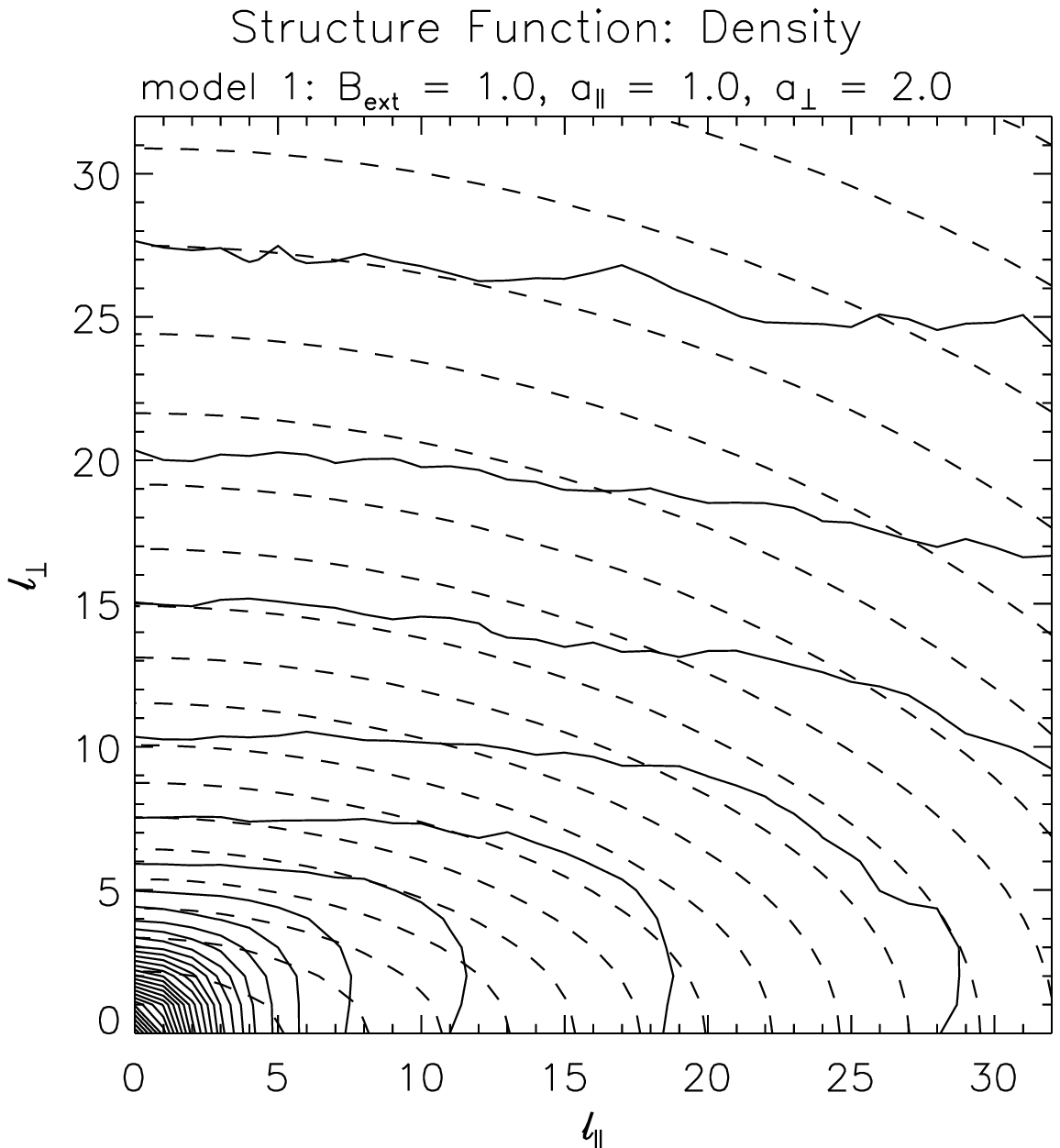}
 \includegraphics[width=0.32\textwidth]{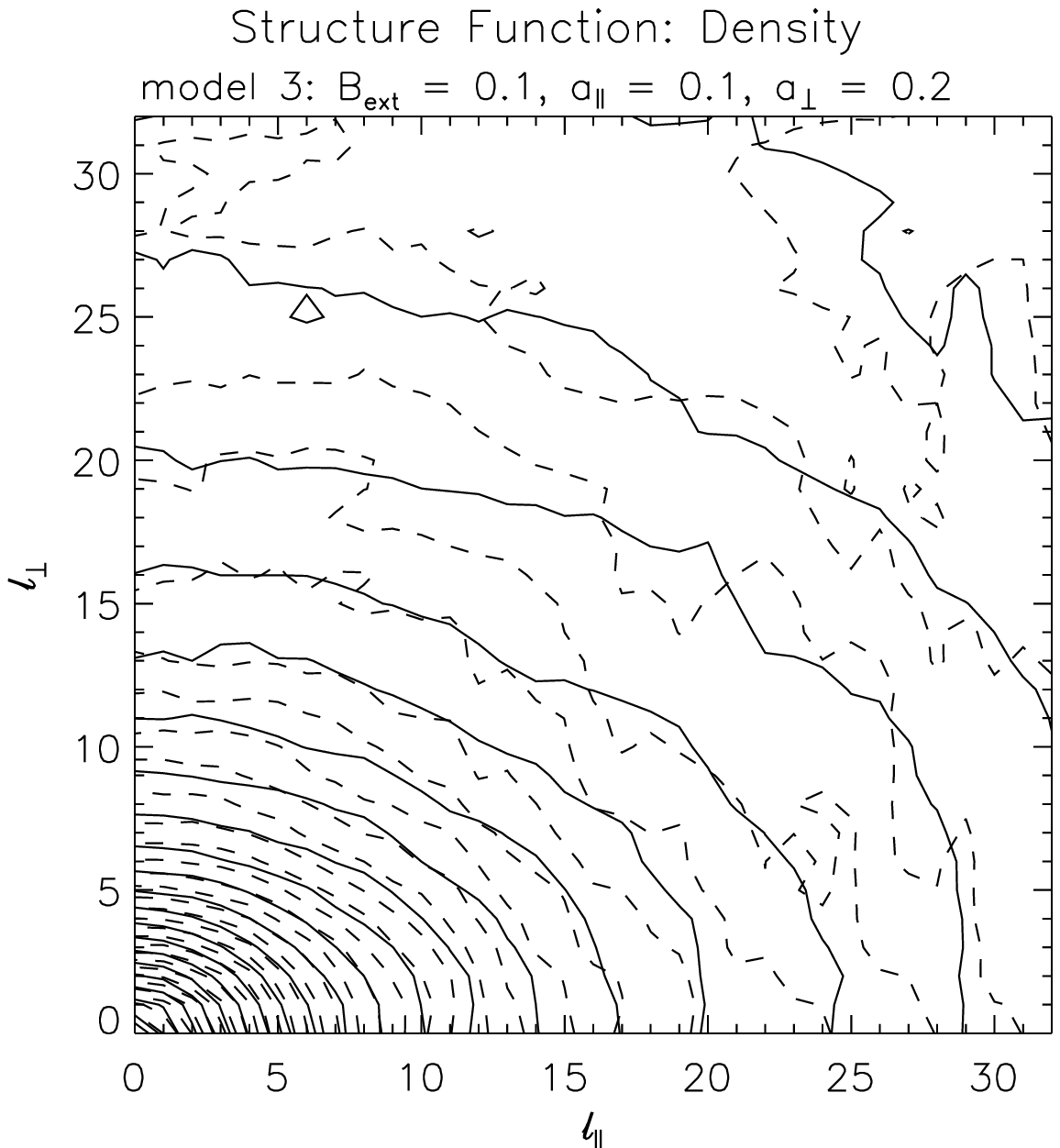}
 \includegraphics[width=0.32\textwidth]{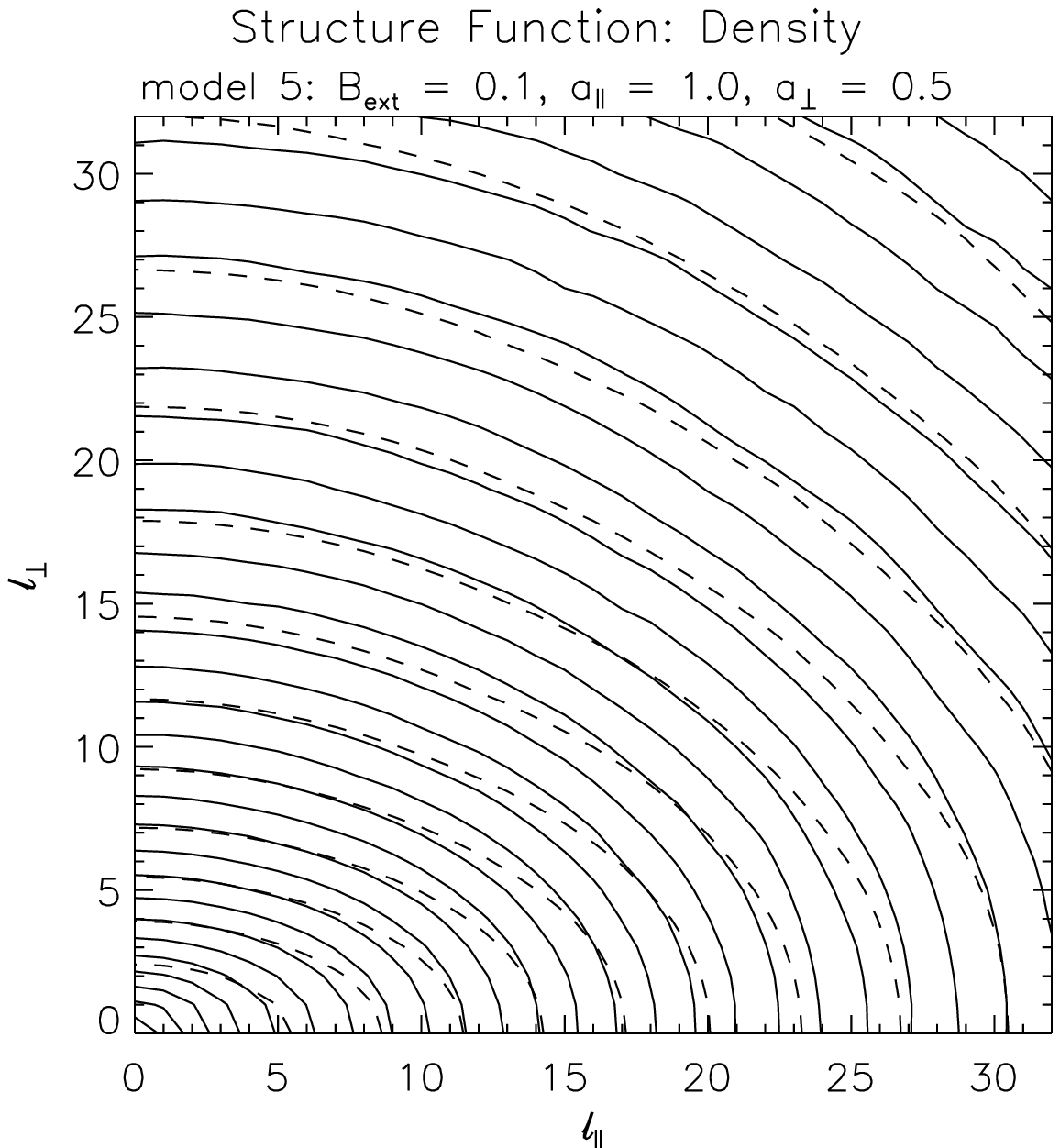}
 \includegraphics[width=0.32\textwidth]{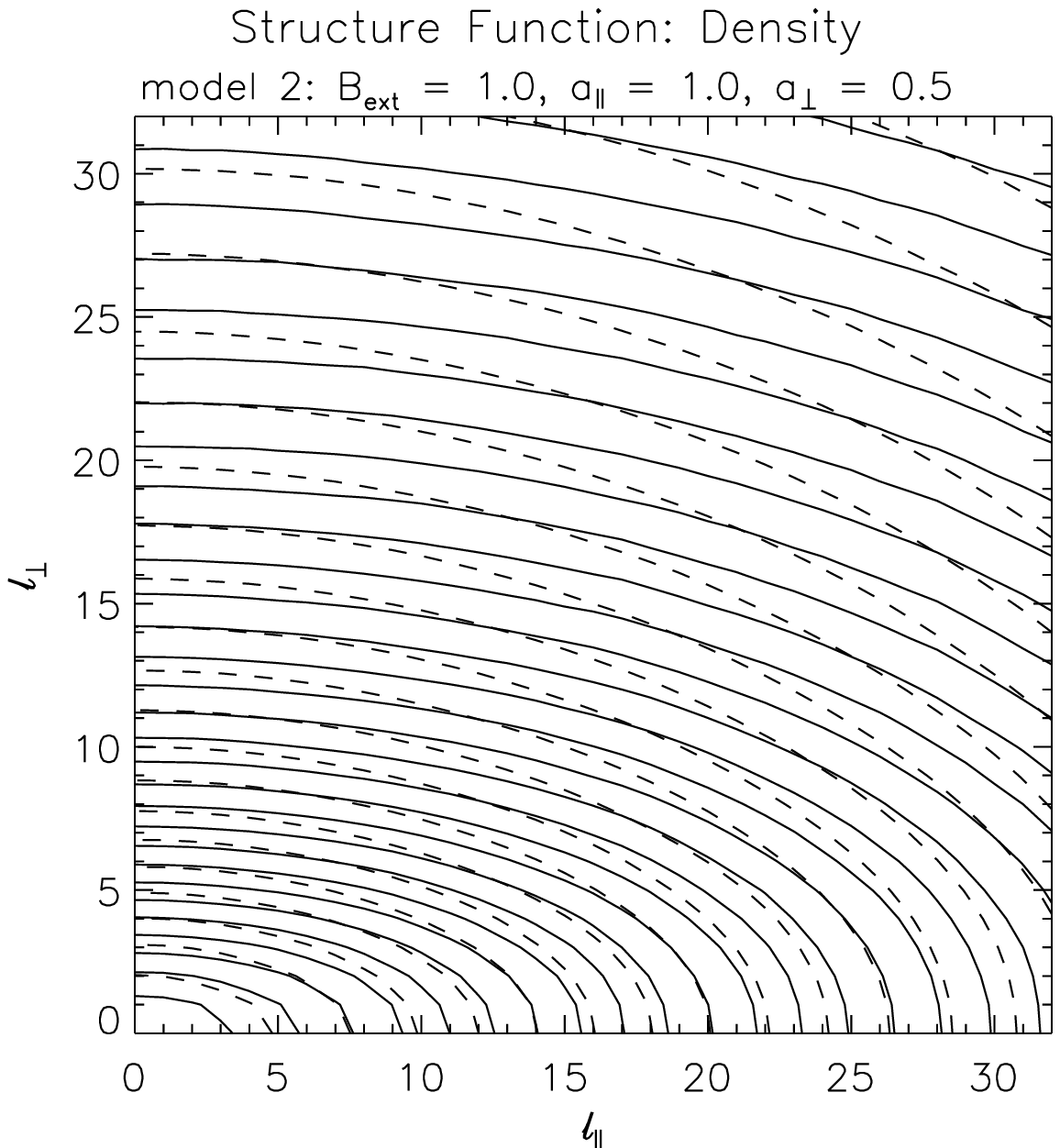}
 \includegraphics[width=0.32\textwidth]{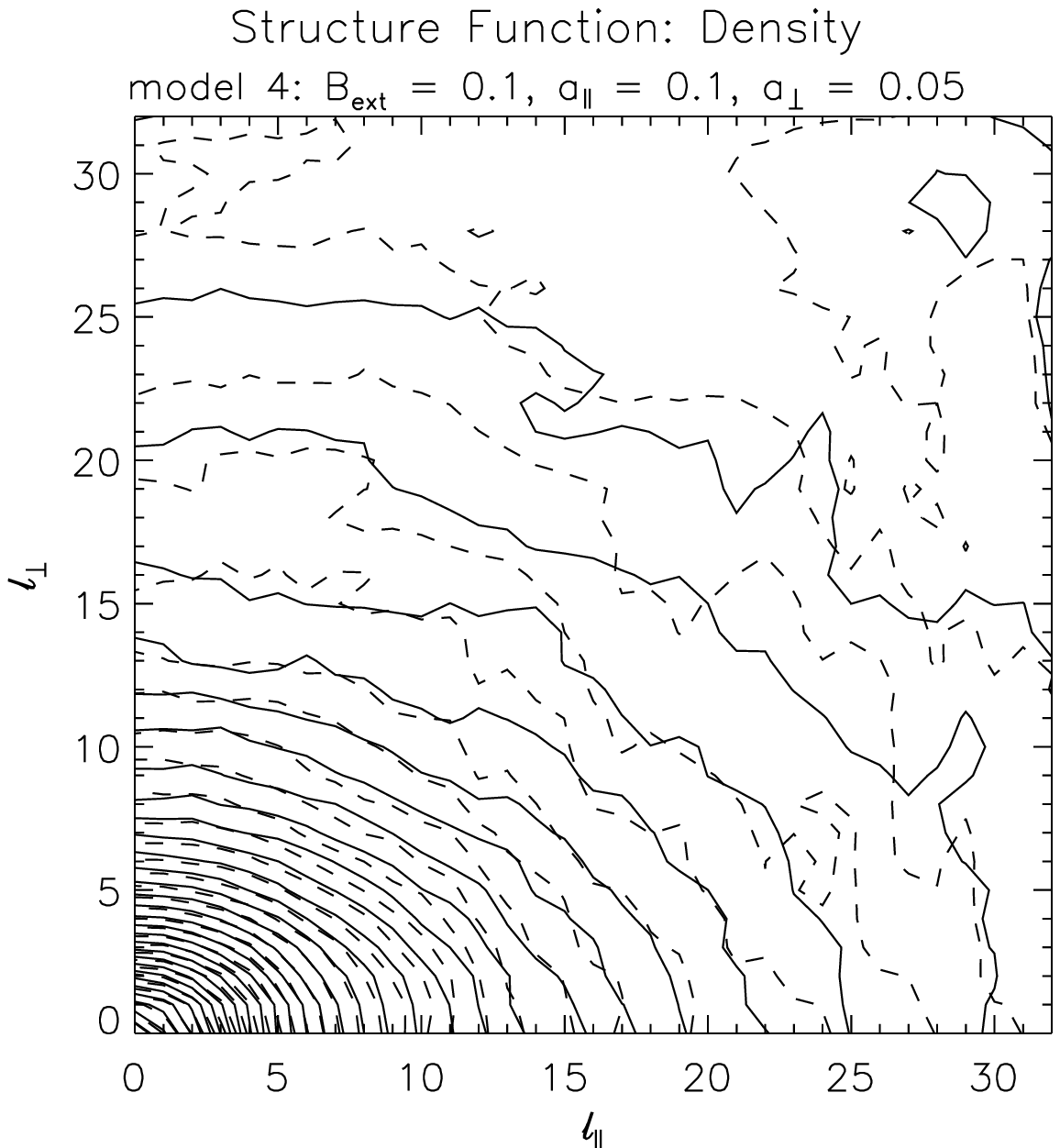}
 \includegraphics[width=0.32\textwidth]{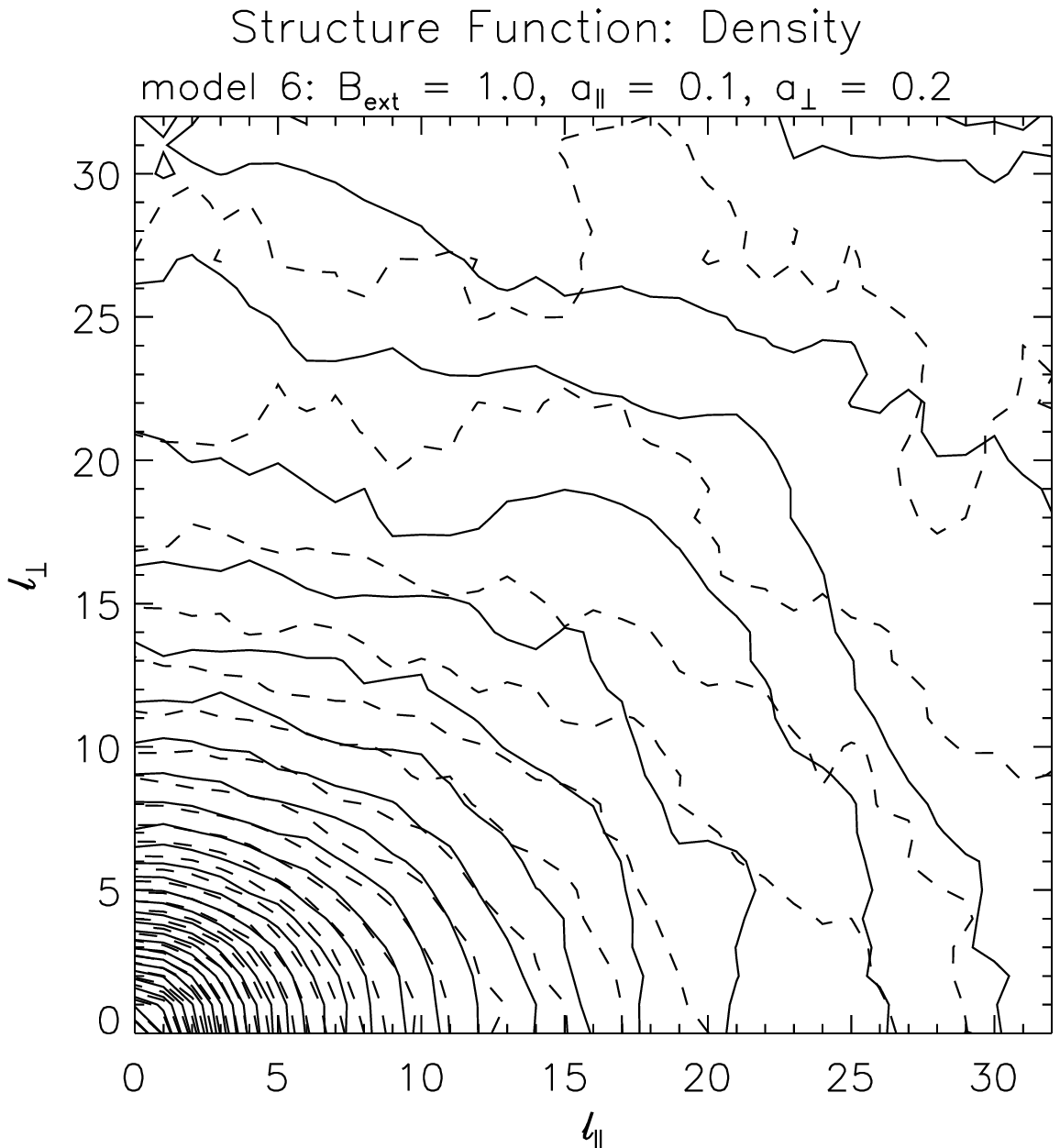}
 \caption{Structure functions of the density in the local reference frame for the studied models. In the left column we show models 1 and 2, inte middle column models 2 and 3, and in the right column, models 5 and 6, according to Table~\ref{tb:models}.  For each case both, the CGL-MHD (solid lines) and MHD (dashed lines) models are shown for comparison. \label{fig:dens_sf}}
\end{figure*}

\begin{figure*}[tbh]
 \center
 \includegraphics[width=0.32\textwidth]{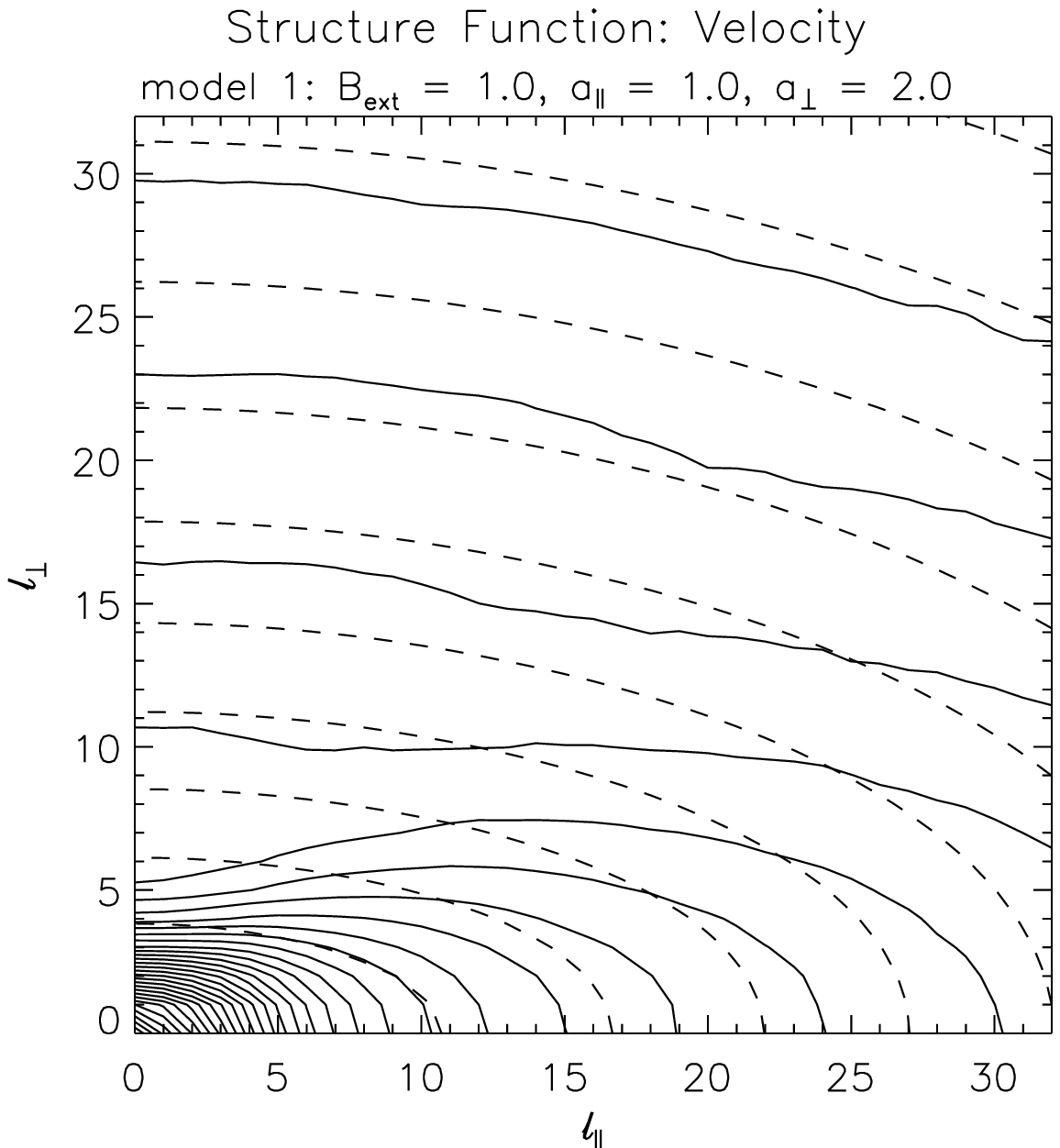}
 \includegraphics[width=0.32\textwidth]{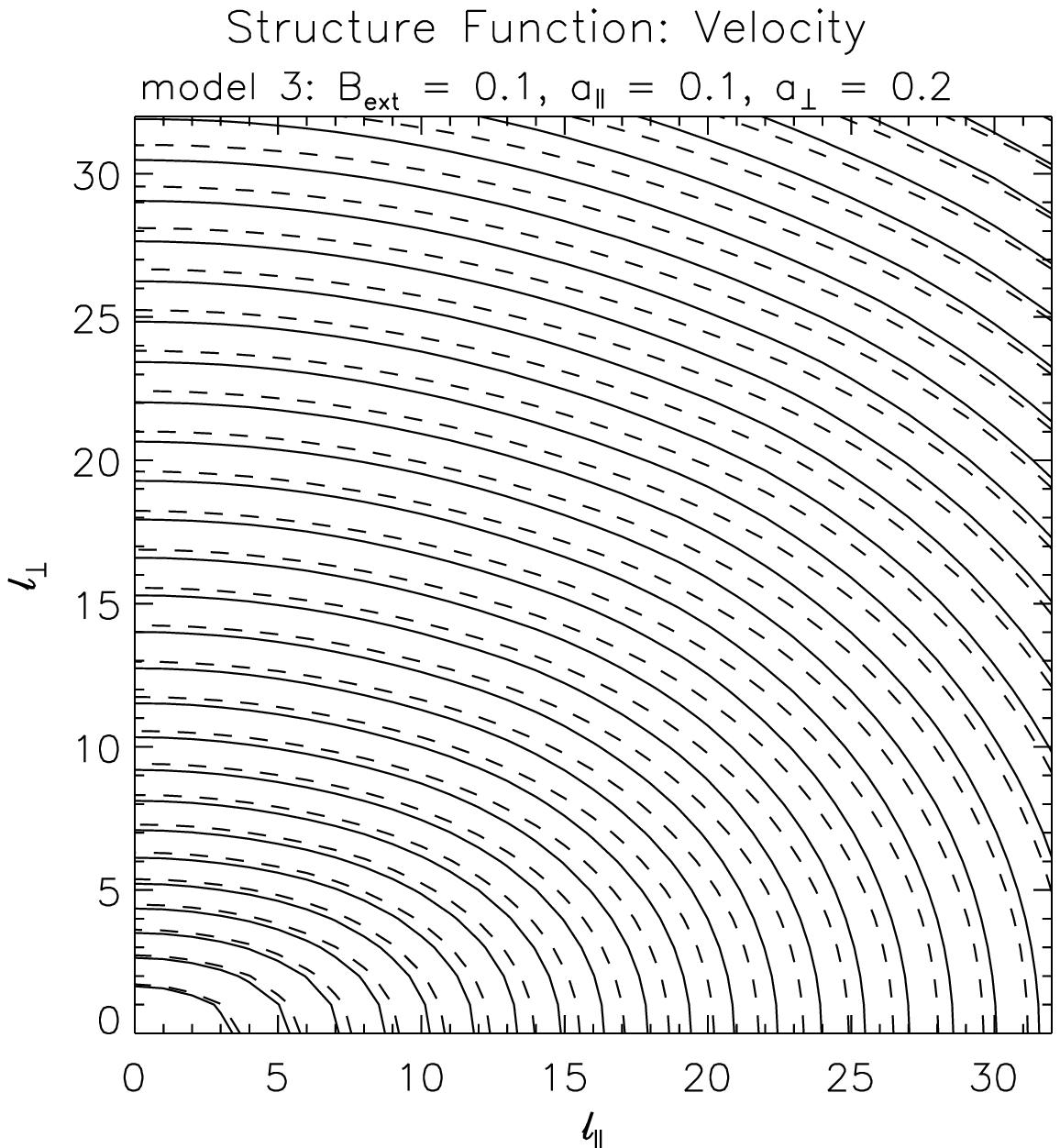}
 \includegraphics[width=0.32\textwidth]{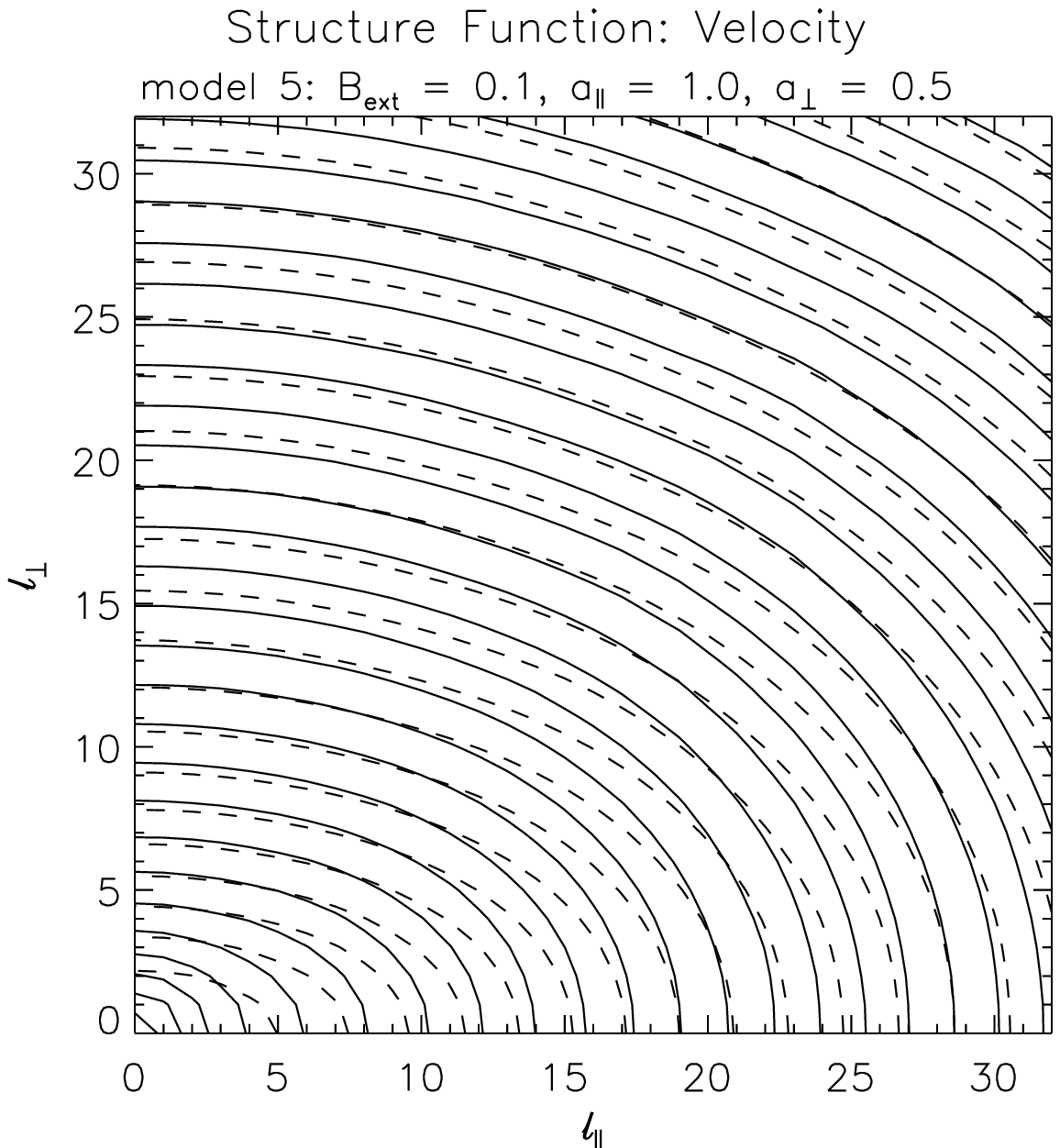}
 \includegraphics[width=0.32\textwidth]{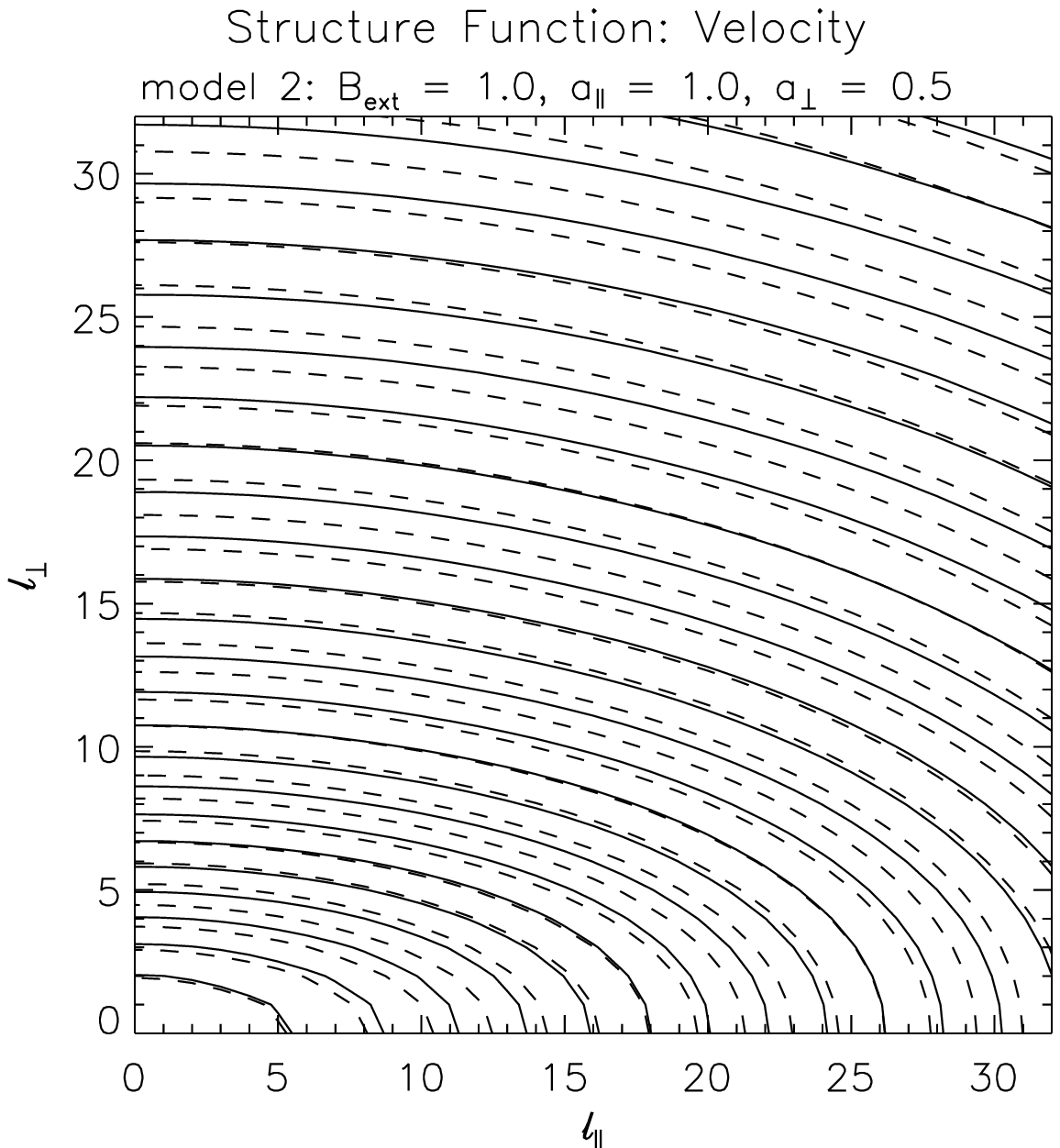}
 \includegraphics[width=0.32\textwidth]{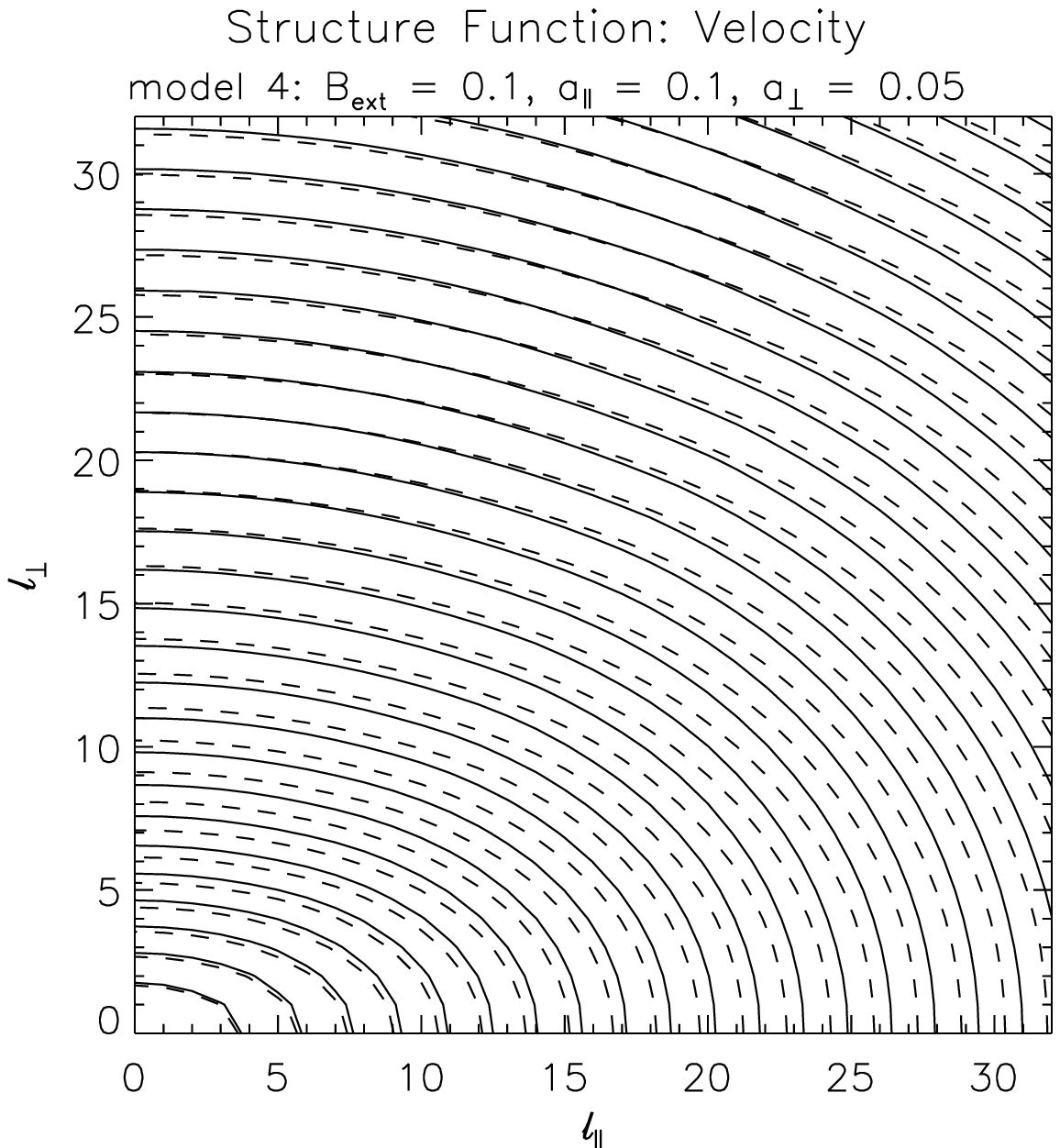}
 \includegraphics[width=0.32\textwidth]{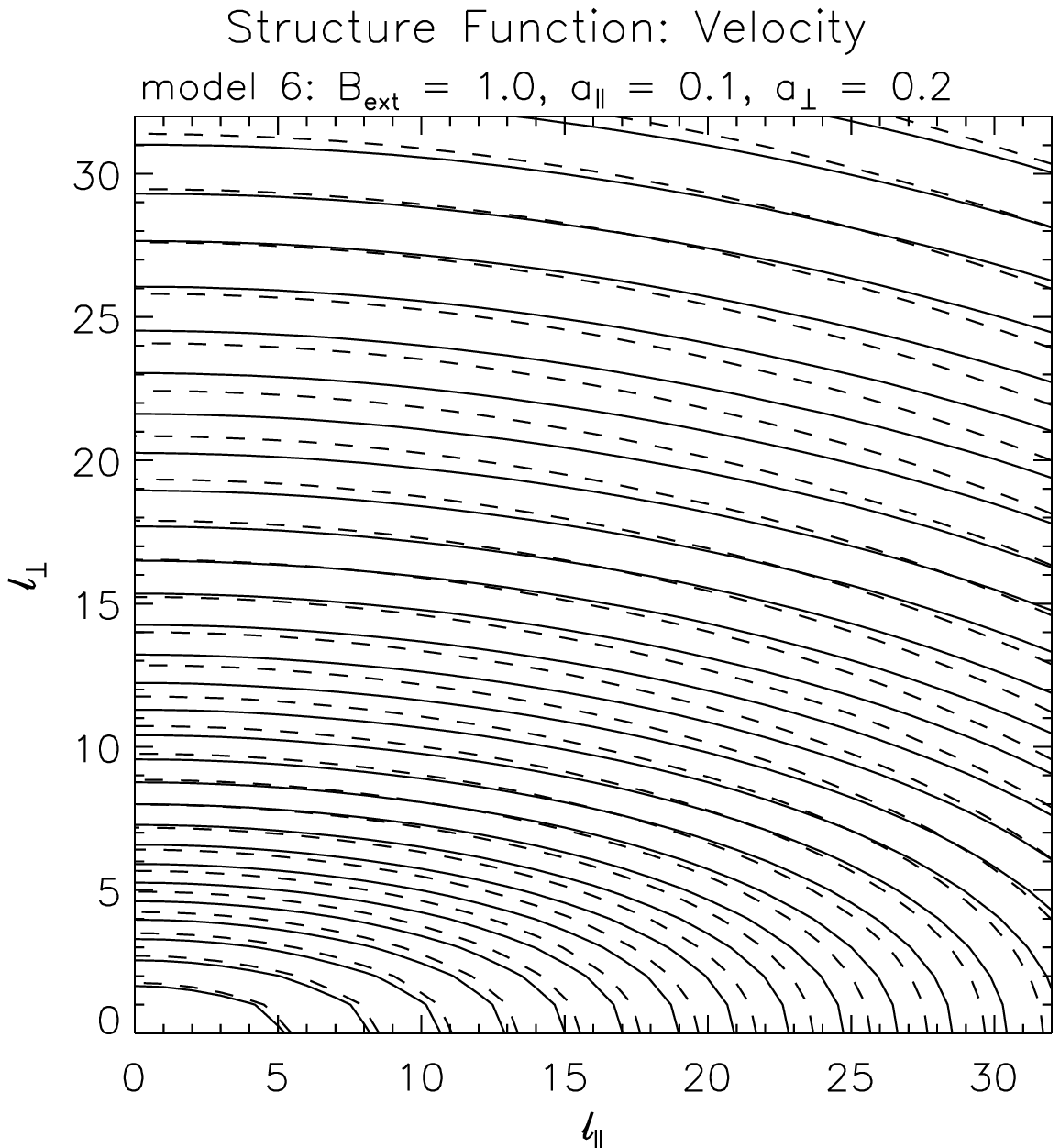}
\caption{Structure functions of velocity in the local reference frame for the studied models. In the left column we show models 1 and 2, inte middle column models 2 and 3, and in the right column, models 5 and 6, according to Table~\ref{tb:models}.  For each case both, the CGL-MHD (solid lines) and MHD (dashed lines) models are shown for comparison. \label{fig:velo_sf}}
\end{figure*}

The obtained SFs for density and velocity fluctuations are shown in Figures
~\ref{fig:dens_sf} and \ref{fig:velo_sf}, respectively. In all plots, the solid
lines represent the results obtained from the MHD simulations, while the dashed
lines were obtained from the CGL-MHD simulations. The numbers refer to each
model, as described in Table~\ref{tb:models}.

Similar to the previous calculations, Model 3 shows no difference between the
two theoretical approaches, independent on the scale. The same result is
obtained for Model 6, which also showed similar spectra for the CGL-MHD and MHD
models. Surprisingly, the structure functions of Model 4, which showed no
differences in PDF and spectra, present more isotropic maps in the CGL-MHD
model. The same behavior is obtained for Models 2 and 5. These are the models
presenting $p_\parallel/p_\perp = 2$, i.e. the firehose instability. Here, the
firehose instability is responsible for changes in the magnetic field topology,
tangling the field lines, resulting in an increase of the perpendicular pressure
in the local reference frame. Obviously, this effect is over estimated in these
calculations because of the double-isothermal approximation. Otherwise, the
magnetic field topology would not change drastically, but the interchange of
energy would cause a reduction of parallel pressure anyway. In this sense, the
result would be the same, i.e. the firehose instability is responsible for a
isotropization of the fluctuations with respect to the magnetic field lines.
Model 1, on the other hand, presented a larger anisotropy for the CGL-MHD
model. Here, as $p_\parallel/p_\perp = 0.5$, the free energy is mostly converted
to kinetic pressure. The mirror instability is responsible for an increase in
the acceleration of the plasma along the field lines (as already seen in the PDF
of velocity), increasing the anisotropy of the fluctuations. The result is more
elongated structures, mostly at small and intermediate scales, as also noticed
in the power spectra.

\section{Discussion}
\label{sec:discuss}

\subsection{Time Evolution of Unstable Systems}

In order to understand the evolution of the system during the growth and
saturation of the instabilities, we calculated the instability condition for the
Alfv\'en and the compressible modes for each cell of the computational domain.
The dispersion relation of these modes reveal the cells where the instability
condition is fulfilled.

\begin{figure*}
 \center
 \includegraphics[width=0.45\textwidth]{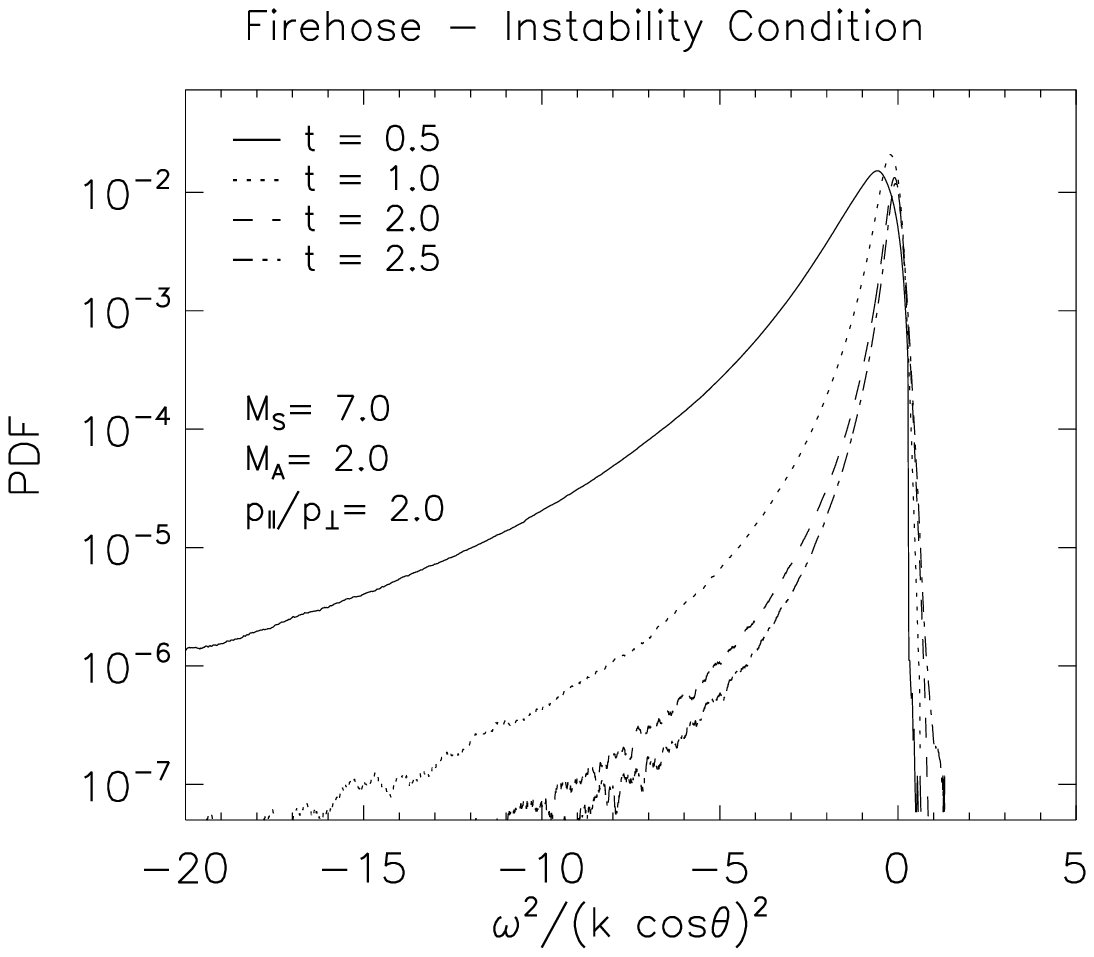}
 \includegraphics[width=0.45\textwidth]{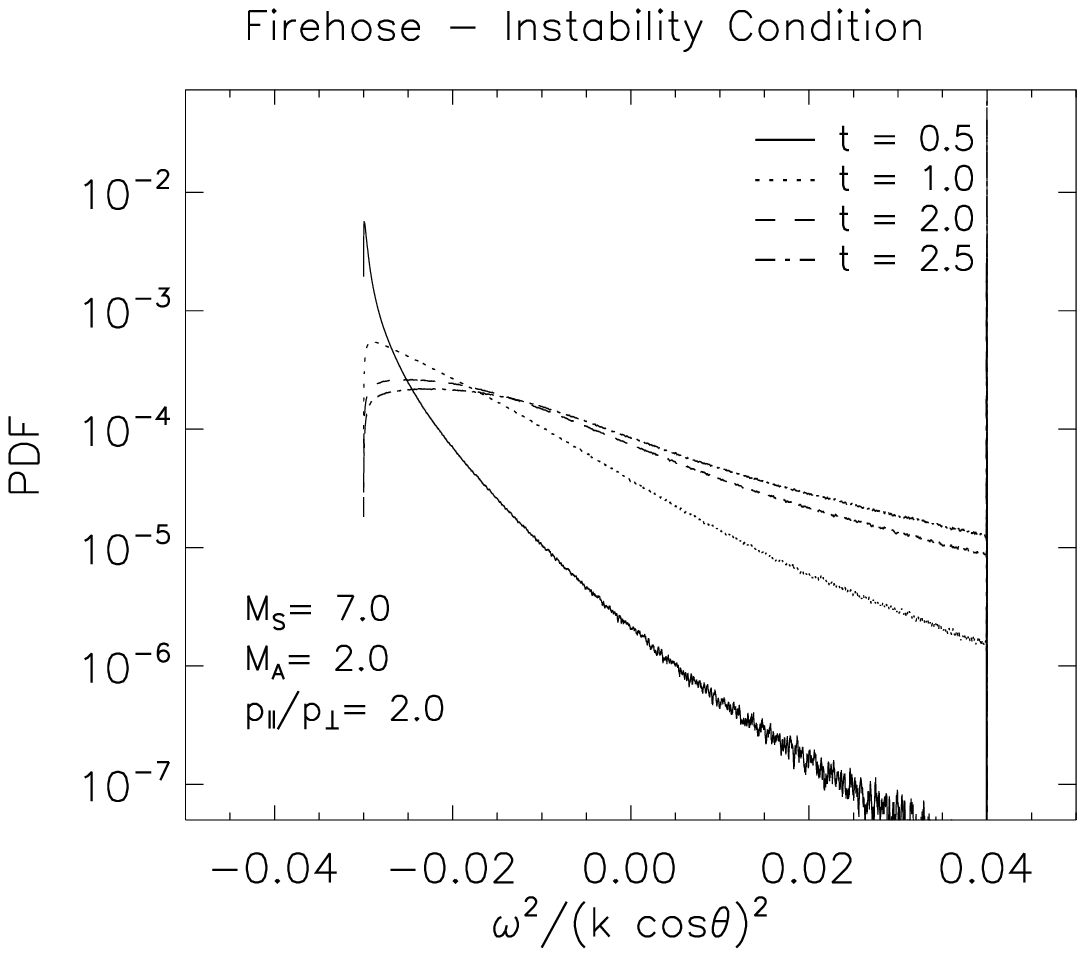}
 \caption{Probability distribution function of stability condition for incompressible Alfvenic (left) and the slow magnetosonic (right) waves. \label{fig:stab_cond}}
\end{figure*}

As a general behavior, we found that the turbulence increases the range of
unstable cells, even for the case with initially stable cells (Model 6).  This
process is illustrated in Figure~\ref{fig:stab_cond} where we show the time
evolution of the dispersion relation of Model 4 (similar distributions were
obtained for all models).  Here, we plot the histograms of the stability
condition for different times.  The Alfv\'en and magnetosonic modes are
independently shown in the left and right plots, respectively.  Unstable cells
populate the negative range of the dispersion relation.  The time is shown in
units of the dynamical timescale $\tau_d \sim \delta V/L$.

As the turbulence is injected in the simulation domain, fluctuations of density
and magnetic field change the local characteristic speeds and, as a consequence,
the stability conditions.  This process is fast compared to the dynamical
timescale of the system ($t < 0.5 \tau_d$).  As a result, the dispersion
relation spread over a larger range including the negative values of
$(\omega/k)^2$.  Then, the instabilities start to grow and saturate, at
different timescales for different scales.  The saturation of the instability
brings the cells towards positive values of $(\omega/k)^2$.  At $t \sim 2.5$,
most of cells have already reached the saturation condition for both modes.  We
believe that the few still unstable conditions are related to the large scale
fluctuations, which evolve slow and has higher saturation values.  It also seems
that the firehose instability of Alfv\'en modes saturates faster than the
magnetosonic mode, though the differences between the modes could be also be
related to the driving mechanism.  We plan to address this possibility in a
future work.

In the present calculations we did not perform any wave mode decomposition and,
therefore we were unable to analyze the time evolution of the individual modes.
Also, for this reason we are unable to characterize the time evolution for each
wavelength (scale).  Needless to say that this is an interesting subject for
future studies on CGL-MHD turbulence.

\section{Conclusions}
\label{sec:conclusions}

In this work we presented the first extensive statistical analysis of
three-dimensional simulations of turbulence in collisionless plasma.  We studied
the case of double-isothermal closure of the CGL-MHD equations in order to
compare our results with previous isothermal simulations of MHD turbulence.  We
performed simulations with different characteristic sonic and Alfvenic Mach
numbers, as well as different pressure anisotropies to account for both firehose
and mirror instabilities.  As main results we showed that:
\begin{itemize}
\item we obtained firehose/mirror unstable conditions in all simulations.  The
unstable conditions may be created, even for stable initial conditions, as a
consequence of the driving and evolutions of turbulent cascade;

\item the supersonic and super-Alfvenic models showed no significant differences
when compared to standard MHD models.  Basically, strong turbulence is able to
destroy the changes in the topology resulted from the instabilities;

\item the PDF's of density showed broadened profiles for the subsonic and
sub-Alfvenic cases.  The PDF's of velocity showed changes for sub-Alfvenic
models.  Specifically, we obtained an increase on the number high velocity flows
in subsonic models, while its decrease for supersonic turbulence;

\item the spectra of density and velocity showed an increase of power in small
scales for subsonic models;

\item the structure functions of velocity and density fluctuations revealed that
the firehose instability tends to isotropize the fluctuations regarding the
local reference frame, i.e. along the magnetic field lines.  On the other hand,
the mirror instability increases the elongation of the fluctuations along the
magnetic field lines;

\item the dynamical timescale ($\tau_d \sim \delta V/L$), may also be a good
estimate for the saturation timescale of the instability growth of most of the
wavelengths.  This fast evolution of the system implies in interesting physical
processes in, e.g. interchange of energy and acceleration of cosmic rays in the
collisionless plasma at intracluster medium of galaxies.  The growth rate of
long wavelengths may be much larger than $\tau_d$.
\end{itemize}

\ack

G.K., D.F.G. and A.L. thank the financial support of the NSF (No.\ AST0307869),
the Center for Magnetic Self-Organization in Astrophysical and Laboratory
Plasmas and the Brazilian agencies FAPESP (No.\ 06/57824-1 and 07/50065-0) and
CAPES (No.\ 4141067).

\appendix

\section{Linearization of the Double-Isothermal CGL-MHD Equations}

In this appendix we derive the dispersion relation for the double-isothermal
CGL-MHD equations.  We start from the isotropic case $a_\parallel^2 =
a_\perp^2$, since it is the case of ideal MHD, and then extend the analysis
including the anisotropy term by introducing the pressure tensor.  In this way
we can distinguish the role of the anisotropy pressure directly.

The ideal double-isothermal CGL-MHD equations with the pressure isotropy
assumption $p_\parallel \equiv a_\parallel^2 \rho = p_\perp \equiv a_\perp^2
\rho$ results in $\mathsf{P} = a_\perp^2 \rho I$ and can be written in the
non-conservative form as follows
\begin{eqnarray}
 \pder{\rho}{t} + \vc{v} \cdot \nabla \rho + \rho \nabla \cdot \vc{v} & = & 0, \label{eq:continuity} \\
 \rho \pder{\vc{v}}{t} + \rho \vc{v} \cdot \nabla \vc{v} + a_\perp^2 \nabla \rho - \frac{1}{4 \pi} \left( \nabla \times \vc{B} \right) \times \vc{B} & = & 0, \label{eq:motion} \\
 \pder{\vc{B}}{t} - \nabla \times \left( \vc{v} \times \vc{B} \right) & = & 0, \label{eq:induction}
\end{eqnarray}
where $\rho$ is the density, $\vc{v}$ is the velocity field, $\vc{B}$ is the
magnetic field, and $a_\perp$ and $a_\parallel$ are the sound speeds in the
perpendicular and parallel directions with respect to $\vc{B}$, respectively.

We assume that all variables can be separated into the uniform and fluctuating
components, i.e., $\rho \rightarrow \rho_0 + \delta \rho$, $\vc{v} \rightarrow
\vc{v}_0 + \delta \vc{v}$, $\vc{B} \rightarrow \vc{B}_0 + \delta \vc{B}$.  We
also assume the lack of uniform flow, i.e., $\vc{v}_0 = 0$.  Substituting the
variable separation in equations (\ref{eq:continuity})-(\ref{eq:induction}) and
removing all non-linear terms leads to the following set of equations
\begin{eqnarray}
 \pder{\delta \rho}{t} + \rho_0 \nabla \cdot \delta \vc{v} & = & 0, \label{eq:continuity_linear} \\
 \rho_0 \pder{\delta \vc{v}}{t} + a_\perp^2 \nabla \delta \rho - \frac{1}{4 \pi} \left( \nabla \times \delta \vc{B} \right) \times \vc{B}_0 & = & 0, \label{eq:motion_linear} \\
 \pder{\delta \vc{B}}{t} - \nabla \times \left( \delta \vc{v} \times \vc{B}_0 \right) & = & 0, \label{eq:induction_linear}
\end{eqnarray}

Introducing the variable perturbation of the form of $\delta f(\bf{x},t) \propto
\exp \left[ i \left( \bf{k} \cdot \bf{x} - \omega t  \right) \right]$ results in
the above set of equations represented in the Fourier space
\begin{eqnarray}
 - \omega \delta \rho + \rho_0 \left( \vc{k} \cdot \delta \vc{v} \right) & = & 0, \label{eq:continuity_fourier} \\
 - \omega \rho_0 \delta \vc{v} + a_\perp^2 \delta \rho \vc{k} - \frac{1}{4 \pi} \left( \vc{k} \times \delta \vc{B} \right) \times \vc{B}_0 & = & 0, \label{eq:motion_fourier} \\
 - \omega \delta \vc{B} - \vc{k} \times \left( \delta \vc{v} \times \vc{B}_0 \right) & = & 0, \label{eq:induction_fourier}
\end{eqnarray}

Multiplying equation~(\ref{eq:motion_fourier}) by $\omega$, dividing by $\rho_0$
and substituting equations~(\ref{eq:continuity_fourier}) and
(\ref{eq:induction_fourier}) we obtain the dispersion relation
\begin{eqnarray}
- \omega^2 \delta \vc{v} + & a_\perp^2 \left( \vc{k} \cdot \delta \vc{v} \right) \vc{k} + \frac{1}{4 \pi \rho_0} \left\{ \left[ \vc{k} \left( \vc{B}_0 \cdot \vc{B}_0 \right) - \vc{B}_0 \left( \vc{B}_0 \cdot \vc{k} \right) \right] \left( \vc{k} \cdot \delta \vc{v} \right) \right. \nonumber \\ & - \left. \vc{k} \left( \vc{k} \cdot \vc{B}_0 \right) \left( \vc{B}_0 \cdot \delta \vc{v} \right) + \left( \vc{k} \cdot \vc{B}_0 \right) \left( \vc{B}_0 \cdot \vc{k} \right) \delta \vc{v} \right\} = 0, \label{eq:disspersion}
\end{eqnarray}

Without loosing the generality we can assume that the mean magnetic field is
parallel to the X direction, i.e. $\vc{B}_0 = B_0 \hat{x}$, and $\vc{k}$ lays in
the XY-plane under an angle $\theta$ with the respect to $\vc{B}_0$, i.e.
$\vc{k} = k \left( \cos \theta \hat{x} + \sin \theta \hat{y} \right)$.  We also
introduce the Alfv\'en speed $c_A \equiv B_0 / \sqrt{4 \pi \rho_0}$.
Substituting these assumptions and diving the dispersion relation by $k^2$ we
can express it in a matrix form
\begin{equation}
 \fl \small
 \mathsf{A} \ \delta \vc{v} =
 \left(
 \begin{array}{ccc}
 - \frac{\omega^2}{k^2} + a_\perp^2 \cos^2 \theta & a_\perp^2 \sin \theta \cos \theta & 0 \\
 a_\perp^2 \sin \theta \cos \theta & - \frac{\omega^2}{k^2} + a_\perp^2 \sin^2 \theta + c_A^2 & 0 \\
 0 & 0 & - \frac{\omega^2}{k^2} + c_A^2 \cos^2 \theta
 \end{array}
 \right)
 \left(
 \begin{array}{c}
   \delta v_x \\
   \delta v_y \\
   \delta v_z
 \end{array}
 \right) = 0.
\end{equation}
Determinant of the matrix $\mathsf{A}$ gives the dispersion relation
\begin{equation}
 \fl \qquad
 \det \mathsf{A} = \left( - \frac{\omega^2}{k^2} + c_A^2 \cos^2 \theta \right) \left[ \frac{\omega^4}{k^4} - \frac{\omega^2}{k^2} \left( a_\perp^2 + c_A^2 \right) + c_A^2 a_\perp^2 \cos^2 \theta \right] = 0,
\end{equation}
with the eigenvalues
\begin{equation}
 \left( \frac{\omega^2}{k^2} \right)_A = c_A^2 \cos^2 \theta,
\end{equation}
corresponding to the Alfv\'en wave and
\begin{equation}
 \left( \frac{\omega^2}{k^2} \right)_{f,s} = \frac{1}{2} \left[ a_\perp^2 + c_A^2 \pm \sqrt{\left( a_\perp^2 + c_A^2 \right)^2 - 4 a_\perp^2 c_A^2 \cos^2 \theta} \right]
\end{equation}
corresponding the fast and slow magnetosonic waves, respectively.

In the next step we include the pressure anisotropy term in the motion equation (\ref{eq:motion}) which for the double-isothermal approximation can be written as
\begin{eqnarray}
 \nabla \cdot & \left[ \left( p_\parallel - p_\perp \right) \hat{b} \hat{b} \right] = \left( a_\parallel^2 - a_\perp^2 \right) \nabla \cdot \left( \rho \hat{b} \hat{b} \right) = \nonumber \\ & \left( a_\parallel^2 - a_\perp^2 \right) \left\{ \hat{b} \left( \hat{b} \cdot \nabla \right) \rho + \frac{\rho}{B} \left( \hat{b} \cdot \nabla \right) \vc{B} - \frac{2 \rho}{B} \left[ \left( \hat{b} \cdot \nabla \vc{B} \right) \cdot \hat{b} \right] \hat{b} \right\} . \label{eq:anisotropy}
\end{eqnarray}
Substituting the variables separation in this term we obtain
\begin{eqnarray}
 \fl \quad \nabla \cdot \delta \left[ \left( p_\parallel - p_\perp \right) \hat{b} \hat{b} \right] = \nonumber \\ \fl \qquad \left( a_\parallel^2 - a_\perp^2 \right) \left\{ \hat{b}_0 \left( \hat{b}_0 \cdot \nabla \right) \delta \rho + \frac{\rho_0}{B_0} \left( \hat{b}_0 \cdot \nabla \right) \delta \vc{B} - \frac{2 \rho_0}{B_0} \left[ \left( \hat{b}_0 \cdot \nabla \delta \vc{B} \right) \cdot \hat{b}_0 \right] \hat{b}_0 \right\} , \label{eq:anisotropy_linear}
\end{eqnarray}
where $\hat{b}_0 \equiv \vc{B}_0 / B_0$, and the first term corresponds to the
perturbation of the density, and two remaining terms correspond to the
perturbation of magnetic field.  Introducing the perturbed variables of the form
as before, i.e. $\delta f(\bf{x},t) \propto \exp \left[ i \left( \bf{k} \cdot
\bf{x} - \omega t  \right) \right]$, the term (\ref{eq:anisotropy_linear}) can
be written in Fourier space as follows,
\begin{equation}
 \fl \qquad
 \vc{k} \cdot \delta \left[ \left( p_\parallel - p_\perp \right) \hat{b} \hat{b} \right] = \left( a_\parallel^2 - a_\perp^2 \right) \left( \hat{b}_0 \cdot \vc{k} \right) \left\{ \hat{b}_0 \delta \rho + \frac{\rho_0}{B_0} \delta \vc{B} - \frac{2 \rho_0}{B_0} \left( \hat{b}_0 \cdot \delta \vc{B} \right) \hat{b}_0 \right\} .
\end{equation}
Next, we multiply it by $\omega$ and substitute equations
(\ref{eq:continuity_linear}) and (\ref{eq:induction_linear}) to obtain simpler
relation
\begin{equation}
 \omega \vc{k} \cdot \delta \left[ \left( p_\parallel - p_\perp \right) \hat{b} \hat{b} \right] = \rho_0 \left( a_\parallel^2 - a_\perp^2 \right) \left( \vc{k} \cdot \hat{b}_0 \right)^2 \left[ 2 \hat{b}_0 \left( \hat{b}_0 \cdot \delta \vc{v} \right) - \delta \vc{v} \right] .
\end{equation}
Finally, assuming $\vc{B}_0 = B_0 \hat{x}$ and $\vc{k} = k \left( \cos \theta \hat{x} + \sin \theta \hat{y} \right)$ the term reduces to a very simple expression
\begin{equation}
 \frac{\omega}{k^2} \frac{1}{\rho_0} \vc{k} \cdot \delta \left[ \left( p_\parallel - p_\perp \right) \hat{b} \hat{b} \right] = \left( 2 \delta v_x \hat{b}_0 - \delta \vc{v} \right) \left( a_\parallel^2 - a_\perp^2 \right) \cos^2 \theta .
\end{equation}
Rewriting this term in a matrix form we obtain the contribution to the matrix
$\mathsf{A}$ resulting from the presence of pressure anisotropy
\begin{equation}
 \fl \small
 \Delta \mathsf{A} \ \delta \vc{v} =
 \left(
 \begin{array}{ccc}
 \left( a_\parallel^2 - a_\perp^2 \right) \cos^2 \theta & 0 & 0 \\
 0 & - \left( a_\parallel^2 - a_\perp^2 \right) \cos^2 \theta & 0 \\
 0 & 0 & - \left( a_\parallel^2 - a_\perp^2 \right) \cos^2 \theta
 \end{array}
 \right)
 \left(
 \begin{array}{c}
   \delta v_x \\
   \delta v_y \\
   \delta v_z
 \end{array}
 \right).
\end{equation}
The matrix $\mathsf{A} + \Delta \mathsf{A}$ takes form
\begin{equation}
 \fl \small \quad
 \left(
 \begin{array}{ccc}
 - \frac{\omega^2}{k^2} + a_\parallel^2 \cos^2 \theta & a_\perp^2 \sin \theta \cos \theta & 0 \\
 a_\perp^2 \sin \theta \cos \theta & - \frac{\omega^2}{k^2} + c_A^2 + a_\perp^2 - a_\parallel^2 \cos^2 \theta & 0 \\
 0 & 0 & - \frac{\omega^2}{k^2} + \left[ c_A^2 - \left( a_\parallel^2 - a_\perp^2 \right) \right] \cos^2 \theta
 \end{array}
 \right),
\end{equation}
and its determinant gives the dispersion relation for the CGL-MHD equations
\begin{eqnarray}
 \fl \small
 \det \mathsf{A} = \left\{ - \frac{\omega^2}{k^2} + \left[ c_A^2 - \left( a_\parallel^2 - a_\perp^2 \right) \right] \cos^2 \theta \right\} \cdot \nonumber \\ \fl \qquad \left\{ \frac{\omega^4}{k^4} - \frac{\omega^2}{k^2} \left( c_A^2 + a_\perp^2 \right) + \left[ a_\parallel^2 \left( c_A^2 + a_\perp^2 - a_\parallel^2 \cos^2 \theta \right) - a_\perp^4 \sin^2 \theta \right] \cos^2 \theta \right\} = 0,
\end{eqnarray}
with the eigenvalues
\begin{equation}
 \left( \frac{\omega^2}{k^2} \right)_A = \left[ c_A^2 - \left( a_\parallel^2 - a_\perp^2 \right) \right] \cos^2 \theta, \label{eq:disspersion_alfven}
\end{equation}
corresponding to the Alfv\'en wave and
\begin{equation}
 \left( \frac{\omega^2}{k^2} \right)_{f,s} = \frac{1}{2} \left\{ b^2 \pm \sqrt{b^4 - 4 \left[ a_\parallel^2 \left( b^2 - a_\parallel^2 \cos^2 \theta \right) - a_\perp^4 \sin^2 \theta \right] \cos^2 \theta} \right\}, \label{eq:disspertion_fast}
\end{equation}
corresponding the fast and slow magnetosonic waves, respectively. where $b^2
\equiv a_\perp^2 + c_A^2$.  The derived dispersion relations determine the
stability conditions for all characteristic waves and growth rates for the
firehose and mirror instabilities.

\References
\item[] Chew G~F, Goldberger M~L and Low F~E 1956 {\it Royal Soc. of London Proc. Ser. A} {\bf 236} p~112
\item[] Einfeldt B, Munz C~D, Roe P~L and Sj\"ogreen B 1991 {\it J. Comput. Phys.} {\bf 92} 273
\item[] Ensslin T~A and Vogt C 2006 {\it Astron. Astrophys. J.} {453} 447
\item[] Goldreich P and Sidhar S 1995 {\it Astrophys. J.} {\bf 438} 763
\item[] Hasegawa A 1969, {\it Phys. Fluids} {\bf 12} 2642
\item[] Hau L-N 2002 {\it Phys. Plas.} {\bf 9} 2455
\item[] Hau L-N and Sonnerup U~O 1993 {\em Geophys. Rev. Let.} {\bf 20} 1763
\item[] Hau L-N and Wang B-J 2007 {\it Nonlin. Processes Geophys.} {\bf 14} 557
\item[] Howes G~G et al. 2006 {\it Astrophys. J.} {\bf 651} 590
\item[] Kowal G, Lazarian A and Beresnyak A 2007 {\it Astrophys. J.} {\bf 658} 423
\item[] Kowal G, Lazarian A, Vishniac E~T and Otmianowska-Mazur K 2009 {\it Astrophys. J.} {\bf 700} 63
\item[] Kowal G and Lazarian A 2010 {\it Astrophys. J.} {\bf 720} 742
\item[] Passot T and Sulem P~L 2006 {\it J. Geophys. Res.} {\bf 111} 4203
\item[] Press W~H, Teukolsky S~A, Vetterling W~T, and Flannery B~P 1992 {\it Numerical recipes in C (2nd ed.): the art of scientific computing} (New York: Cambridge University Press)
\item[] Quest K~B and Shapiro V~D 1996 {\it J. Geophys. Res.} {\bf 101} 24 p~457
\item[] Schekochihin A~A et al. 2005 {\it Astrophys. J.} {\bf 629} 139
\item[] Schekochihin A~A et al. 2008 {\it Phys. Rev. Lett.} {\bf 100} 081301
\item[] Sharma P, Hammett G~W and Quataert E 2003 {\it Astrophys. J.} {\bf 596} 1121
\item[] Sharma P, Hammett G W, Quataert E and Stone J~M 2006 {\it Astrophys. J.} {\bf 637} 952
\item[] T\'oth G. 2000 {\it J. Comput. Phys.} {\bf 161} 605
\item[] Wang B-J and Hau L-N 2003 {\it J. Geophys. Res.} {\bf 108} 1463
\endrefs

\end{document}